\documentclass[manuscript]{acmart}

\usepackage{amsmath}
\usepackage{algorithmic}
\usepackage{graphicx}
\usepackage{textcomp}
\usepackage{xcolor}

\usepackage{multirow}
\usepackage{tabularx}
\usepackage{array}
\usepackage{adjustbox}
\usepackage{makecell}
\usepackage{longtable}

\usepackage{xr}
\usepackage[most]{tcolorbox}
\usepackage{enumitem}
\usepackage[normalem]{ulem}
\usepackage{etoolbox}
\usepackage{placeins}

\newcolumntype{Y}{>{\raggedright\arraybackslash}X}

\setlist{nolistsep}

\AtBeginDocument{%
  }

\setcopyright{acmlicensed}
\copyrightyear{2018}
\acmYear{2018}
\acmDOI{XXXXXXX.XXXXXXX}

\acmJournal{JDS}
\acmVolume{37}
\acmNumber{4}
\acmArticle{111}
\acmMonth{8}

\begin{document}

\title{Towards Energy-Efficient Large Language Model Inference: A Systematic Literature Review}

\author{Sabiya Banu Masthan Ali}
\email{smasthanali@algomau.ca}
\author{Syed Muhammad Danish}
\email{syed.danish@algomau.ca}
\affiliation{%
  \institution{Algoma University}
  \city{Brampton}
  \country{Canada}
}

\author{Rupinder Kaur}
\email{rupinder.kaur.ece@torontomu.ca}
\orcid{XXX-XXXX}
\affiliation{%
  \institution{Toronto Metropolitan University}
  \country{Canada}
}
\author{Marin Litoiu}
\email{mlitoiu@yorku.ca}
\orcid{XXX-XXXX}
\affiliation{%
  \institution{York University}
  \country{Canada}
}

\author{Arghavan Asad}
\email{arghavan.asad@algomau.ca}
\orcid{XXX-XXXX}
\affiliation{%
  \institution{Algoma University}
  \city{Brampton}
  \country{Canada}
}

\renewcommand{\shortauthors}{Sabiya, Syed, Rupinder, Marin, and Arghavan}
%%
%% Article type: Research, Review, Discussion, Invited or Position
\acmArticleType{Review}
%%
%% Links to code and data
\acmCodeLink{https://github.com/borisveytsman/acmart}
\acmDataLink{htps://zenodo.org/link}
%%
%% Authors' contribution
\acmContributions{BT and GKMT designed the study; LT, VB, and AP
  conducted the experiments, BR, HC, CP and JS analyzed the results,
  JPK developed analytical predictions, all authors participated in
  writing the manuscript.}
%%
%% Sometimes the addresses are too long to fit on the page.  In this
%% case uncomment the lines below and fill them accodingly.
%%
%% \authorsaddresses{Corresponding author: Ben Trovato,
%% \href{mailto:trovato@corporation.com}{trovato@corporation.com};
%% Institute for Clarity in Documentation, P.O. Box 1212, Dublin,
%% Ohio, USA, 43017-6221}

\begin{abstract}
Large language model (LLM) inference is becoming a significant contributor to both operational cost and environmental impact in modern AI deployments. Energy consumption during inference depends on multiple factors across the pipeline, including input and output length, model design, deployment-level decisions, and hardware characteristics. This paper presents a systematic literature review (SLR) of energy-efficient and sustainable LLM inference. We analyze selected primary studies to identify techniques proposed to reduce energy consumption across different levels of the inference pipeline, including input, model, deployment, and hardware. We further examine the trade-offs between energy consumption, latency, and accuracy, highlighting inconsistencies in how these relationships are evaluated. In addition, we review the tools, datasets, and metrics used for energy measurement, covering both software-based and hardware-based approaches. This survey provides a comprehensive synthesis of existing literature, offering insights into the effectiveness of different techniques in terms of energy savings and their associated impact on accuracy. It highlights the trade-offs between energy efficiency and performance, and identifies limitations in current evaluation practices. Overall, this work aims to support future research toward energy-efficient and environmentally responsible deployment of LLMs. Our repository of relevant references is maintained on GitHub \footnote{https://github.com/sabiyabanu829/Sustainable-LLM-INference-Survey/tree/main}.
\end{abstract}


\maketitle

\section{Introduction}
Large Language Models (LLMs) have gained significant popularity due to their strong generative capabilities in tasks such as reasoning, summarization \cite{niu2023psqs,DBLP:conf/dac/0009WWZLTY24}, and text generation in natural language processing \cite{DBLP:conf/ssdbm/NiuZS025}. However, their real-world deployment is increasingly constrained by the high energy consumption of inference, which directly impacts operational costs and sustainability goals \cite{DBLP:conf/hotcarbon/ChienLNRSW23,jiang2024chatbots_lifecycle}. The environmental implications of large-scale deep learning systems have been widely recognized, motivating increased attention toward energy-efficient and sustainable AI deployment \cite{strubell2019energy}. According to a report by Amazon Web Services (AWS) \cite{hutt2019inferentia}, inference accounts for nearly 90\% of the total energy consumption over the model lifecycle. Unlike training, which occurs during development or fine-tuning, inference is continuously executed whenever users interact with LLM-based systems. This persistent demand makes inference the primary contributor to computational cost, latency, and energy usage. As a result, measuring and optimizing inference efficiency is essential for reducing both operational expenses and the environmental impact of AI systems \cite{yu2022orca}.

Most modern LLMs are based on the Transformer architecture \cite{vaswani2017attention} and generate text autoregressively through two main inference stages: prefill and decode \cite{leviathan2022speculative}. During prefill, the model processes the input prompt in parallel and builds the key--value (KV) cache \cite{agrawal2024sarathi_serve}, making this stage primarily compute-intensive. During decode, tokens are generated sequentially while reusing the KV cache. Each new token attends to previously processed tokens, increasing memory access \cite{dao2022flashattention} and bandwidth requirements \cite{shazeer2019fast}. Consequently, prefill and decode exhibit distinct computational, memory, and energy characteristics \cite{kwon2023pagedattention}. Inference energy is further influenced by input length \cite{eliopoulos2025pruning}, model size \cite{lin2026qwen}, attention mechanisms \cite{zhang2025attention}, numerical precision \cite{yuan2026understanding}, batching \cite{pang2025optimizing}, scheduling \cite{ma2025multi}, accelerator type \cite{li2024large}, and kernel efficiency \cite{hsu2024liger}. Thus, energy-efficient inference requires considering interactions among input characteristics, model design, deployment decisions, and hardware constraints.

A wide range of techniques has been proposed to reduce LLM inference energy, including input-level methods such as prompt optimization and token reduction, model-level approaches such as quantization and pruning, and deployment-level techniques such as batching, scheduling, and hardware acceleration. Although existing surveys review efficient LLM inference and related optimization techniques \cite{cheng2025survey, zhou2024survey, wan2024efficient, yuan2024llm_inference_unveiled, bai2401beyond, ding2025efficiency, xu2024survey, liu2026efficient, miao2025towards}, relatively limited attention has been given to energy-efficiency and energy consumption as primary objectives rather than general efficiency. This motivates a systematic investigation of how energy is reduced, evaluated, and balanced against performance during LLM inference.

To address this gap, we present an SLR that identifies, collects, and organizes existing research on the energy efficiency of LLM inference. Our goal is to provide a structured and comprehensive understanding of how energy consumption is analyzed and reduced in modern LLM systems. Specifically, this study examines three key aspects of the literature. First, we investigate the range of techniques proposed to reduce energy consumption during LLM inference across different levels of the pipeline, including input, model, deployment, and hardware levels. This analysis categorizes existing approaches and highlights how energy reduction strategies are applied in practice, while identifying where the LLM inference pipeline can be optimized to improve energy efficiency. Second, we analyze the trade-offs between energy consumption, latency, and accuracy. Although many studies report improvements in efficiency, the relationship between these metrics is often not consistently evaluated. This creates a significant gap in understanding how energy savings affect performance and output quality, and which techniques improve energy efficiency without degrading accuracy. Finally, we examine the tools, datasets, and metrics used to measure energy consumption in LLM inference. This analysis can help researchers and practitioners select appropriate optimization techniques and evaluation methods for developing, deploying, and assessing energy-efficient LLM inference systems, while identifying gaps that require further investigation.

To conduct this study, we follow a systematic and well-defined methodology for identifying and analyzing relevant literature. Initially, a set of candidate studies is collected through database searches and reference exploration. These studies are then filtered through multiple screening stages using clearly defined inclusion and exclusion criteria to ensure relevance and quality. For the selected papers, data is collected using structured extraction forms to maintain consistency. Finally, the analysis is used to identify key trends, major findings, and limitations within the existing body of research. In this review, \textit{sustainable LLM inference} refers to the environmental and operational aspects of inference-time energy consumption, power demand, resource utilization, and associated carbon impacts. Broader social, economic, and embodied hardware impacts are outside the scope.

\noindent\textbf{Contributions: }By synthesizing the existing body of research, this work provides a structured overview of techniques proposed to reduce energy consumption in LLM inference, along with the models and system configurations evaluated and the methodologies used for energy measurement. The analysis further highlights the trade-offs between energy consumption, latency, and accuracy, as well as the limitations present in current studies. Through this structured synthesis, key research gaps are identified, and directions for future work are outlined toward more sustainable and energy-efficient LLM inference.

%The main contributions of this work are as follows:

%\begin{itemize}
%    \item We provide a phase-aware analysis of energy consumption during LLM inference by separating the process into prefill and decode stages. We also present a structured overview of existing techniques for reducing inference energy, covering factors such as prompt optimization, model configuration, decoding strategies, memory usage, and system-level optimizations, etc., along with their practical implications.

 %   \item We analyze the trade-offs associated with these techniques in terms of energy efficiency, latency, and output quality. This supports informed decision-making for their use in real-world settings.
%
%    \item We provide insights into energy-efficient LLM inference from multiple perspectives, including model behavior, inference strategies, system-level design, and hardware utilization. We identify key research gaps and outline future directions. This analysis provides a comprehensive understanding of the challenges and opportunities in designing energy-efficient LLM inference systems.
%\end{itemize}

The remainder of this paper is organized as follows. Section 2 reviews existing surveys and related work on energy-efficient and sustainable LLM inference. Section 3 describes the SLR methodology. Section 4 synthesizes the findings to answer the research questions. Section 5 discusses the future research directions. Section 6 presents the limitation of this study. Finally, Section 7 concludes the paper by summarizing the main findings and contributions.

\vspace{-0.4cm}
\section{Related Work}

\label{sec:rw}
As a first step, we searched for existing systematic literature reviews and surveys related to LLMs, energy-efficient AI, and LLM inference optimization across major digital libraries such as Compendex, Inspec, Scopus, and Google Scholar.
We identified several recent surveys and review papers covering related areas, including compression-based methods such as quantization, pruning, and architectural simplification \cite{zhu2023survey_compression_llms, park2024compression_language_models, wang2024compression_efficient_inference_llms, tang2024transformer_compression_survey},  speculative decoding and execution acceleration \cite{xia2024speculative_decoding_survey, zhang2024speculative_execution_survey,xia2024unlocking}, LLM inference scheduling \cite{heisler2025llm}, hardware acceleration \cite{li2024large, li2024hardware, koilia2024hardware}, edge-based LLMs \cite{cai2026efficient}, LLM inference engines \cite{park2025survey, pan2025survey}, and studies on prompt optimization, decoding strategies, serving systems, and memory-efficient inference \cite{dong2022survey_in_context_learning, ning2023skeleton_of_thought, jin2024adaptive_skeleton_graph_decoding, liu2024apar, cai2024medusa, kwon2023pagedattention, zheng2023sglang, chen2023frugalgpt}. These surveys primarily focus on different aspects of LLM efficiency, often emphasizing specific techniques to improve scalability and performance. In contrast, this work adopts an end-to-end perspective, examining energy consumption across the entire LLM inference pipeline, including input, model, system, and hardware levels.

Several prior surveys have explored efficiency in LLMs from different perspectives. A number of studies provide broad taxonomies of efficient LLM techniques, organizing methods across multiple optimization dimensions \cite{cheng2025survey, zhou2024survey}. These studies categorize techniques across model, data, and system levels, offering structured overviews of approaches such as model compression, architectural optimization, and inference acceleration. More recent surveys further extend
these taxonomies by providing comprehensive overviews of resource-efficient LLMs and efficient inference techniques \cite{wan2024efficient}. In addition, other studies adopt a more holistic view of efficiency across the LLM lifecycle \cite{yuan2024llm_inference_unveiled, bai2401beyond, ding2025efficiency, xu2024survey}, analyzing multiple stages including architecture design, training, inference, and system-level optimization, often supported by frameworks or bottleneck analyses. Furthermore, some surveys focus on specialized aspects of efficiency, such as reasoning efficiency in large reasoning models and system-level serving optimizations for low-latency deployment scenarios \cite{liu2026efficient, miao2025towards}.

\noindent\textbf{Novelty of this survey: }Existing surveys provide valuable coverage of efficient LLM inference, but most focus on computational efficiency, model compression, or system-level acceleration rather than energy consumption as a central objective. As summarized in Table~\ref{tab:survey_comparison}, prior surveys typically cover only a subset of the input, model, deployment, and hardware optimization levels, with limited attention to techniques explicitly targeting inference energy reduction. They also provide limited analysis of energy--performance trade-offs and rarely synthesize the tools, datasets, and metrics used for energy evaluation. Moreover, existing surveys generally lack an energy-centered taxonomy. In contrast, this review systematically analyzes energy optimization techniques across all four levels, examines their trade-offs with latency and accuracy, synthesizes energy measurement tools, datasets, and metrics, and organizes the literature through an energy-centered taxonomy.
\begin{table}[t!]
\centering
\footnotesize
\setlength{\tabcolsep}{4pt}
\caption{Comparison of existing surveys. \checkmark~indicates explicit coverage, while -- indicates limited or no explicit coverage.}
\label{tab:survey_comparison}
\begin{tabular}{lcccccccc}
\toprule
\multirow{2}{*}{\textbf{Survey}} &
\multicolumn{4}{c}{\textbf{Optimization Levels}} &
\multirow{2}{*}{\shortstack{\textbf{Energy Optimization}\\\textbf{Techniques}}} &
\multirow{2}{*}{\shortstack{\textbf{Energy--Performance}\\\textbf{Trade-offs}}} &
\multirow{2}{*}{\shortstack{\textbf{Tools, Datasets}\\\textbf{\& Metrics}}} &
\multirow{2}{*}{\shortstack{\textbf{Energy-Centered}\\\textbf{Taxonomy}}} \\
\cmidrule(lr){2-5}
&
\textbf{Input} &
\textbf{Model} &
\textbf{System} &
\textbf{Hardware} &
&
&
\\
\midrule

\cite{cheng2025survey}
& -- & \checkmark & -- & -- & -- & -- & -- & -- \\

\cite{zhou2024survey}
& \checkmark & \checkmark & \checkmark & -- & -- & -- & -- & -- \\

\cite{wan2024efficient}
& -- & \checkmark & \checkmark & -- & -- & -- & -- & -- \\

\cite{yuan2024llm_inference_unveiled}
& -- & \checkmark & \checkmark & \checkmark & -- & -- & -- & -- \\

\cite{bai2401beyond}
& -- & \checkmark & \checkmark & -- & -- & -- & -- & -- \\

\cite{ding2025efficiency}
& -- & \checkmark & -- & -- & -- & --  & -- & -- \\

\cite{xu2024survey}
& \checkmark & \checkmark & \checkmark & -- & -- & -- & -- & -- \\

\cite{liu2026efficient}
& \checkmark & -- & -- & -- & -- & -- & -- & -- \\

\cite{miao2025towards}
& -- & \checkmark & \checkmark & -- & -- & --  & --& -- \\

\textbf{Ours}
& \checkmark & \checkmark & \checkmark & \checkmark &
\checkmark & \checkmark &  \checkmark & \checkmark\\

\bottomrule
\end{tabular}
\end{table}
%\noindent\textbf{Novelty of this survey: }Existing surveys provide valuable coverage of efficient LLM inference, but most primarily focus on computational efficiency, model compression, or system-level acceleration rather than energy consumption as a central objective. As summarized in Table~\ref{tab:survey_comparison}, prior surveys typically cover only a subset of the input, model, system, and hardware optimization levels, with limited attention to techniques explicitly targeting inference energy reduction. They also provide limited analysis of energy--performance trade-offs and rarely synthesize the tools, datasets, and metrics used for energy evaluation. Moreover, existing surveys generally lack an energy-centered taxonomy that organizes optimization techniques from a sustainability perspective. In contrast, this review systematically analyzes energy optimization techniques across all four levels of the inference pipeline, examines their trade-offs with latency and accuracy, synthesizes current energy measurement tools, datasets, and metrics, and organizes the literature through an energy-centered taxonomy. Together, these contributions provide a unified view of how inference energy is reduced, evaluated, and balanced against performance across the LLM inference pipeline.

\vspace{-0.3cm}
\section{Research Methodology}
Based on the identified gaps, we designed our SLR by following the guidelines proposed by Kitchenham \cite{kitchenham2002preliminary}. First, we developed a review protocol that defined the main components of the study, including the research questions, search strategy, search string, selected digital libraries, inclusion and exclusion criteria, iterative study selection process, and data extraction strategy for classifying the primary studies. We applied, evaluated, and refined this process over several cycles until the final protocol was established. The finalized protocol is described in the rest of this section.
\begin{table}[htbp]
\centering
\small
\caption{Research Questions}
\label{tab:rqs}
\begin{tabularx}{\columnwidth}{lX}
\toprule
\textbf{RQ} & \textbf{Description} \\
\midrule
\textbf{RQ1} & What techniques are proposed to reduce energy consumption during LLM inference, and at which levels (input, model, system, and hardware) are they applied? \\
\textbf{RQ2} & What trade-offs exist between energy consumption, latency, and accuracy in LLM inference? \\
\textbf{RQ3} & What tools, datasets, and metrics are used to measure energy consumption in LLM inference? \\
\bottomrule
\end{tabularx}
\end{table}

\vspace{-0.7cm}
\subsection{Research Questions (RQs)}
The following RQs were defined to guide the systematic literature review, as summarized in Table~\ref{tab:rqs}. \textit{RQ1} identifies techniques for reducing energy consumption during LLM inference and classifies them across input, model, deployment, and hardare levels. Its goal is to develop a structured taxonomy of optimization approaches and highlight commonly explored and underexplored areas. \textit{RQ2} examines trade-offs among energy consumption, latency, and accuracy to understand how energy optimizations affect performance and output quality. It helps identify techniques that provide favorable efficiency--performance trade-offs. \textit{RQ3} analyzes the tools, datasets, and metrics used to evaluate LLM inference energy, including software- and hardware-based measurement approaches and energy evaluation metrics.

\vspace{-0.3cm}
\subsection{Search Strategy}
\begin{figure}[t!]
    \centering
    \includegraphics[width=0.6\linewidth]{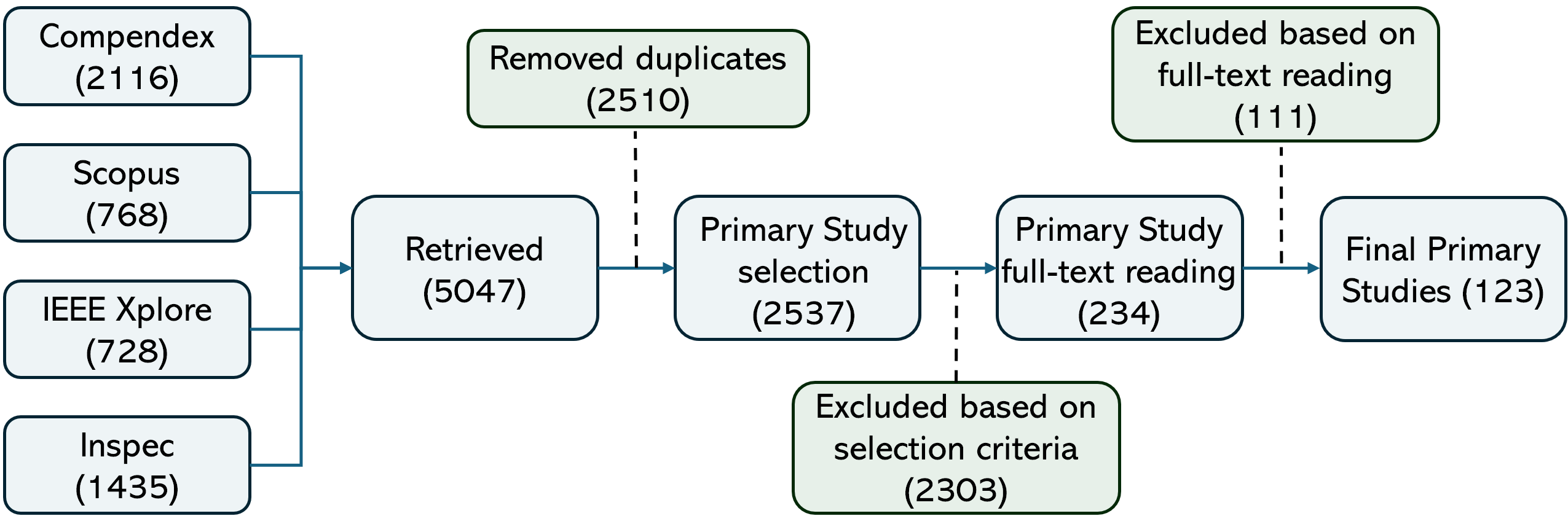}
    \caption{Overview of the study selection process}
    \label{fig:prisma}
\end{figure}

 The search process and publication counts are shown in Fig.~\ref{fig:prisma}. Guided by the RQs, we adopt an iterative search strategy that queries digital libraries using Boolean combinations of relevant keywords. The keywords are refined after each iteration to improve coverage, and the process continues until all manually identified publications are retrieved.

The search string was refined iteratively to ensure that optimization-based studies were included. The initial query mainly focused on optimization-related terms such as optimization, reduction, and mitigation. Although this approach retrieved most relevant studies, only 13 out of 15 benchmark papers were identified, indicating that some measurement-based studies were missed. The benchmark studies \cite{verma2025energy, mandusia2025green, fu2025moa, potkins2024improve, ardakani2025llmpi, patel2024splitwise, mariaraj2025smart, argerich2024measuring, everman2023carbon, li2025cim, he2024hlstransform, monteiro2026reducing, wang2025mcbp, wang2025low, du2024energy} were identified through manual searches in Google Scholar and consisted of known relevant studies in the area. These studies were then used to validate and refine the search strategy.

\begin{table*}[t]
  \centering
  \small
  \captionsetup{font=small}
  \caption{Search String}
  \begin{tabular}{p{0.99\textwidth}}
    \hline
    \begin{footnotesize}
   ( "large language model*" OR LLM OR "small language model*" OR SLM OR "foundation model*" OR "generative AI" ) AND ( inference OR serving OR decod* OR deployment OR "model execution" OR "on-device" OR "resource-constrained" OR "resource-limited" ) AND ( energy OR "energy consumption" OR "power consumption" OR "energy efficiency" OR "carbon footprint" OR sustainability OR "compute efficiency" OR "energy-intensive" OR "lifecycle consumption" OR "environmental impact" ) AND ( optimiz* OR reduction OR improvement OR minimization OR mitigation OR "energy-aware" OR "cost-aware" OR measurement OR "measurement-based" OR benchmarking OR "empirical study" OR "systematic analyses" )
   \end{footnotesize}
   \\
    \hline
  \end{tabular}
  
  \label{table:searchstring}
\end{table*}
To improve coverage, additional terms were added to capture empirical and deployment-related work. These included “on-device”, “energy-efficient”, “resource-limited”, “energy-intensive”, and “environmental impact”. These terms are commonly used in studies that examine real-world usage and energy behavior rather than proposing optimization techniques. In addition, terms such as systematic analyses were included to capture studies focused on evaluation or those using broader terminology. This refinement improved the inclusion of benchmark studies while maintaining relevant results. Overall, the final search string, shown in Table. \ref{table:searchstring} achieves a balance between recall and precision and includes both technique-based and measurement-based studies. We searched four databases: IEEE Xplore, Compendex, Inspec, and Scopus, and managed the retrieved publications using an Excel spreadsheet. The same search string was used in IEEE Xplore and Engineering Village, with Compendex and Inspec selected through database filters, while the Scopus string was adapted to its syntax while preserving the same search terms and intent.

\begin{table}[htbp]
\centering
\small
\caption{Inclusion and Exclusion Criteria}
\label{tab:inc_exc}
\begin{tabularx}{\columnwidth}{X}
\toprule
\textbf{Inclusion Criteria} \\
\midrule
\textbf{--} The study focuses on LLMs or generative models. \\
\textbf{--} The study addresses inference-time energy consumption, power usage, resource utilization, or related environmental impacts. \\
\textbf{--} The study proposes or evaluates an energy-efficient technique at the input, model, deployment, or hardware level. \\
\textbf{--} The study provides empirical evaluation or implementation results relevant to LLM inference. \\
\midrule
\textbf{Exclusion Criteria} \\
\midrule
\textbf{--} The study focuses only on training time energy or efficiency without addressing inference. \\
\textbf{--} The study does not involve LLMs, Transformers, or generative models (e.g., CNN-only or traditional ML approaches). \\
\textbf{--} The study does not evaluate or discuss energy, power, or sustainability aspects. \\
\textbf{--} The study is purely theoretical or conceptual without empirical evaluation or implementation. \\
\textbf{--} The study focuses solely on benchmarking or comparative evaluation without proposing or evaluating an energy-efficient optimization technique. \\
\textbf{--} The publication is a doctoral symposium paper, editorial, position paper, or poster. \\
\textbf{--} The study is a survey, review, or tutorial without original experimental results. \\
\textbf{--} The full text is not accessible or is not written in English. \\
\bottomrule
\end{tabularx}
\end{table}
\vspace{-0.3cm}
\subsection{Inclusion and Exclusion Criteria}
Initially, we aimed to include all relevant studies, covering optimization-based work. To ensure the quality and relevance of the selected studies, inclusion and exclusion criteria were applied iteratively throughout the screening process. The inclusion and exclusion criteria are presented in Table~\ref{tab:inc_exc}. During the review, it was observed that several studies claim efficiency improvements but do not provide empirical evaluation of energy consumption. Therefore, an exclusion criterion was added to remove studies that do not include quantitative energy evaluation. Furthermore, since this research area is relatively recent and rapidly evolving, we also included preprints from arXiv to ensure broader coverage of recent developments.

\vspace{-0.3cm}
\subsection{Study Selection Procedure}
The study selection procedure followed established SLR guidelines and PRISMA 2020 reporting recommendations \cite{kitchenham2007guidelines,page2021prisma}. Applying the search strings reported in Table~\ref{table:searchstring} retrieved 5,047 records. After removing 2,510 duplicates, 2,537 unique records remained for title and abstract screening. Before screening the complete set, three researchers conducted a pilot study to ensure a consistent interpretation of the inclusion and exclusion criteria. A random sample of 100 records was independently assessed by the reviewers, and inter-rater agreement was measured using Fleiss' Kappa \cite{fleiss1981measurement}. The first round resulted in a Kappa value of 0.378, indicating limited agreement. Most disagreements arose from different interpretations of terms such as \emph{efficient serving}, \emph{efficient LLM inference}, pruning, and other optimization-related concepts, as well as the treatment of benchmarking studies reporting latency or general efficiency without explicitly considering energy consumption. The reviewers discussed these differences and refined the inclusion and exclusion criteria accordingly.

To verify the refined criteria, a second independent random sample of 100 records, disjoint from the first sample, was assessed by the reviewers. This round produced an inter-rater agreement of 0.825, indicating strong agreement and confirming a more consistent interpretation of the selection criteria. The remaining 2,537 records were then evenly distributed between two researchers, who independently applied the refined criteria to their assigned studies. This process resulted in 234 publications being selected for full-text assessment. Inter-rater agreement was again evaluated during full-text screening using a sample of 30 studies, yielding a Fleiss' Kappa value of 0.70, indicating substantial agreement. Disagreements and uncertain cases were resolved through discussion among the reviewers, with a third experienced researcher consulted when consensus could not be reached. The remaining studies were then evenly distributed between same two researchers, with each researcher performing the full-text assessment of their assigned studies using the agreed criteria. Following full-text assessment, 123 studies satisfied the final eligibility criteria and were retained as primary studies. No study was excluded solely on the basis of methodological quality, as the objective was to provide broad coverage of an emerging research area. Instead, differences in experimental reporting and methodological completeness were considered during data extraction, evidence synthesis, and interpretation of the findings.

\subsection{Data Extraction}

After study selection, we performed data extraction using the structured form shown in Table~\ref{tab:dataform}. Similar to the study selection process, data extraction was conducted collaboratively by multiple researchers. Before extracting data from all primary studies, two researchers independently extracted information from a common sample of 10 studies using the initial extraction form. The extracted results were compared in joint discussions to identify differences in interpretation, inconsistencies in categorization, and ambiguous fields. These discussions were used to refine the extraction guidelines and establish a consistent interpretation of each field.

Once agreement on the extraction procedure was reached, the primary studies were randomly distributed among two researchers for full data extraction. To reduce errors associated with single-researcher extraction, an independent reviewer independently cross-checked 15 randomly selected samples of the extracted data for each reviewer. Any discrepancies, ambiguous cases, or unclear categorizations identified during extraction or cross-checking were discussed jointly and resolved by consensus. Only information explicitly reported in the primary studies was recorded; unavailable information was marked as \textit{Not Specified} rather than inferred. The finalized dataset was then used for data synthesis and analysis across the three RQs.

%We conducted a snowballing \cite{jalali2012systematic} search to identify additional primary studies. This involved reviewing both reference lists (backward snowballing) and citing papers (forward snowballing). Both approaches were applied to the 15 selected studies, using Google Scholar to retrieve citations. The same selection criteria were applied to all references and citations. This process identified 3 additional studies, bringing the final set to 18 primary studies. Fig. \ref{fig:protocol} illustrates how the initial 2,015 publications were reduced to these 18 studies used for data extraction.

\begin{table}[t]
\centering
\caption{Data extraction form.}
\label{tab:dataform}
\small
\begin{tabular}{p{2.2cm} p{10cm}}
\toprule
\textbf{Item} & \textbf{Extracted Information} \\
\midrule

Optimization level &
Classifies the study as input-, model-, deployment-, or hardware-level optimization. \\

Stage impact &
Identifies whether the method affects prefill, decode, or both inference stages. \\

%Model type &
%Records whether the evaluated models are SLMs, LLMs, or both. \textcolor{blue}{Can we remove model type?} \\ 

Model category &
Groups models by size: small ($\leq$7B), medium (8--30B), or large ($>$30B). \\

Model family &
Identifies the model architectures or families evaluated in the study. \\

Energy metrics &
Captures reported metrics such as energy, energy@k, energy/token, power, or emissions. \\

Measurement tool &
Identifies tools used for measurement, such as CodeCarbon, pyNVML, RAPL, or others. \\

External hardware &
Records whether external energy measurement devices were used, where applicable. \\

Dataset considered &
Captures the datasets, benchmarks, or workloads used for evaluation. \\

Energy reduction &
Records reported energy savings or reduction values, where available. \\

Trade-off reported &
Identifies trade-offs involving energy, accuracy, and latency. \\

\bottomrule
\end{tabular}
\end{table}

\subsection{Data Extraction Form}
After selecting the primary studies, we designed a data extraction form, shown in Table~\ref{tab:dataform}, to systematically collect information relevant to all three RQs. The form captures the optimization level and inference stage, model characteristics, energy metrics and measurement approaches, datasets, reported energy reductions, and trade-offs with accuracy or latency. These fields enabled consistent comparison across studies and supported the identification of optimization trends, evaluation practices, and research gaps. Table~\ref{tab:data_optimization_levels} presents the optimization-level categories used for RQ1, based on where the proposed techniques are applied: input, model, deployment, hardware, or multiple levels. The detailed energy-reduction and trade-off analyses supporting RQ2 are presented in the main paper. To improve readability, detailed RQ3 extraction results, including model characteristics, energy metrics, measurement tools, external measurement hardware, and datasets or benchmarks, are provided in the Appendix. 

\begin{table}[t]
\centering
\caption{Optimization-level categories}
\label{tab:data_optimization_levels}

\small
\renewcommand{\arraystretch}{1.05}
\setlength{\tabcolsep}{3pt}

\begin{tabular}{p{2cm}p{12cm}p{0.5cm}}
\hline

\textbf{Optimization Level} &
\textbf{Primary Studies (PS)} &
\textbf{Total} \\
\hline

Input &
\cite{masciari2026toon, verma2025energy, tanaka2024adaptive, solovyeva2026towards, podder2026green, guo2026effectiveness}
& 6 \\
\hline

Model &
\cite{shriaarthy2025eco,mandusia2025green,fu2025moa, rajput2024benchmarking,nik2026impact,qin2024mjsd, qin2025dynamic, potkins2024improve,niu2025tokenpowerbench,2025promptstopower,delavande2026understandingefficiency, ardakani2025llmpi, husom2025sustainable,shi2025systematic,hooper2025fgmp}
& 15 \\
\hline

Deployment &
\cite{walkowiak2025energy, patel2024splitwise, tian2026cache, zhang2024carbon, stojkovic2025dynamollm, mariaraj2025smart,  ziller2026greenserv, yuvarani2026switching, chen2026onewlaw, chadli2025pearl, almeida2025idle, lee2025tokens, potatmitis-etal-2025-cache-saver, wiesner2025carbon, xue2025hybridserve_icdcs, argerich2024measuring, kakolyris2025throttllem, stojkovic2024towards,moore2025slit, wheeler-etal-2025-reducing,everman2023carbon, gar2026carbonaware, wang2025storellm,  jareno2026energyefficient,ps2024inference,li2026ecothink}
& 26 \\
\hline

Hardware & 
\cite{ samsi2023words_to_watts,kakolyris2024slo,li2025cim,zhang2025energy,singh2025scalable,ortega2024pim,chen2025reis,li2025immutable,zhao2023sustainable,zhu2025leveraging,ji2026flash, wang2025decoupled, yin2025tereffic, qiao2026tellme, he2024hlstransform, jayanth2026fast}
& 16 \\
\hline

Multi-level Optimization &
\cite{ abstreiter2026sometimes,maliakel2025characterizing, he2025freesh,shi2025disaggregated,  liu2025greenllm, fernandez2025energyconsiderations, wu2026crosskv, wang2025wages, xu2026daypq, yubeaton2025huff, zhang2024mixpe, juneja2026halo, moradifirouzabadi2025end, lee2025prime, xu2025unicaim, fu2025h, wu2025kllm, wang2026harmonia, liang2026accllm, crafton2025finding, kim20254, monteiro2026reducing, wang2025mcbp, wang2025low,wang2025lad, du2024energy, xu2024llamaf, fang2025anda, kang2025dynamic, jiang2026rapid, scherer2024deeploy, fan2025ellie, feng2026ellm, xu2026spactndp, basit2026biscale, xu2025camel, appuswamy2024northpole, yang2025hades, wang2025leap, moon2025hybe,  fan2025sparse, wang2025asymserve,xu2025llmnpu, xia2025kelle, yu2025voltanallm, li2025ecoserve, lu2025thermalaware, gosmar2026prompt, li2024sprout, kalzhan2025energyefficient, hoxha2025deployment, khan2025optimizing, peng2024m2cache, raha2025llmnpu, tanaka2026aellm, belkahla2025measurements, qiao2025tellme, xie2025energyaware, hisaharo2024optimizing, nguyen2025towards} 
& 60 \\
\hline

\end{tabular}
\end{table}

\section{Results and Synthesis}

\section*{\textbf{RQ1: What techniques are proposed to reduce energy consumption
during LLM inference, and at which levels are they applied?}}

\subsection{Input Factors}

\subsubsection{\textbf{Prompt Optimization}}

The core idea of prompt compression is to reduce the length of the input prompt while preserving the quality of the generated output. By decreasing the number of input tokens, the model processes fewer tokens during the prefill stage, leading to reduced latency and lower energy consumption.
 Energy-efficient prompt engineering practices in \cite{podder2026green} demonstrated varying degrees of energy savings, with overall reductions in GHG (Green House Gas) emissions ranging from 39\% to 48\%. In \cite{verma2025energy}, the authors released a taxonomy of energy sensitive prompt elements that preserve accuracy while significantly
lowering energy cost and observed that green prompts reduce energy usage by 20–45\% at iso-accuracy and achieves up to 1.6× improvement in Tokens-per-Joule.

\subsubsection{\textbf{Output Length Control}}

Output length control aims to reduce energy consumption by limiting unnecessary token generation. One approach is \emph{format controls}, which enforce structured outputs such as JSON, XML, YAML, or fixed templates to reduce verbosity. TOON \cite{masciari2026toon} shows that structured output formats reduce token generation by 15.3--49.4\% and decoding emissions by 76.9--87.6\%.

\subsubsection{\textbf{Stop Criteria}}

Stop criteria reduce energy consumption by terminating generation once sufficient output has been produced. One approach is \emph{rule-based semantic stopping}, where generation ends when a predefined condition indicating task completion is satisfied. For example, authors in \cite{solovyeva2026towards} applies babbling suppression to code generation by executing unit tests during decoding and stopping generation once all tests pass. This approach reduces unnecessary token generation and achieves energy savings of up to 89\%.

\subsubsection{\textbf{Context Construction}}

Context construction aims to reduce inference cost by controlling the information provided to the LLM before generation. This includes selecting relevant context, retrieving external knowledge when necessary, reusing previous responses, and limiting unnecessary tool usage. 
\paragraph{Token Pruning}
Token pruning reduces computation by removing less important input tokens before they are processed by later transformer layers. This reduces computational cost, memory usage, and latency, especially for long sequences. ANTS \cite{tanaka2024adaptive} dynamically estimates token importance and retains only the most relevant tokens for further processing. By reducing the number of processed tokens, ANTS  consumed significantly less energy across multiple tasks, leading to an average reduction of
19.6\% in energy usage compared to the baseline model.

%\begin{figure}[t!]
%    \centering
%    \includegraphics[width=\columnwidth]{Figures/Prompt-compression.jpg}
%     \caption{Effect of prompt compression on inference energy through input token reduction during the prefill phase.}
%    \label{fig:input_factors_overview}
%\end{figure}

\paragraph{RAG Activation and Retrieval Control}

Retrieval scope and depth can be controlled through \emph{retrieval capping}, which limits the number of retrieved passages by selecting only the top-$k$ most relevant results and discarding the rest. In \cite{guo2026effectiveness}, the authors report that increasing the similarity threshold to 0.78 reduces energy by 20.7\%, and to 0.88 reduces energy by 51.0\% with accuracy trade-offs at aggressive thresholds.

\subsection{Model Factors}
\subsubsection{\textbf {Model Size and Architecture Controls}}

\paragraph{Architecture Choice (Dense/MoE, Layer/Width Variants)}

The goal of selecting an efficient architecture is to meet the required quality and minimize computation and memory. In a Mixture-of-Experts (MoE) model, the work per token can be reduced because only a small subset of experts is activated for each token. TokenPowerBench~\cite{niu2025tokenpowerbench} reported that Mixtral-8\texttimes7B (MoE) uses roughly similar energy per token as a dense 8B, yet achieves quality closer to a much larger dense model; importantly, sparse routing (only two experts per token) reduces energy per token by about 2--3\texttimes\ compared to dense models with similar accuracy.

\paragraph {Mixture of Attention Spans}
Mixture of Attention Spans (MoA) assigns different sliding-window attention lengths to individual attention heads and transformer layers based on their attention patterns. This improves inference efficiency while maintaining long-context modeling capabilities.
The authors in \cite{fu2025moa} show that MoA increases the effective context length by 3.9× while maintaining the same average sliding-window length. MoA improves retrieval accuracy by 1.5–7.1× over uniform sliding-window attention and reduces energy per output token by 8.7–10× with lower GPU power.

\subsubsection{\textbf{Numerical Precision and Quantization Controls}}

Numerical precision and quantization techniques reduce inference energy by lowering the precision of model computations, weights, or activations, or by improving how quantized models are stored and executed. Table~\ref{tab:quantization_controls} summarizes the main approaches and their reported benefits.

\begin{table*}[h!]
\centering
\footnotesize
\caption{Numerical Precision and Quantization Techniques for Energy-Efficient LLM Inference}
\label{tab:quantization_controls}

\begin{tabular}{p{3.0cm} p{5.5cm} p{5.7cm}}
\toprule
\textbf{Technique} & \textbf{Mechanism} & \textbf{Example Reported Findings} \\
\midrule

Mixed-Precision Inference (FP16/BF16)
&
Uses lower-precision floating-point formats instead of FP32 for most computations, reducing computational and memory overhead.
&
Reduces GPU energy consumption by up to 4$\times$ during prefill, while decode energy per token remains largely unchanged due to memory-bound execution \cite{delavande2026understandingefficiency}. \\

\midrule

Post-Training Quantization (PTQ)
&
Quantizes model weights and, in some cases, activations after training using formats such as INT8, INT4, weight-only, or W8A8.
&
Post-training quantization (q3/q4) reduces inference energy per token by up to 79\% compared with FP16 on a Raspberry Pi \cite{husom2025sustainable}. \\

\midrule

Activation-Aware Quantization (AWQ)
&
Uses activation patterns to identify important weights and preserve them during low-bit quantization.
&
AWQ INT4 and related quantized variants consume approximately 2--2.5$\times$ less energy per prompt than unquantized LLaMA 3.1 8B at the same output length \cite{2025promptstopower}. \\

\midrule

Hybrid / Mixed-Precision Quantization
&
Applies lower precision to most model components while retaining higher precision for sensitive layers or blocks.
&
QServe W4A8KV4 achieves up to 30\% lower energy per token \cite{shi2025systematic}. FGMP reduces inference energy by 14\% and weight memory by 30\% with less than 1\% perplexity degradation \cite{hooper2025fgmp}. \\

\midrule

Adaptive Quantization
&
Dynamically selects numerical precision during inference according to input complexity or energy conditions.
&
Adaptive selection among INT4, INT8, FP16, and FP32 achieves up to 59\% energy savings \cite{mandusia2025green}. \\

\midrule

Model Representation (GGML/GGUF)
&
Uses optimized representations for storing and executing quantized models, reducing memory traffic and improving runtime efficiency.
&
GGML and GGUF consume 308~mWh and 318~mWh, respectively. GGUF achieves 342.11 tokens/s and 3.42 tokens/mWh \cite{rajput2024benchmarking}. \\

\bottomrule
\end{tabular}
\end{table*}

\subsubsection{\textbf{Activation Pruning / Freezing}}

These techniques reduce inference cost by avoiding unnecessary activation computations inside the model. In activation pruning, low-magnitude activations are masked or skipped because they contribute little to the final output. In activation freezing, activations that are close to zero or change minimally are reused instead of being recomputed. Eco-Transformers \cite{shriaarthy2025eco} combines dynamic weight masking and activation freezing to improve transformer efficiency without modifying the model architecture. This approach reduced CO$_2$ emissions by up to 56.1\% and inference latency by upto 47.0\%, while maintaining output quality.

\subsubsection{\textbf{Token and Layer Reduction Controls}}

\paragraph{Dynamic Token Compression (DTC) and Adaptive Layer Pruning(ALP)} Dynamic token compression compresses the redundant or low-entropy tokens before inference, while ALP skips unnecessary layers based on input complexity.
Potkins et al. \cite{potkins2024improve} propose Dynamic Token Compression (DTC) and ALP to reduce redundant token processing and selectively skip unnecessary layers during inference. This approach reduces power consumption by 38.0\% and inference time by 36.6\%, while improving perplexity, BLEU score, and hallucination rate.

\subsubsection{\textbf{Decoding Strategy Controls}}

Decoding strategy controls influence how tokens are selected during generation and can significantly affect inference cost. 

\paragraph{Speculative Decoding (Draft-and-Verify)}

Speculative decoding improves efficiency by using a smaller draft model to propose multiple tokens, which are then verified by a larger target model. Authors in \cite{qin2024mjsd} propose a joint draft-and-verify approach that accepts multiple draft tokens jointly to reduce expensive large-model decoding iterations and achieves 23.6\% lower energy consumption than vanilla speculative decoding.

\paragraph{Sampling Policy Control (Top-k, Top-p, Temperature)}

Sampling policy control restricts the set of candidate tokens during generation to reduce decoding complexity. Techniques such as top-k, top-p (nucleus sampling), and temperature scaling balance randomness and computational cost. Authors in \cite{nik2026impact} show that sampling-based decoding strategies such as top-p can provide competitive quality with lower GPU energy than beam search in some tasks. On GSM8K, top-p reduced energy consumption from 14.44 Wh to 12.78 Wh, with only a small decrease in exact-match accuracy from 0.87 to 0.84.

\subsection{Deployment Factors}
\subsubsection{\textbf{Batching and Scheduling}}

Batching and scheduling techniques improve inference efficiency by controlling how requests are grouped and executed on available hardware resources. Table~\ref{tab:batching_scheduling} summarizes the main approaches and their reported benefits.

\begin{table*}[t]
\centering
\small
\caption{Batching and Scheduling Techniques for Energy-Efficient LLM Inference}
\label{tab:batching_scheduling}

\begin{tabular}{p{2.5cm} p{5.5cm} p{6.0cm}}
\toprule
\textbf{Technique} & \textbf{Mechanism} & \textbf{Example Reported Findings} \\
\midrule

Continuous Batching
&
Dynamically adds and removes requests between decoding steps to improve GPU utilization and reduce sequential serving overhead.
&
Text Generation Inference (TGI) reduces mean energy consumption by 12.5$\times$ compared with sequential serving and by up to 100$\times$ when combined with fixed inter-arrival shaping \cite{delavande2026understandingefficiency}. \\

\midrule

Dynamic Batching
&
Groups incoming requests into dynamically sized batches according to workload conditions, improving utilization and reducing per-request overhead.
&
A CPU-based inference engine combining dynamic batching, memory-tiled KV-cache management, and NUMA-aware scheduling achieves 18--22$\times$ higher token-generation throughput and 48.9\% lower power consumption than an A100 GPU-based setup \cite{ps2024inference}. \\

\midrule

Layered Prefill Scheduling
&
Partitions prefill computation across groups of transformer layers and interleaves prefill with decoding, reducing repeated expert-weight loading and memory traffic.
&
Layered Prefill improves TTFT, end-to-end latency, and serving efficiency for MoE models while reducing per-token energy by up to 22\% \cite{lee2025tokens}. \\

\bottomrule
\end{tabular}
\end{table*}

\subsubsection{\textbf{Elastic Provisioning Controls}}

Elastic provisioning controls reduce inference energy by dynamically adjusting serving resources (e.g., GPUs) to match workload demand. 

\paragraph{Traffic-Aware Autoscaling (Scale Up/Down)}

Traffic-aware autoscaling dynamically adjusts the number of active serving instances based on workload demand and system utilization. By scaling resources up during peak periods and down during low demand, it reduces idle energy consumption while maintaining service-level objectives. throttLL'eM \cite{kakolyris2025throttllem} employs SLO-aware autoscaling and reports up to 43.8\% lower energy consumption than NVIDIA Triton, with average savings of 24.7\%.

\begin{table*}[t!]
\centering
\small
\caption{Routing and Model Selection Techniques for Energy-Efficient LLM Inference}
\label{tab:routing_model_selection}

\begin{tabular}{p{2.8cm} p{5cm} p{6.2cm}}
\toprule
\textbf{Technique} & \textbf{Mechanism} & \textbf{Example Reported Findings} \\
\midrule

Task-Aware Model Selection
&
Routes requests according to task complexity and resource requirements, assigning simpler queries to smaller models and more difficult queries to larger models.
&
GreenServ uses dynamic context-aware routing and reduces cumulative energy consumption by 31\% while improving accuracy by 22\% \cite{ziller2026greenserv}. \\

\midrule

Confidence-Based Routing
&
Processes a query first with a smaller model and forwards it to a larger model only when confidence falls below a predefined threshold.
&
HYBRIDSERVE reports up to 19.8$\times$ lower energy consumption than state-of-the-art DNN serving systems \cite{xue2025hybridserve_icdcs}. Another approach reduces environmental cost by 39\% while improving response quality by 3.3\% \cite{wheeler-etal-2025-reducing}. \\

\midrule

Carbon-Aware Scheduling / Workload Allocation
&
Routes requests or allocates workloads according to carbon intensity, renewable-energy availability, hardware efficiency, electricity cost, and performance requirements.
&
GAR jointly considers carbon emissions, accuracy, and latency when routing queries, reducing CO$_2$ emissions by 74\% compared with always using the largest model \cite{gar2026carbonaware}. \\

\bottomrule
\end{tabular}
\end{table*}
\paragraph{Instance Scaling / Idle Instance Shutdown}

Scale-to-zero techniques reduce energy consumption by shutting down or consolidating underutilized serving instances during low-traffic periods. While this can introduce cold-start latency, it enables significant savings by minimizing idle capacity. DynamoLLM \cite{stojkovic2025dynamollm} achieves a similar effect through dynamic server consolidation, reducing service-level energy consumption by approximately 4.1\% while meeting latency SLOs.

\subsubsection{\textbf{Memory Management}}

\paragraph{KV-cache Reuse/Prefix Caching}

KV-cache reuse improves inference efficiency by reusing previously computed key-value states instead of recomputing them. Prompt caching avoids repeated prefill computation, while KV-cache sharing enables multiple requests to access the same cached states until their outputs diverge. GreenCache \cite{tian2026cache} is a carbon-aware KV-cache management framework that dynamically allocates cache resources based on workload and latency requirements. By balancing the benefits of cache reuse against SSD storage overhead, it reduces carbon emissions by an average of 15.1\% and up to 25.3\% for Llama-3 70B workloads.

\paragraph{Pre-stored Attention Matrices}

Pre-stored attention matrices reduce inference cost by reusing previously computed attention-related information instead of recomputing it for frequently occurring tokens. StoreLLM \cite{wang2025storellm} pre-stores and reuses KVQ matrices across inference requests, replacing repeated KVQ computation with storage access. This approach achieves up to 1.45$\times$ energy savings compared with state-of-the-art LLM inference systems.

\subsubsection {Semantic Caching} Semantic caching reduces the number of LLM inference calls by reusing responses to identical or semantically similar queries. Across five reasoning strategies, five benchmarks, and three LLMs, \cite{potatmitis-etal-2025-cache-saver} yields roughly 25\% cost savings and 35\% CO$_2$ reduction on average, and achieves up to \(\sim\)60\% cost/carbon reduction by eliminating redundant generations in benchmarking and ablation settings.

\subsubsection{\textbf{Routing \& Model Selection Controls}}

Routing and model selection techniques reduce inference energy by directing requests to suitable models or execution configurations based on query complexity, confidence, or carbon intensity. Table~\ref{tab:routing_model_selection} summarizes the main approaches and their reported benefits.

\subsubsection{\textbf{Hardware-Aware Scheduling and Resource Allocation}}

Hardware-aware scheduling and resource allocation improve inference efficiency by matching workloads to appropriate hardware resources and dynamically managing resource allocation based on performance and energy requirements. These approaches optimize where and how inference requests are executed across heterogeneous or shared hardware environments. Authors in \cite{almeida2025idle} show that hybrid routing across NVIDIA H100 and RTX 4090 GPUs can balance throughput, latency, cost, and sustainability under SLO constraints.

\subsubsection{\textbf{Parallelism \& Disaggregated Serving}}

Parallelism and disaggregated serving improve inference efficiency by optimizing how computation is distributed across multiple devices. 

\paragraph{Right-Sized Parallelism}

Right-sized parallelism dynamically selects the minimum number of devices required to satisfy latency requirements. Using too many devices can increase communication and synchronization overhead, leading to additional energy consumption and underutilized resources, while using too few devices may increase execution time and energy per request. DynamoLLM \cite{stojkovic2025dynamollm} dynamically switches among TP2, TP4, and TP8 configurations based on workload characteristics, reducing energy consumption by approximately 7\%, operational carbon emissions by 38\%, and customer costs by 61\% while meeting latency SLOs.

\paragraph{Disaggregated Serving}

Disaggregated serving separates different stages of LLM inference across different GPUs or machines to improve resource utilization. Since the prefill stage is usually compute-intensive and the decoding stage is more memory-intensive, assigning them to suitable hardware can reduce cost, power consumption, and latency. Splitwise \cite{patel2024splitwise} assigns prompt computation and token generation to different machines, achieving up to 1.4$\times$ higher throughput at 20\% lower cost and 2.35$\times$ higher throughput under the same cost and power budgets.

\subsection{\textbf{Hardware / Platform Power Management Controls}}

%Hardware and platform-level power management techniques reduce inference energy by controlling how computing resources operate during execution. 

\subsubsection{\textbf{Hardware Power Management}}

\paragraph{DVFS / Clock Scaling}

Dynamic Voltage and Frequency Scaling (DVFS) adjusts processor or GPU operating frequencies to match workload requirements. Since prefill is typically compute-bound while decode is more memory-bound, different frequency settings can be applied to balance energy consumption and latency. Zhang et al. \cite{zhang2025energy} showed that adjusting GPU frequency according to workload conditions can significantly improve energy efficiency by 70.83\% with performance degradation of 2.18–12.49\% compared to running the GPU at its maximum frequency. 

\paragraph{Power Capping}

Power capping limits the maximum power draw of GPUs or other accelerators during inference. By restricting peak power consumption, it can reduce energy usage and thermal overhead, although excessive power limits may increase execution time. Samsi et al. \cite{samsi2023words_to_watts} shows that reducing the power cap of NVIDIA A100 GPUs from 250W to 175W lowers inference energy by 23.21\%, while a 150W cap achieves 32.76--34.66\% energy savings at the cost of increased runtime. Zhao et al. \cite{zhao2023sustainable} demonstrated that  GPU power capping reduced LLaMA-65B inference energy by 22–24\% with only a 5–8\% performance degradation, indicating that moderate power limits can improve energy efficiency with relatively small latency overheads.

\subsubsection{\textbf{Hardware Accelerators}}

Hardware accelerators reduce inference energy through architectural specialization, improved data movement, and workload-specific execution. Table~\ref{tab:hardware_accelerators} summarizes the main accelerator types and their reported benefits.

\begin{table*}[t!]
\centering
\small
\caption{Hardware Accelerators for Energy-Efficient LLM Inference}
\label{tab:hardware_accelerators}

\begin{tabular}{p{2.2cm} p{5.1cm} p{6.7cm}}
\toprule
\textbf{Accelerator Type} & \textbf{Mechanism} & \textbf{Example Reported Findings} \\
\midrule

CIM / PIM
&
Performs computation closer to the stored data to reduce data movement between memory and processing units.
&
A DRAM-based PIM architecture achieved 21.9$\times$ higher energy efficiency and 1.3$\times$ higher throughput \cite{singh2025scalable}. Related CIM/PIM approaches also report improved energy efficiency for Transformer and generative AI workloads \cite{li2025cim,wang2025low}. \\

\midrule

In-Storage Processing (ISP)
&
Executes parts of the inference or retrieval pipeline within or near storage devices, reducing data transfers among storage, memory, and processors \cite{ji2026flash}.
&
An in-storage retrieval system for RAG achieved 13$\times$ higher performance and 55$\times$ higher energy efficiency without sacrificing quality \cite{chen2025reis}. \\

\midrule

ASIC / Dataflow Accelerators
&
Uses specialized hardware and optimized dataflows to improve tensor execution, increase data reuse, and reduce memory-access overhead.
&
A custom ASIC that embeds LLM weights directly in hardware achieved 49.6$\times$ higher energy efficiency while operating at only 1.13~W device power \cite{li2025immutable}. \\

\midrule

TPU / NPU
&
Uses specialized tensor and matrix-processing engines together with optimized memory hierarchies and workload-aware execution.
&
A CIM-MXU-based TPU approach achieved up to 27.3$\times$ lower MXU energy consumption and improved inference performance by up to 44.2\% for LLMs. \\

\midrule

FPGA
&
Uses reconfigurable dataflows, fine-grained parallelism, and workload-specific hardware structures for efficient LLM execution.
&
An FPGA accelerator for ternary LLMs achieved up to 19$\times$ higher power efficiency than Jetson Orin Nano and 8$\times$ higher power efficiency than NVIDIA A100 \cite{yin2025tereffic}. Additional FPGA-based approaches \cite{qiao2026tellme,he2024hlstransform,jayanth2026fast}  demonstrate significant energy savings while maintaining competitive inference performance. \\

\bottomrule
\end{tabular}
\end{table*}

\subsection{\textbf{Cross-Level Optimization Strategies}}

Cross-level optimization jointly optimizes multiple layers of the LLM inference stack, including input, model, deployment, and hardware. By coordinating decisions across these layers, such approaches can reduce computation, improve resource utilization, and achieve better energy efficiency than isolated optimizations. Table~\ref{tab:cross_level_optimization} summarizes the main strategies.

\begin{table*}[h!]
\centering
\footnotesize
\caption{Cross-Level Optimization Strategies for Energy-Efficient LLM Inference}
\label{tab:cross_level_optimization}

\begin{tabular}{p{2.0cm} p{6.5cm} p{5.7cm}}
\toprule
\textbf{Strategy} & \textbf{Joint Optimization} & \textbf{Example Reported Findings} \\
\midrule

Input--Deployment
&
Combines prompt optimization with runtime inference management, dynamically adapting prompt or generation strategies according to energy, carbon intensity, and response-quality requirements \cite{gosmar2026prompt,li2024sprout}.
&
SPROUT dynamically selects generation directives based on grid carbon intensity and generation quality to produce concise responses and reduce decoding-related carbon emissions \cite{li2024sprout}. \\

\midrule

Input--Model
&
Jointly reduces input information and adapts the model architecture to efficiently process the optimized input.
&
A MobileVLMv2 framework combining image-token pruning with cross-modal sliding-window attention reduces runtime energy by 51.9\% with a 3.5 percentage-point decrease in accuracy \cite{kalzhan2025energyefficient}. \\

\midrule

Model--Deployment
&
Combines model-level optimization with deployment strategies according to workload demands and available system resources \cite{feng2026ellm,khan2025optimizing,tanaka2026aellm}.
&
A framework combining model quantization with local inference reduces energy consumption and carbon emissions by up to 55\% while maintaining comparable accuracy \cite{khan2025optimizing}. \\

\midrule

Deployment--Hardware
&
Coordinates serving strategies, scheduling, request routing, and hardware controls such as frequency scaling and accelerator management
\cite{maliakel2025characterizing,kang2025dynamic,scherer2024deeploy,
liu2025greenllm,he2025freesh,wang2025wages,lu2025thermalaware,
xu2025camel,fan2025ellie,basit2026biscale,li2025ecoserve,
moon2025hybe,jiang2026rapid,wang2025leap,wang2025asymserve,
yu2025voltanallm,hoxha2025deployment,nguyen2025towards}.
&
VoltanaLLM jointly optimizes request routing and GPU frequency for disaggregated prefill--decode inference, reducing end-to-end energy by up to 36.3\% while maintaining high SLO attainment \cite{yu2025voltanallm}. \\

\midrule

Model--Deployment--Hardware
&
Jointly coordinates model execution, inference scheduling, and heterogeneous hardware allocation to reduce computation, communication overhead, and carbon emissions
\cite{shi2025disaggregated,fernandez2025energyconsiderations,
peng2024m2cache,belkahla2025measurements,xie2025energyaware,
hisaharo2024optimizing}.
&
GreenLLM combines speculative decoding, disaggregated serving, and heterogeneous GPU deployment, reducing total carbon emissions by up to 40.6\% while meeting latency SLOs \cite{shi2025disaggregated}. \\

\midrule

Model--Hardware
&
Adapts model optimization to hardware characteristics such as processing units, memory hierarchy, numerical precision, and power-management mechanisms
\cite{xu2025llmnpu,abstreiter2026sometimes,juneja2026halo,
monteiro2026reducing,fan2025sparse,raha2025llmnpu,lee2025prime,
yubeaton2025huff,zhang2024mixpe,moradifirouzabadi2025end,
kim20254,du2024energy,wang2025low,yang2025hades,
appuswamy2024northpole,fu2025h,wang2025lad,wang2025mcbp,
xu2025unicaim,liang2026accllm,wu2026crosskv,wu2025kllm,
xu2024llamaf,qiao2025tellme,xia2025kelle,wang2026harmonia,
xu2026daypq,fang2025anda,crafton2025finding,xu2026spactndp}.
&
HALO jointly optimizes quantization and hardware execution using circuit-level characteristics and DVFS, achieving 51\% energy savings and an average 270\% improvement in inference performance with minimal accuracy degradation \cite{juneja2026halo}. \\

\bottomrule
\end{tabular}
\end{table*}

\begin{tcolorbox}[
colback=gray!15,
colframe=gray!60,
boxrule=0.5pt,
arc=2mm,
left=2mm,
right=2mm,
top=1mm,
bottom=1mm
]
\textbf{Summary:} Energy-efficient LLM inference research spans input, model, deployment, and hardware levels, with many studies also adopting multi-level optimization. Among single-level approaches, deployment-level optimization is most common, followed by hardware, model, and input-level optimization. Multi-level optimization appears in 60 studies, highlighting growing interest in cross-layer approaches, while input-level optimization remains comparatively underexplored.
\end{tcolorbox}

\section*{\textbf{RQ2: What trade-offs exist between energy consumption, latency, and
accuracy in LLM inference?}}

Addressing this RQ requires comparing the trade-offs reported across the selected studies. We systematically examined how energy-saving techniques affect inference latency and model accuracy across different optimization levels.

\paragraph*{\textbf{Input-Level Trade-off Analysis}}

Table~\ref{tab:input_tradeoff} summarizes the trade-offs associated with input-level optimization techniques. Among the six reviewed studies, all reported energy savings, five (83.3\%) evaluated accuracy, and only three (50\%) reported latency. Consequently, only 3 (50\%) studies assessed energy consumption, latency, and accuracy simultaneously, highlighting that comprehensive trade-off evaluations remain limited.

Overall, the reviewed studies indicate that reducing energy consumption does not necessarily require sacrificing accuracy. Semantic stopping \cite{solovyeva2026towards} achieved energy savings of up to 89\% with only minor accuracy variation, prompt optimization \cite{verma2025energy} reduced energy consumption by 20--45\% without reported accuracy loss, and adaptive token pruning \cite{tanaka2024adaptive} reduced energy consumption by 19.6\% while improving model performance. These results suggest that eliminating redundant computation is generally more effective than reducing model capability.

Energy consumption also appears to be more strongly correlated with latency than with accuracy. Techniques that reduce the amount of processed information or generated tokens typically lower both energy consumption and response time. For example, RAG capping \cite{guo2026effectiveness} reduced energy consumption and latency by 60\% and 77\%, respectively, while adaptive token pruning achieved reductions of 19.6\% and 25\%. Although only three studies reported latency, none observed an increase in latency while reducing energy consumption.

The primary trade-off arises when energy savings are achieved by limiting information available to the model. Techniques such as semantic stopping, prompt optimization, and adaptive token pruning preserve relevant information and therefore maintain or improve accuracy. In contrast, RAG capping \cite{guo2026effectiveness} and format controls \cite{masciari2026toon} reduce computation by restricting contextual information or generation, leading to degraded output quality. Notably, larger energy savings do not necessarily imply greater accuracy loss: semantic stopping achieved the largest reported energy reduction while preserving accuracy, whereas RAG capping produced smaller energy savings but substantially reduced accuracy.

Overall, semantic stopping, prompt optimization, and adaptive token pruning provide the most favorable trade-offs between energy consumption, latency, and accuracy. Among them, adaptive token pruning represents the most favorable approach by simultaneously reducing energy consumption and latency while improving accuracy.

\begin{tcolorbox}[
colback=gray!15,
colframe=gray!60,
boxrule=0.5pt,
arc=2mm,
left=2mm,
right=2mm,
top=1mm,
bottom=1mm
]
\textbf{Summary:} Input-level optimization generally reduces energy consumption by eliminating redundant computation. Energy consumption is more strongly associated with latency than accuracy, with energy reductions typically accompanied by lower latency. Accuracy degradation primarily occurs when contextual information or generated content is restricted.
\end{tcolorbox}

\begin{table}[t]
\centering
\caption{Input-Level Trade-off Analysis}
\label{tab:input_tradeoff}
\small

\setlength{\tabcolsep}{3pt}
\begin{tabular}{p{0.55cm} p{3cm} p{1.2cm} p{3cm} p{3cm} p{0.7cm}}
\toprule
\textbf{Ref.} &
\textbf{Technique} &
\textbf{Energy} &
\textbf{Carbon} &
\textbf{Accuracy} &
\textbf{Latency} \\
\midrule

\cite{solovyeva2026towards}
& Semantic Stopping
& $\downarrow$89\%
& -
& $\pm$2\%
& - \\

\cite{guo2026effectiveness}
& RAG Capping
& $\downarrow$60\%
& -
& $\downarrow$11--71\%
& $\downarrow$77\% \\

\cite{podder2026green}
& Prompt Design
& $\downarrow$39\%
& $\downarrow$48\%
& -
& - \\

\cite{verma2025energy}
& Prompt Optimization
& $\downarrow$20--45\%
& -
& Maintained
& - \\

\cite{tanaka2024adaptive}
& Token Pruning
& $\downarrow$19.6\%
& -
& Slightly improved
& $\downarrow$25\% \\

\cite{masciari2026toon}
& Format Controls
& $\downarrow$15--49\%
& $\downarrow$77--88\%
& Decreased
& Mixed \\

\bottomrule
\end{tabular}
\end{table}

\paragraph*{\textbf{Model-Level Trade-off Analysis}}

Model-level optimization improves the sustainability of LLM inference by reducing the computational and memory requirements of the model itself. Table~\ref{tab:model_tradeoff} summarizes the reported trade-offs among energy efficiency, accuracy, and latency. Among the fifteen reviewed studies, ten (66.7\%) reported direct energy reductions, three (20.0\%) evaluated energy efficiency improvements, one (6.7\%) focused on carbon emission reduction, and one (6.7\%) reported power reduction. Consequently, all studies assessed at least one sustainability-related metric. Furthermore, twelve studies (80.0\%) evaluated accuracy, twelve (80.0\%) reported latency, and eleven (73.3\%) jointly assessed energy, accuracy, and latency, enabling comprehensive trade-off analysis.

The relationship between energy consumption and latency varies across optimization techniques. Methods that reduce computational complexity or memory traffic generally improve both energy efficiency and inference speed. For example, PTQ \cite{husom2025sustainable} reduced energy consumption by 79\% and latency by 69\%, adaptive quantization \cite{mandusia2025green} reduced energy by 59\% while improving latency, and self-pruning with activation freezing \cite{shriaarthy2025eco} reduced CO$_2$ emissions by 34.2--56.1\% together with latency by 12.1--47.0\%. In contrast, techniques that introduce additional routing or verification overhead may reduce energy at the expense of response time. MoA attention \cite{fu2025moa} achieved the largest energy reduction (8.7--10$\times$) but increased latency by 6.6--8.2$\times$, while speculative beam decoding \cite{qin2025dynamic} reduced energy by 1.8--2.5$\times$ with a latency increase of 1.5--1.9$\times$.

\begin{table}[ht!]
\centering
\caption{Model-Level Trade-off Analysis}
\label{tab:model_tradeoff}
\small

\begin{tabular}{
p{0.6cm}
p{4cm}
>{\centering\arraybackslash}p{4cm}
>{\centering\arraybackslash}p{2.5cm}
>{\centering\arraybackslash}p{2cm}}
\toprule

\textbf{Ref.} &
\textbf{Technique} &
\textbf{Energy / Carbon / Power} &
\textbf{Accuracy} &
\textbf{Latency} \\

\midrule

\multicolumn{5}{c}{\textbf{A. Energy Reduction}} \\
\midrule

\cite{fu2025moa}
& MoA Attention
& $\downarrow$8.7--10$\times$
& $<$1\% loss
& $\uparrow$6.6--8.2$\times$ \\

\cite{delavande2026understandingefficiency}
& Mixed Precision Inference
& $\downarrow$4$\times$
& --
& $\downarrow$10$\times$ \\

\cite{niu2025tokenpowerbench}
& Architecture Choice
& $\downarrow$2--3$\times$
& --
& -- \\

\cite{2025promptstopower}
& AWQ
& $\downarrow$2--2.5$\times$
& --
& -- \\

\cite{qin2025dynamic}
& Speculative Beam Decoding
& $\downarrow$1.8--2.5$\times$
& Maintained
& $\uparrow$1.5--1.9$\times$ \\

\cite{husom2025sustainable}
& PTQ
& $\downarrow$79\%
& 0.17--0.45 drop
& $\downarrow$69\% \\

\cite{mandusia2025green}
& Adaptive Quantization
& $\downarrow$59\%
& Maintained
& Improved \\

\cite{shi2025systematic}
& MP Quantization
& $\downarrow$30\%
& $\downarrow$92\%
& $\downarrow$70\% \\

\cite{qin2024mjsd}
& Speculative Decoding
& $\downarrow$23.6\%
& $\uparrow$43\%
& $\uparrow$1.26$\times$ \\

\cite{hooper2025fgmp}
& FGMP
& $\downarrow$14\%
& $<$1\% loss
& -- \\

\midrule

\multicolumn{5}{c}{\textbf{B. Carbon Reduction}} \\
\midrule

\cite{shriaarthy2025eco}
& Self-Pruning + Activation Freezing
& $\downarrow$34.2--56.1\% CO$_2$
& Maintained
& $\downarrow$12.1--47.0\% \\

\midrule

\multicolumn{5}{c}{\textbf{C. Power Reduction}} \\
\midrule

\cite{potkins2024improve}
& Token Compression + Layer Pruning
& Power $\downarrow$38.0\%
& Improved
& $\downarrow$36.6\% \\

\midrule

\multicolumn{5}{c}{\textbf{D. Energy-Efficiency Improvement}} \\
\midrule

\cite{ardakani2025llmpi}
& k-Quantization / BitNet Q1.58
& $\uparrow$4--60$\times$ energy efficiency
& Maintained
& Improved \\

\cite{rajput2024benchmarking}
& GGUF/GGML 4-bit Quantization
& 308--318 mWh; best tokens/mWh
& Minimal Loss
& 290--342 tok/s \\

\cite{nik2026impact}
& Sampling Policy Control
& 3.83--12.78 Wh
& Maintained
& 1.29--4.18 s/query \\

\bottomrule
\end{tabular}
\end{table}

Accuracy is generally preserved when optimization removes redundant computation rather than task-relevant information. Adaptive quantization \cite{mandusia2025green}, speculative beam decoding \cite{qin2025dynamic}, FGMP \cite{hooper2025fgmp}, and token compression with layer pruning maintain or even improve accuracy while reducing energy consumption. In contrast, aggressive mixed-precision quantization and excessive pruning can significantly degrade performance by compressing sensitive weights or removing important tokens, layers, or activation patterns.

Overall, the results indicate that energy consumption is more strongly associated with latency than with accuracy. Both energy and latency are primarily determined by computational complexity and memory access, whereas accuracy depends on preserving task-relevant information. Among the reviewed techniques, adaptive quantization, self-pruning with activation freezing, token compression with layer pruning, and GGUF/GGML quantization \cite{rajput2024benchmarking} demonstrate the most favorable approaches by reducing energy while maintaining or improving accuracy and latency. In contrast, MoA attention and speculative beam decoding provide substantial energy savings with little or no accuracy degradation but are less suitable for latency-sensitive applications because of their additional overhead.

\begin{tcolorbox}[
colback=gray!15,
colframe=gray!60,
boxrule=0.5pt,
arc=2mm,
left=2mm,
right=2mm,
top=1mm,
bottom=1mm
]
\textbf{Summary:} Model-level optimization generally reduces energy consumption by lowering computational complexity. Energy is more strongly associated with latency than accuracy, while accuracy is largely preserved when redundant computation is removed. However, aggressive quantization and pruning may degrade accuracy, and techniques requiring additional routing or verification stages can increase latency despite improving energy efficiency. 
\end{tcolorbox}

\begin{table*}[ht!] 
\centering 
\caption{Deployment-Level Trade-off Analysis: Energy Reduction}
\label{tab:deployment_energy} 

\small

\begin{tabular}{ 
p{0.55cm} 
p{3.3cm} 
p{4cm} 
p{2.5cm} 
p{3cm}} 
\toprule 
\textbf{Ref.} & \textbf{Technique} & \textbf{Energy} & \textbf{Accuracy} & \textbf{Latency} \\ 
\midrule 

\cite{walkowiak2025energy} 
& Continuous batching 
& $\downarrow$42$\times$ energy/token 
& -- & $\uparrow$3.6$\times$ gen. time \\ 

\cite{xue2025hybridserve_icdcs} 
& Cascade routing 
& $\downarrow$up to 19.8$\times$ 
& 90.5\%$\rightarrow$89.8\% 
& $\uparrow$1.7--8.9$\times$ throughput \\ 

\cite{moore2025slit} 
& SLIT scheduling 
& Cost $\downarrow$96\% vs Helix; $\downarrow$99\% vs Splitwise 
& Maintained 
& TTFT $\downarrow$81\%; $\downarrow$73\% \\ 

\cite{yuvarani2026switching} 
& Context routing 
& $\downarrow$67.5\% 
& 93.6\% retained 
& $\downarrow$68\% \\ 

\cite{mariaraj2025smart} 
& Dynamic routing 
& $\downarrow$62\% 
& 99.1\% 
& $\downarrow$52--55\% \\ 

\cite{kakolyris2025throttllem} 
& Autoscaling 
& $\downarrow$43.8\% 
& Maintained 
& Meets p99 SLOs \\ 

\cite{li2026ecothink}
& Complexity-aware query routing 
& $\downarrow$40.4\% avg.; up to $\downarrow$81.9\%
& 89.6--90.7\%; retained
& 148.6 tok/s ($\uparrow$55.1\% vs. Llama-3.1-8B) \\

\cite{ziller2026greenserv} 
& Task-aware routing
& $\downarrow$31\% 
& $\uparrow$22\% 
& 7.77 ms overhead \\ 

\cite{potatmitis-etal-2025-cache-saver} 
& Caching 
& $\downarrow$25\% cost 
& Maintained 
& $\downarrow$up to 6$\times$ \\

\cite{lee2025tokens} 
& Layered prefill 
& $\downarrow$up to 22\% 
& Maintained 
& TTFT $\downarrow$up to 70\%; E2E $\downarrow$41\% \\ 

\cite{chadli2025pearl} 
& Energy routing 
& Energy $\downarrow$$>$18\%; MMLU: 0.77 Wh/query; ARC-C: 0.95 Wh/query 
& MMLU 58.17\%; ARC-C 74.66\% 
& 0.20--25.5 s \\ 

\cite{stojkovic2024towards} 
& GPU scaling
& $\downarrow$up to 15\%; 1.6 GHz uses $<$80\% energy vs 2.0 GHz 
& Maintained 
& TTFT/TBT under SLOs \\ 

\cite{jareno2026energyefficient} 
& Compiler optimization 
& $\downarrow$13.5\% 1B Q4; $\downarrow$5.3\% 3B Q4; $\approx\downarrow$12\% 1B Q8 
& Maintained 
& Runtime $\downarrow$19.5\%; $\downarrow$5.3\%; $\approx\downarrow$11.5\% \\ 

\cite{stojkovic2025dynamollm} 
& Parallelism 
& $\downarrow$7\% 
& -- 
& Meets SLOs 

\\ 
\bottomrule 
\end{tabular} 
\end{table*}

\begin{table*}[t]
\centering
\caption{Deployment-Level Trade-off Analysis: Carbon, Power, and Energy-Efficiency}
\label{tab:deployment_carbon_power}
\small
\setlength{\tabcolsep}{1.0pt}
\renewcommand{\arraystretch}{1.15}

\begin{tabular}{
p{0.6cm}
p{3cm}
p{6cm}
p{2.5cm}
p{2.5cm}}
\toprule
\textbf{Ref.} &
\textbf{Technique} &
\textbf{Result} &
\textbf{Accuracy} &
\textbf{Latency} \\
\midrule

\multicolumn{5}{c}{\textbf{A. Carbon Reduction}}\\
\midrule

\cite{gar2026carbonaware}
& Carbon routing
& \makecell[l]{
$\downarrow$74\% CO$_2$;
2.2--12.0 Wh/1k tokens
}
& 0.737 macro acc.
& 703 ms \\

\cite{wheeler-etal-2025-reducing}
& Environmental-cost routing
& $\downarrow$39\% CO$_2$
& $\uparrow$3.3\% quality
& 0.0976 s \\

\cite{tian2026cache}
& KV-cache reuse
& \makecell[l]{
$\downarrow$15.1\% avg.;
up to $\downarrow$25.3\%
}
& --
& $>$90\% SLO attainment \\

\cite{wiesner2025carbon}
& Carbon scheduling
& \makecell[l]{
QoR=1: 121 MWh;
QoR=0.5: 87 MWh;\\
QoR=0: 53 MWh;
$\downarrow$up to 10\% CO$_2$
}
& Maintained
& -- \\

\cite{zhang2024carbon}
& Carbon scheduling
& \makecell[l]{
$\downarrow$1,652 lb CO$_2$/day;
per 10M requests
}
& --
& $\downarrow$8.71--38.78\% \\

\cite{almeida2025idle}
& Heterogeneous GPU
& \makecell[l]{
3.1$\times$ energy-efficient/token;
15.7 vs 48.4 gCO$_2$/M tokens
}
& --
& \makecell[l]{
$<$55 ms;
200--500 ms
} \\

\midrule
\multicolumn{5}{c}{\textbf{B. Power / Efficiency}}\\
\midrule

\cite{ps2024inference}
& Dynamic batching
& Power $\downarrow$48.9\%
& Maintained
& $\uparrow$18--22$\times$ \\

\cite{chen2026onewlaw}
& FleetOpt
& \makecell[l]{
$\approx$2.5$\times$ Tokens/Watt;
$\approx$4.25$\times$ Tokens/Watt
}
& Maintained
& P99 TTFT $\leq$500 ms \\

\cite{patel2024splitwise}
& Disaggregated serving
& \makecell[l]{
$\uparrow$2.35$\times$ throughput;
$\downarrow$20\% cost
}
& Maintained
& \makecell[l]{
$\sim$0.8\% overhead;
$\uparrow$1.4$\times$
} \\

\cite{argerich2024measuring}
& Batch tuning
& \makecell[l]{
Up to 20$\times$;
about 3$\times$
}
& --
& \makecell[l]{
Up to 50$\times$;
7$\times$; 95\%
} \\

\cite{everman2023carbon}
& CPU/GPU
& 91\% energy saving vs CPU
& STAAR score
& Up to 17$\times$ speedup \\

\cite{wang2025storellm}
& StoreLLM
& \makecell[l]{
$\downarrow$1.45$\times$;
43.82 mJ/token
}
& 96.87\% retained
& \makecell[l]{
$\uparrow$5.05\%;
$\uparrow$2.87\%
} \\

\bottomrule
\end{tabular}
\end{table*}

\begin{table*}[t!]
\centering
\caption{Hardware-Level Trade-off Analysis}
\label{tab:hardware_tradeoff}
\small

\begin{tabular}{
p{0.6cm}
p{1.5cm}
p{5.3cm}
p{2.5cm}
p{3cm}}
\toprule
\textbf{Ref.} &
\textbf{Technique} &
\textbf{Result} &
\textbf{Acc.} &
\textbf{Latency} \\
\midrule

\multicolumn{5}{c}{\textbf{A. Energy Reduction}}\\
\midrule

\cite{li2025cim}
& CIM/PIM
& \makecell[l]{
$\downarrow1.199\times10^{-3}\times$--
$3.396\times10^{-5}\times$ CPU
}
& --
& \makecell[l]{
$\downarrow0.026\times$--
$0.0035\times$ CPU
} \\

\cite{zhu2025leveraging}
& CIM-MXU TPU
& $\downarrow$27.3$\times$ MXU energy
& --
& \makecell[l]{
$\uparrow$44.2\% LLM perf.;
$\uparrow$33.8\% DiT perf.
} \\

\cite{ortega2024pim}
& CIM/PIM
& \makecell[l]{
$\downarrow$10--20$\times$ energy/token;
$\downarrow$6.9--13.4$\times$ energy/query
}
& --
& \makecell[l]{
$\uparrow$25-45\% QPS;
$\uparrow$2.23-2.75$\times$ tokens/s
} \\

\cite{kakolyris2024slo}
& SLO-DVFS
& $\downarrow$22.8--45.5\%
& Maintained
& \makecell[l]{
Maintains p95/p99
latency SLOs
} \\

\cite{samsi2023words_to_watts}
& GPU cap
& $\downarrow$23.21--34.66\%
& --
& $\uparrow$6.23--21.70\% runtime \\

\cite{zhao2023sustainable}
& Power cap
& $\downarrow$22--24\%
& --
& $\uparrow$5.2--7.4\% latency \\

\cite{ji2026flash}
& ISP
& $\downarrow$37.73\% read energy
& --
& Maintained \\

\cite{wang2025decoupled}
& Stage-DVFS
& \makecell[l]{
$\downarrow$35\% energy at low-load prefill;\\
$\downarrow<10$\% at medium/high prefill
}
& --
& $\downarrow\sim$10\% throughput \\

\midrule
\multicolumn{5}{c}{\textbf{B. Power}}\\
\midrule

\cite{he2024hlstransform}
& HLS FPGA
& \makecell[l]{
9 W avg.;
12 W max
}
& \makecell[l]{
PPL 2.9679;
+0.04\% PPL
}
& \makecell[l]{
57.11 tok/s;
17.51 ms/token
} \\

\cite{yin2025tereffic}
& Ternary FPGA
& \makecell[l]{
$\uparrow$19$\times$ power efficiency vs Jetson;\\
$\uparrow$8$\times$ vs A100
}
& --
& \makecell[l]{
16,300 tok/s;\\
727 tok/s for 2.7B
} \\

\midrule
\multicolumn{5}{c}{\textbf{C. Energy-Efficiency Improvement}}\\
\midrule

\cite{chen2025reis}
& ISP
& \makecell[l]{
$\uparrow$55--157$\times$
}
& Maintained
& \makecell[l]{
$\uparrow$13--112$\times$ speedup
} \\

\cite{li2025immutable}
& ASIC
& $\uparrow$49.6$\times$
& --
& -- \\

\cite{singh2025scalable}
& CIM/PIM
& 21.9$\times$
& --
& \makecell[l]{
$\uparrow$1.3$\times$ speedup 
} \\

\cite{qiao2026tellme}
& FPGA
& \makecell[l]{
$\uparrow$17.9$\times$ (LLaMAF);
$\uparrow$26$\times$ (MEADOW)
}
& PPL 12.79
& \makecell[l]{
25 tok/s (decode);\\
143 tok/s (prefill);
TTFT 0.45--0.96 s
} \\

\cite{jayanth2026fast}
& Sparse FPGA
& $\uparrow$4.5$\times$
& Maintained
& $\uparrow$1.2--2.5$\times $speedup \\

\cite{zhang2025energy}
& Queue DVFS
& $\uparrow$70.83\%
& --
& \makecell[l]{
$\uparrow$2.18--12.49\%
performance loss
} \\

\bottomrule
\end{tabular}
\end{table*}

\paragraph*{\textbf{Deployment-Level Trade-off Analysis}}

Tables~\ref{tab:deployment_energy} and~\ref{tab:deployment_carbon_power} summarize the trade-offs associated with deployment-level optimization techniques. Among the twenty-six reviewed studies, fourteen (53.84\%) reported direct energy reductions, six (23.07\%) reported carbon reductions, two (7.69\%) reported power reductions, and four (15.38\%) evaluated energy efficiency improvements. Consequently, all studies assessed at least one energy-related metric. In addition, twenty studies (76.92\%) evaluated accuracy, twenty-five (96.15\%) reported latency, and nineteen (73.07\%) jointly assessed energy, accuracy, and latency, providing a comprehensive basis for trade-off analysis.

Deployment-level optimization generally preserves model quality (accuracy) because it improves inference management, scheduling, and resource utilization without modifying model parameters. Techniques such as caching, autoscaling, compiler optimization, GPU scaling, layered prefill scheduling, disaggregated serving, and KV-cache reuse achieve energy gains while maintaining accuracy. For example, context-aware routing \cite{yuvarani2026switching} reduced energy consumption by 67.5\% while retaining 93.6\% response quality, dynamic routing \cite{mariaraj2025smart} reduced energy by 62\% while maintaining 99.1\% accuracy, and environmentally aware routing \cite{wheeler-etal-2025-reducing} reduced carbon emissions by 39\% while improving response quality by 3.3\%.

The relationship between energy consumption and latency is more variable. Many techniques reduce both energy and latency by improving resource utilization or eliminating redundant computation. For example, context-aware routing \cite{yuvarani2026switching} reduced energy consumption and latency by 67.5\% and 68\%, respectively, while dynamic routing \cite{mariaraj2025smart} achieved reductions of 62\% and 52--55\%. Similarly, layered prefill scheduling \cite{lee2025tokens} reduced energy by up to 22\% while improving TTFT by up to 70\%, and caching \cite{potatmitis-etal-2025-cache-saver} improved latency by up to 6$\times$. In contrast, techniques based on batching or resource consolidation may reduce energy at the expense of response time. Continuous batching \cite{walkowiak2025energy} reduced energy per token by 42$\times$ but increased generation time by 3.6$\times$, while dynamic batching \cite{ps2024inference} reduced power consumption by 48.9\% but increased latency by 18--22$\times$. These results indicate that deployment-level optimization primarily trades energy efficiency against latency, whereas accuracy is largely preserved unless requests are intentionally routed to smaller or lower-quality models.

Overall, energy consumption is strongly associated with latency than with accuracy at the deployment level. Among the reviewed techniques, context-aware routing, dynamic routing, caching, layered prefill scheduling, compiler optimization, GPU scaling, and KV-cache reuse provide the most favorable approaches by improving energy-efficiency while maintaining accuracy and acceptable latency. In contrast, continuous and dynamic batching are better suited for throughput-oriented workloads where higher latency is acceptable.

\begin{tcolorbox}[
colback=gray!15,
colframe=gray!60,
boxrule=0.5pt,
arc=2mm,
left=2mm,
right=2mm,
top=1mm,
bottom=1mm
]
\textbf{Summary:} Deployment-level optimization generally preserves model accuracy because it optimizes the serving infrastructure rather than the model. Energy consumption is more strongly associated with latency than accuracy. Routing and caching often reduce both energy and latency, whereas batching-based methods trade lower energy consumption for higher latency.
\end{tcolorbox}

\paragraph*{\textbf{Hardware-Level Trade-off Analysis}}

Hardware-level optimization improves the energy-efficiency of LLM inference through hardware-aware execution, specialized accelerators, and power management techniques that reduce energy consumption, power usage, and data movement. Table~\ref{tab:hardware_tradeoff} summarizes their impacts on energy, accuracy, and latency. Among the sixteen reviewed studies, eight (50.0\%) reported direct energy reductions, two (12.5\%) reported power-related results, and six (37.5\%) evaluated energy-efficiency improvements. In addition, five studies (31.25\%) evaluated accuracy, while fifteen (93.75\%) reported latency, throughput, or speedup. However, only five studies (31.25\%) jointly assessed energy, accuracy, and latency, limiting comprehensive trade-off analysis.

The relationship between energy consumption and latency depends on the underlying hardware optimization strategy. Specialized accelerators generally improve both energy efficiency and inference performance by executing LLM operations more efficiently. For example, CIM/PIM-based accelerators \cite{li2025cim,ortega2024pim} achieved substantial energy reductions together with lower latency, higher throughput, and improved token generation rates. Similarly, in-storage processing \cite{chen2025reis} minimized data movement, providing significant energy savings and up to $112\times$ speedup, while FPGA-, TPU-, and ASIC-based accelerators improved efficiency through hardware parallelism and customized data-paths.

In contrast, power management techniques such as DVFS and GPU power capping reduce energy by lowering processor voltage and frequency, often increasing execution time or reducing throughput. GPU power capping \cite{samsi2023words_to_watts,zhao2023sustainable} reduced energy consumption by 22--35\% while increasing runtime by 5--22\%, whereas stage-aware and queue-based DVFS \cite{wang2025decoupled,zhang2025energy} improved energy efficiency at the expense of throughput. These findings indicate that, unlike specialized accelerators, power management techniques primarily trade latency for lower energy consumption.

Accuracy degradation is uncommon because most hardware-level techniques optimize the execution platform rather than modifying model parameters. Specialized accelerators and hardware-aware execution perform the same computations more efficiently, preserving inference quality. For example, SLO-aware DVFS \cite{kakolyris2024slo}, in-storage processing \cite{chen2025reis}, and FPGA-based accelerators \cite{qiao2026tellme,he2024hlstransform,jayanth2026fast} maintained comparable model quality while substantially improving energy efficiency.

Overall, energy consumption is more strongly associated with latency than with accuracy at the hardware level. Among the reviewed techniques, CIM/PIM acceleration, CIM-MXU TPU acceleration, in-storage processing, and FPGA-based accelerators demonstrate the most favorable approaches by substantially reducing energy consumption while maintaining inference quality and, in many cases, improving performance.

\begin{tcolorbox}[
colback=gray!15,
colframe=gray!60,
boxrule=0.5pt,
arc=2mm,
left=2mm,
right=2mm,
top=1mm,
bottom=1mm
]
\textbf{Summary:} Hardware-level optimization improves energy-efficiency without modifying model parameters, so accuracy is generally preserved. Energy consumption is more strongly linked to latency than accuracy. Specialized accelerators often reduce both energy consumption and latency, whereas power management techniques typically exchange lower energy consumption for higher latency or lower throughput.
\end{tcolorbox}

\begin{table*}[t]
\centering
\caption{Input--Deployment Co-optimization Trade-off Analysis}
\label{tab:input_deployment}
\small
\setlength{\tabcolsep}{3pt}

\begin{tabular}{p{1cm} p{6.5cm} p{1.5cm} p{1.5cm} p{1.5cm} p{1.5cm}}
\toprule
\textbf{Ref.} &
\textbf{Technique} &
\textbf{Energy} &
\textbf{Carbon} &
\textbf{Accuracy} &
\textbf{Latency} \\
\midrule

\cite{gosmar2026prompt}
&
Prompt optimization + Semantic cache
&
$\downarrow$41.6\%
&
--
&
--
&

9.2$\rightarrow$3.8 s
\\

\cite{li2024sprout}
&
Prompt opt. + Runtime and carbon aware scheduling
&
--
&
$\downarrow$40--60\%
&
Maintained
&
--
\\

\bottomrule
\end{tabular}
\end{table*}

\begin{table}[t!]
\centering
\caption{Input--Model Co-optimization Trade-off Analysis}
\captionsetup{font=footnotesize}
\label{tab:input_model}
\small

\setlength{\tabcolsep}{3pt}
\renewcommand{\arraystretch}{1.12}

\begin{tabular}{p{0.7cm} p{5cm} p{1.3cm} p{1cm} p{1.2cm} p{1cm}}
\toprule
\textbf{Ref.} &
\textbf{Technique} &
\textbf{Energy} &
\textbf{Carbon} &
\textbf{Accuracy} &
\textbf{Latency} \\
\midrule

\cite{kalzhan2025energyefficient}
&
Token pruning + KV cache
&
$\downarrow$51.9\%
&
--
&
$\downarrow$3.5 pts
&
--
\\

\bottomrule
\end{tabular}
\end{table}

\begin{table*}[t!]
\centering
\caption{Model--Deployment Co-optimization Trade-off Analysis}
\label{tab:model_deployment}
\small

\setlength{\tabcolsep}{3pt}
\renewcommand{\arraystretch}{1.12}

\begin{tabular}{p{1cm} p{4.5cm} p{2.3cm} p{1.5cm} p{2.3cm} p{2cm}}
\toprule
\textbf{Ref.} &
\textbf{Technique} &
\textbf{Energy} &
\textbf{Carbon} &
\textbf{Accuracy} &
\textbf{Latency} \\
\midrule

\cite{feng2026ellm}
&
Struct.-guided distributed decoding
&
$\uparrow$9--12$\times$ efficiency
&
--
&
Maintained
&
$\downarrow$74--87.7\%
\\

\cite{khan2025optimizing}
&
INT4 quantization + Local inference
&
$\downarrow$55\%
&
$\downarrow$55\%
&
Slightly improved
&
--
\\

\cite{tanaka2026aellm}
&
Adaptive cross-layer optimization
&
$\downarrow$52\%
&
--
&
$\downarrow$0.3 pts
&
$\downarrow$48\%
\\

\bottomrule
\end{tabular}
\end{table*}

\section*{Cross-Layer Trade-off Analysis}

Since many studies combine optimization techniques across multiple levels, discussing each co-optimization category independently would result in considerable overlap and repetitive observations. Therefore, we organize the synthesis according to the primary optimization focus, input-centric, model-centric, and hardware-centric. This perspective provides a more holistic comparison of the trade-offs among energy consumption, latency, and accuracy, while illustrating how shifting optimization closer to the execution hardware influences overall inference efficiency. Among the 60 primary studies, model--hardware co-optimization was most common (30 studies, 50.0\%), followed by deployment--hardware (18, 30.0\%). Model--deployment--hardware and model--deployment appeared in 6 (10.0\%) and 3 (5.0\%) studies, respectively, while input--deployment and input--model remained largely unexplored, with only 2 (3.3\%) and 1 (1.7\%) studies.

\paragraph*{\textbf{Input-Centric Co-optimization}}
Input-centric co-optimization (Table. \ref{tab:input_deployment} \& \ref{tab:input_model}) demonstrates that the trade-off between energy, latency, and accuracy depends more on the optimization strategy than the magnitude of energy reduction. Deployment-assisted techniques, such as semantic caching and carbon-aware scheduling, consistently reduce energy or carbon emissions while maintaining accuracy and improving latency. In contrast, combining input optimization with model-level techniques, such as token pruning and KV-cache optimization, achieves larger energy savings but introduces small accuracy degradation because informative tokens may be removed. Overall, latency generally improves alongside energy reduction, whereas accuracy is affected only when optimization removes task-relevant information. It should also be noted that relatively few techniques focus on input-level co-optimization with model and deployment strategies.

\paragraph*{\textbf{Model-Centric Co-optimization}}
Model-centric co-optimization (Tables~\ref{tab:model_deployment}, \ref{tab:cross_layer_tradeoff}, \ref{tab:model_hardware_energy_power}, and \ref{tab:model_hardware_efficiency}) combines model optimization with deployment- and hardware-level techniques to jointly improve computational and execution efficiency. The reviewed studies show that reducing energy consumption is generally accompanied by lower inference latency, higher throughput, or improved energy efficiency, making execution performance the metric most consistently aligned with energy optimization. This trend is particularly evident in model--hardware co-optimization, where quantization, pruning, sparse attention, KV-cache optimization, specialized accelerators, PIM/CIM architectures, and hardware-aware data formats achieve substantial improvements in tokens/J or energy efficiency while simultaneously increasing throughput or reducing latency. In contrast, accuracy exhibits comparatively limited degradation. Most studies either maintain prediction quality or report only small losses despite substantial energy and performance improvements. For example, model--deployment co-optimization achieves up to 52\% energy reduction with only a 0.3 percentage-point accuracy decrease, while many model--hardware approaches report accuracy losses below 1--3\% or characterize the degradation as minimal. A small number of studies even report improved task accuracy after optimization. Overall, model-centric co-optimization reduces computational and memory overhead, enabling substantial energy and performance gains while largely preserving model accuracy.
\begin{table*}[t]
\centering
\caption{Model, Deployment and Hardware Co-optimization Trade-off Analysis}
\label{tab:cross_layer_tradeoff} 
\small
\setlength{\tabcolsep}{1.8pt}
\renewcommand{\arraystretch}{1.14}

\begin{tabular}{
p{0.6cm}
p{5.3cm}
>{\centering\arraybackslash}p{3.5cm}
>{\centering\arraybackslash}p{2cm}
>{\centering\arraybackslash}p{3cm}}
\toprule
\textbf{Ref.} &
\textbf{Technique} &
\textbf{Energy / Carbon / Power} &
\textbf{Accuracy} &
\textbf{Latency / Throughput} \\
\midrule

\multicolumn{5}{c}{\textbf{A. Energy Reduction}} \\
\midrule

\cite{fernandez2025energyconsiderations}
&
Serving optimization
&
$\downarrow$up to 73\%
&
--
&
(TP latency) $\downarrow$40.16--61.34\%
\\

\cite{belkahla2025measurements}
&
4-bit AWQ + MARLIN kernel
&
$\downarrow$54.4\%
&
Maintained
&
latency $\downarrow$29.4--51.9\%
\\

\cite{hisaharo2024optimizing}
&
Cluster + GPT-Neo optimization
&
$\downarrow$47.6\%
&
GLUE $\uparrow$5\%
&
$\downarrow$37.5\%
\\

\cite{xie2025energyaware}
&
Compression + quantization + scheduling
&
$\downarrow$31\%
&
Maintained
&
$\downarrow$27\%
\\

\midrule
\multicolumn{5}{c}{\textbf{B. Carbon Reduction}} \\
\midrule

\cite{peng2024m2cache}
&
Model Parallel inference + multi-level cache
&
$\downarrow$7.67$\times$
&
Minimal loss
&
Speedup $10.51\times$
\\

\cite{shi2025disaggregated}
&
Disaggregated serving + speculative decoding
&
$\downarrow$31.3--40.6\%
&
Maintained
&
Meets 90\% TTFT/TPOT SLOs
\\

\bottomrule
\end{tabular}
\end{table*}

\begin{table*}[t]
\centering
\caption{Model--Hardware Co-optimization Trade-off Analysis: Energy and Power Reduction}
\label{tab:model_hardware_energy_power}
\footnotesize
\setlength{\tabcolsep}{2pt}
\renewcommand{\arraystretch}{1.12}

\begin{tabular}{
p{0.7cm}
p{4cm}
>{\centering\arraybackslash}p{5cm}
>{\centering\arraybackslash}p{2cm}
>{\centering\arraybackslash}p{3cm}}
\toprule
\textbf{Ref.} &
\textbf{Technique} &
\textbf{Energy / Power} &
\textbf{Accuracy} &
\textbf{Latency} \\
\midrule

\multicolumn{5}{c}{\textbf{A. Energy Reduction}} \\
\midrule

\cite{xu2025llmnpu}
& Mobile NPU offloading
& Energy $\downarrow$30.7$\times$ avg.; up to $\downarrow$59.5$\times$
& $<$1\% loss
& Prefill $\uparrow$22.4$\times$ avg.; up to $\uparrow$43.6$\times$ \\

\cite{abstreiter2026sometimes}
& Q4 quantization + edge tuning
& CPU gen. $\downarrow$10--70\%; GPU prefill $\downarrow$10--70\%
& Maintained
&  Orin 15W latency $\downarrow$$>$50\% \\

\cite{juneja2026halo}
& HALO PTQ + DVFS
& Energy $\downarrow$51\%
& Minimal loss
& Perf. $\uparrow$270\%; up to $\uparrow$353\% \\

\cite{monteiro2026reducing}
& Ternary quant. + FPGA HLS
& Energy $\downarrow$51.9\%; power $\downarrow$37.3--41.0\%
& Maintained
& Wall-clock $\downarrow$23.3\% \\

\cite{fan2025sparse}
& Sparse attention + KV mapping
& Attention energy $\downarrow$19--67\%
& Maintained
& Attention latency $\downarrow$19--74\% \\

\cite{raha2025llmnpu}
& SW--HW NPU co-opt.
& SoC energy $\downarrow$14--35\%
& Minimal loss
& TTFT $\downarrow$12--28\%; TPOT $\downarrow$17--47\% \\

\cite{lee2025prime}
& XMC + PIM architecture
& Energy $\downarrow$$>$30\%
& Maintained
& 3-cycle decompression delay \\

\cite{yubeaton2025huff}
& Huffman weight compression
& Energy $\downarrow$up to 26\%
& Maintained
& Latency $\downarrow$up to 31\% \\

\cite{zhang2024mixpe}
& MixPE W4A8
& Energy $\downarrow$1.4$\times$
& Maintained
& Speedup $\uparrow$2.6$\times$ \\

\cite{moradifirouzabadi2025end}
& Runtime cache eviction
& Energy saving $\uparrow$1.52$\times$
& Minimal loss
& Speedup $\uparrow$3.62$\times$ \\

\midrule
\multicolumn{5}{c}{\textbf{B. Power Reduction}} \\
\midrule

\cite{kim20254}
& Bit-layered NUQ accelerator
& Power $\downarrow$40.7\%; 4.21 TFLOPS/W
& Maintained
& -- \\

\cite{du2024energy}
& Spike transformer accelerator
& Energy efficiency $\uparrow$11.3\%; power $\downarrow$13.7\%
& Maintained
& -- \\

\bottomrule
\end{tabular}
\end{table*}

\paragraph*{\textbf{Hardware-Centric Co-optimization}}
Hardware-centric co-optimization (Table~\ref{tab:deployment_hardware_tradeoff}) combines deployment strategies with hardware-aware scheduling, DVFS, workload routing, and specialized accelerators. It shows a strong relationship between energy reduction and execution performance, with most techniques simultaneously reducing energy and latency or improving throughput through better resource utilization and hardware acceleration. Accuracy is generally preserved because these approaches optimize execution without substantially modifying the model, while some studies also report lower carbon emissions. Overall, hardware-centric co-optimization provides a favorable balance among energy efficiency, performance, and accuracy, but often depends on specialized hardware or custom accelerators, which can limit portability across heterogeneous environments.
\begin{tcolorbox}[
colback=gray!15,
colframe=gray!60,
boxrule=0.5pt,
arc=2mm,
left=2mm,
right=2mm,
top=1mm,
bottom=1mm
]
\textbf{Summary:} Cross-layer analysis shows that optimization becomes effective as it moves closer to the execution hardware. Input-centric approaches improve energy efficiency with minimal impact on latency and accuracy, provided context information is preserved. Model-centric optimization offers the best balance between energy efficiency, latency, and accuracy by reducing computational redundancy while maintaining model capability. Hardware-centric optimization delivers the greatest sustainability gains through efficient execution and hardware-aware resource management, with deployment complexity as the primary trade-off rather than model performance.
\end{tcolorbox}

\begin{table*}[t]
\centering
\caption{Model--Hardware Co-optimization Trade-off Analysis: Energy-Efficiency Improvement}
\label{tab:model_hardware_efficiency}
\footnotesize

\renewcommand{\arraystretch}{1.12}

\begin{tabular}{
p{0.6cm}
p{3.5cm}
>{\centering\arraybackslash}p{4cm}
>{\centering\arraybackslash}p{2cm}
>{\centering\arraybackslash}p{3cm}}
\toprule
\textbf{Ref.} &
\textbf{Technique} &
\textbf{Energy / Power} &
\textbf{Acc.} &
\textbf{Latency} \\
\midrule

\multicolumn{5}{c}{\textbf{C. Energy-Efficiency Improvement}} \\
\midrule

\cite{wang2025low}
& RRAM PIM + mixed-precision MAC
& Efficiency $\uparrow$165.1$\times$ vs GPU; power $\downarrow$52\%
& Minimal Loss
& Latency $\uparrow$3.8$\times$ vs GPU \\

\cite{yang2025hades}
& HW speculative decoding
& Tokens/s/W $\uparrow$117.95$\times$ vs A100
& Maintained
& $\uparrow$6.99$\times$ vs A100 \\

\cite{appuswamy2024northpole}
& NorthPole appliance
& Efficiency $\uparrow$72.7$\times$; 42.19 tok/s/W
& Minimal loss
& 2.5$\times$ lower latency \\

\cite{fu2025h}
& Sparse attention + HB accelerator
& Efficiency $\uparrow$6.22--73.48$\times$
& 0.87\% drop
& Speedup $\uparrow$5.20--48.21$\times$ \\

\cite{wang2025lad}
& LAD + accelerator
& Attention efficiency $\uparrow$52.4$\times$; E2E $\uparrow$13.4$\times$
& 97\% ROUGE-1
& Attention $\uparrow$10.7$\times$; E2E $\uparrow$2.3$\times$ \\

\cite{wang2025mcbp}
& Bit-slice accelerator
& Efficiency $\uparrow$31.1$\times$ vs A100; power 2.395 W
& Minimal loss
& Speedup $\uparrow$9.43$\times$ vs A100 \\

\cite{xu2025unicaim}
& KV pruning + CIM design
& Efficiency $\uparrow$5.3--27$\times$
& --
& Speedup $\uparrow$4.2--16.7$\times$ \\

\cite{liang2026accllm}
& 2:4 pruning + $\Lambda$-attention + W2A8KV4 + FPGA
& Efficiency $\uparrow$4.07$\times$ vs FlightLLM; $\uparrow$24.8$\times$ vs A100
& Maintained
& Throughput $\uparrow$2.98$\times$ \\

\cite{wu2026crosskv}
& CrossKV
& Efficiency $\uparrow$6.03$\times$ vs SOTA; $\uparrow$23.43$\times$ vs A100
& 0.1--2\% loss
& Speedup $\uparrow$4.10$\times$ vs SOTA \\

\cite{wu2025kllm}
& Quantization + hardware
& Efficiency $\uparrow$1.44$\times$ avg.; up to $\uparrow$28.20$\times$
& 1.94\% drop
& Speedup $\uparrow$3.00$\times$ avg. \\

\cite{xu2024llamaf}
& W8A8 GQMV + FPGA pipeline
& Power efficiency $\uparrow$6.1$\times$
& PPL +0.57\%
& Speedup $\uparrow$14.3--15.8$\times$ \\

\cite{qiao2025tellme}
& 1.58-bit FPGA accelerator
& 5.2 tokens/J; power $<$5 W
& PPL 12.79
& Decode 25 tok/s; prefill 143 tok/s \\

\cite{xia2025kelle}
& KV-cache + eDRAM co-design
& Energy saving $\uparrow$4.5$\times$
& Negligible
& Speedup $\uparrow$3.9$\times$ \\

\cite{wang2026harmonia}
& BFP quant. + reconfigurable PE
& Efficiency $\uparrow$2.03$\times$ avg.; up to $\uparrow$3.90$\times$
& 0.3\% drop
& Speedup $\uparrow$3.08$\times$ avg. \\

\cite{xu2026daypq}
& DayPQ pruning + FP8 quant.
& Efficiency $\uparrow$3.13$\times$; total power 955 mW
& $\leq$3.8\% degradation
& Speedup $\uparrow$1.18--3.44$\times$ \\

\cite{fang2025anda}
& Activation format + bit-serial HW
& Efficiency $\uparrow$3.1$\times$; power 0.18 mW
& 0.1--1\% loss
& Speedup $\uparrow$2.4$\times$ \\

\cite{crafton2025finding}
& Low-precision MAC formats
& TOPS/W $\uparrow$1.8--2.2$\times$
& $<$1\% loss
& -- \\

\cite{xu2026spactndp}
& Sparse activation on NDP-GPU
& Efficiency $\uparrow$1.45--1.53$\times$
& 98\% maintained
& Speedup $\uparrow$1.92--2.17$\times$ \\

\bottomrule
\end{tabular}
\end{table*}

%A wide variety of other tools, including DCGM, ZeusMonitor, Linux perf, AMD-SMI, and NVIDIA Nsight, appear only in a small number of studies. This suggests that there is currently no widely accepted standard for measuring LLM inference energy. In addition, 42 studies do not clearly report the tools used for measurement, which limits reproducibility and makes comparison across studies difficult. Another important observation is that only a small number of studies use external hardware power meters, while 117 studies rely solely on software-based monitoring. Although software tools are easier to deploy, they may not always capture the actual power consumption of the entire system. These findings highlight the need for standardized and transparent energy measurement methodologies.
\begin{table*}[t]
\centering
\caption{Deployment--Hardware Co-optimization Trade-off Analysis}
\label{tab:deployment_hardware_tradeoff}
\footnotesize
\setlength{\tabcolsep}{2.5pt}
\renewcommand{\arraystretch}{1.12}

\begin{tabular}{
p{0.6cm}
p{2.5cm}
>{\centering\arraybackslash}p{5.9cm}
>{\centering\arraybackslash}p{2.00cm}
>{\centering\arraybackslash}p{3.2cm}}
\toprule
\textbf{Ref.} &
\textbf{Technique} &
\textbf{Energy / Carbon / Power} &
\textbf{Accuracy} &
\textbf{Latency} \\
\midrule

\multicolumn{5}{c}{\textbf{A. Energy Reduction}} \\
\midrule

\cite{kang2025dynamic}
& Tensor offloading + affinity scheduling
& Prefill: $\downarrow$93.6\% QK, $\downarrow$94.0\% SV; Decode: $\downarrow$8.3\%, $\downarrow$50.4\%
& --
& Prefill latency $\downarrow$15.63$\times$, decode speedup $\uparrow$1.09$\times$, $\uparrow$2.01$\times$ \\

\cite{hoxha2025deployment}
& Carbon- and water-aware serving
&$\downarrow$57--59\%; CO$_2$ $\downarrow$78--80\%
& Controlled
& p95 TTFT/TPOT SLOs met \\

\cite{maliakel2025characterizing}
& Model selection + GPU DVFS
&$\downarrow$42\%
& DVFS: no quality loss; routing changes quality
& -- \\

\cite{yu2025voltanallm}
& GPU frequency scaling + request routing
& $\downarrow$36.3\%; power reduced
& Maintained
& Near-perfect TTFT/ITL SLO \\

\cite{liu2025greenllm}
& GPU DVFS + prompt routing
&$\downarrow$up to 34\%;
& --
& No throughput loss; $<$3.5\% SLO violation increase \\

\cite{he2025freesh}
& Carbon-aware routing + scheduling
&$\downarrow$28.6\%; carbon $\downarrow$45.45\%
& --
& Mean E2E latency $\downarrow$8.2\% \\

\cite{wang2025wages}
& GPU sharing + SM repartition + clock tuning
&$\downarrow$26\%
& --
& SLO attainment $\uparrow$up to 4\% \\

\cite{lu2025thermalaware}
& RL thermal-aware scheduling
&Energy $\downarrow$17.46\% air cooling; $\downarrow$18.61\% water cooling; $\sim$6000 tons/year CO$_2$ saving
& Maintained
& Throughput $\uparrow$up to 40.94\% \\

\cite{xu2025camel}
& Batching + DVFS
& EDP $\downarrow$12.4--29.9\%; search energy $\downarrow$27.13--34.43\%
& Maintained
& Search latency $\downarrow$17.93\% / 1.12\% \\

\cite{fan2025ellie}
& Prefill--decode splitting
&$\downarrow$1.8 -- 2.11$\times$
& Maintained
& $\uparrow$ 17--31\% \\

\cite{basit2026biscale}
& Disaggregated serving + DVFS
& Prefill: $\downarrow$up to 39\%; Decode: $\downarrow$up to 48\%
& Maintained
& Maintains TTFT/TPOT SLOs \\

\midrule

\multicolumn{5}{c}{\textbf{B. Carbon Reduction}} \\
\midrule

\cite{li2025ecoserve}
& Carbon-aware provisioning + scheduling
& Carbon $\downarrow$up to 47\%; 1.4--2.2$\times$ total carbon reduction; power reduced
& Maintained
& Maintains performance targets and SLOs \\

\midrule

\multicolumn{5}{c}{\textbf{C. Power Reduction / Energy Efficiency Improvement}} \\
\midrule

\cite{moon2025hybe}
& GPU--NPU hybrid inference
& Energy efficiency $\uparrow$3.9$\times$ for Llama-3; $\uparrow$10.5$\times$ for Phi-3; power $\downarrow$2.9$\times$
& Maintained
& Up to $\uparrow$2.1$\times$ speedup \\

\cite{jiang2026rapid}
& Scheduling + GPU power capping
& Uses 4800 W GPU budget instead of 6000 W; 1.7$\times$ higher QPS/W
& Maintained
& 2$\times$ SLO attainment at peak load \\

\cite{wang2025leap}
& PIM-NoC + dynamic placement
& Energy efficiency $\uparrow$71.94$\times$ vs A100; $\uparrow$24.22$\times$ vs H100; power 10.53 W
& Maintained
& Throughput $\uparrow$2.55$\times$ vs A100 \\

\cite{wang2025asymserve}
& Asymmetric model + SLO scheduler
& $\uparrow$4.46--11.09$\times$ energy efficiency
& Maintained
& TTFT $\downarrow$1.71--1.73$\times$ \\

\midrule

\multicolumn{5}{c}{\textbf{E. Profiling / Characterization}} \\
\midrule

\cite{nguyen2025towards}
& Operational + embodied carbon profiling
& T4 uses 20--28\% less energy at batch size 1; embodied carbon: T4 10.3 kg, RTX6000 Ada 26.6 kg
& --
& T4 is 1.1--2.2$\times$ slower \\

\cite{scherer2024deeploy}
& Compiler-guided SLM deployment
& 490 $\mu$J/token
& --
& 340 tokens/s \\

\bottomrule
\end{tabular}
\end{table*}

\section*{\textbf{RQ3: What tools, datasets, and metrics are used to measure energy
consumption in LLM inference?}}

To answer this research question, we analyzed the evaluation metrics, measurement tools, and benchmark datasets reported in the selected studies. The results show common trends in how energy consumption is measured, while also highlighting several limitations in current evaluation practices.

\noindent\textbf{Evaluation Metrics:} Energy consumption (61 studies) and power consumption (57 studies) are the most frequently reported metrics, showing that most studies focus on measuring the direct resource usage of LLM inference. Although these metrics are closely related, they measure different aspects of system behavior. Power represents the instantaneous energy draw, whereas energy measures the total consumption over time. However, some studies use these terms interchangeably, making comparisons more difficult. Energy efficiency metrics are also widely used (40 studies), suggesting that researchers increasingly recognize that absolute energy values alone are not sufficient for comparing models and optimization techniques. Similarly, the growing use of energy per token (24 studies) and energy per request (18 studies) reflects a shift toward workload-aware evaluation, particularly for autoregressive LLMs where inference cost depends on the number of generated tokens. 

Despite the increasing focus on sustainable AI, environmental metrics remain relatively underused. Only 21 studies report carbon-related metrics, while only 5 studies consider broader infrastructure-level measures such as PUE, WUE, or water consumption. One possible reason is that energy measurements can be collected directly through hardware monitoring tools, whereas environmental metrics often require access to additional operational information. Furthermore, only 6 studies report the Energy-Delay Product, which jointly considers energy consumption and latency. As a result, many studies evaluate energy savings without fully examining their impact on system performance. Overall, current evaluation practices focus mainly on operational energy measurements, while broader sustainability and performance-related metrics receive less attention. Since individual studies often report multiple energy-related metrics, the cumulative number of metric occurrences exceeds the total number of primary studies.

\noindent\textbf{Measurement Tools:} NVML/PyNVML (28 studies) and NVIDIA-SMI (15 studies) are the most widely used tools for energy measurement. This reflects the dominant role of NVIDIA GPUs in LLM inference research. Carbon-aware frameworks such as CodeCarbon (11 studies) and CarbonTracker (2 studies) are also used, although less frequently. Intel RAPL appears in only 7 studies, indicating that most work focuses on GPU energy consumption rather than total system energy. A wide variety of other tools, including DCGM, ZeusMonitor, Linux perf, AMD-SMI, and NVIDIA Nsight, appear only in a few studies, indicating the lack of a widely accepted standard for measuring LLM inference energy. Additionally, 63 studies do not report the measurement tools used, limiting reproducibility and cross-study comparisons. Only 6 studies employ external hardware power meters, while 117 rely solely on software-based monitoring, which may not accurately capture whole-system power consumption. These findings highlight the need for standardized and transparent energy measurement methodologies. Since some studies use more than one measurement tool, the cumulative tool counts exceed the total number of primary studies.

\noindent\textbf{Benchmark Datasets:} Question answering (QA) and reasoning benchmarks are the most commonly used dataset category (29 studies), followed by general NLP benchmarks (21 studies) and production traces or real-world logs (18 studies). Although benchmark datasets provide consistency across studies, they may not fully represent real-world deployment workloads, which often involve longer contexts and more complex interactions. Several important workload categories remain underrepresented. Code generation datasets appear in only 11 studies, while long-context datasets are used in only 6, despite the significant impact of long contexts on KV-cache usage and energy consumption. Similarly, only 2 studies evaluate energy consumption using vision or multimodal datasets. Overall, current benchmarking practices remain heavily focused on traditional text-based tasks and may not adequately reflect emerging LLM applications.

\begin{tcolorbox}[
colback=gray!15,
colframe=gray!60,
boxrule=0.5pt,
arc=2mm,
left=2mm,
right=2mm,
top=1mm,
bottom=1mm
]
\textbf{Summary:} Energy and power consumption dominate evaluation metrics in LLM inference studies, while environmental indicators such as carbon emissions, water usage, and PUE are reported less frequently. GPU-based tools, particularly NVML/PyNVML and NVIDIA-SMI, are the primary measurement platforms, with few studies using hardware power meters or reporting sufficient methodological details. Benchmarking is largely focused on QA and general NLP tasks, with limited coverage of code generation, long-context inference, and multimodal workloads.
\end{tcolorbox}

\section{Future Research Opportunities}

%Building on the findings, we identify several research opportunities for advancing sustainable LLM inference.
\subsection{Cross-Layer and Adaptive Optimization}

\subsubsection{Input-Aware and End-to-End Co-Optimization}

Most existing studies focus on model--hardware and deployment--hardware co-optimization, while approaches involving the input level remain comparatively limited. Only a small number of studies investigate input-centric co-optimization (Tables~\ref{tab:input_deployment} and~\ref{tab:input_model}), despite the potential of techniques such as prompt optimization, token pruning, \& context compression, etc., to reduce unnecessary computation. Future research should therefore investigate frameworks that jointly optimize input-level techniques with model- and deployment-level optimizations. A broader direction is the development of end-to-end co-optimization frameworks spanning the input, model, deployment, and hardware levels. These frameworks could dynamically coordinate prompt processing, model configuration, batching, caching, precision, and hardware selection based on workload characteristics, resource availability, and quality-of-service requirements, improving energy--latency--accuracy trade-offs. This leads to several open RQs: \textit{(RQ1) How can input-level optimization techniques be jointly optimized with model-, deployment-, and hardware-level techniques to reduce LLM inference energy consumption?} \textit{(RQ2) How do different cross-layer optimization strategies influence the trade-offs among energy consumption, latency, and accuracy?} \textit{(RQ3) How can LLM inference systems dynamically select the most suitable co-optimization techniques for different workloads and deployment conditions?}
\subsection{Measurement, Evaluation, and Benchmarking}

\subsubsection{Standardized Benchmarking and Energy Evaluation}

The lack of standardized benchmarks and evaluation protocols limits comparison, reproducibility, and generalizability across studies. Existing evaluations vary in datasets, workloads, model configurations, hardware platforms, and measurement methodologies, while benchmark coverage remains concentrated on QAs, reasoning, and general NLP, with limited representation of code generation, long-context, multimodal, RAG, and agentic workloads. Future research should therefore develop representative benchmark suites that combine controlled microbenchmarks with realistic applications and systematically vary factors such as prompt and output length, batch size, model size, precision, and concurrency. This raises an important RQ: \textit{How can standardized benchmarks be designed to represent diverse and realistic LLM inference workloads across different application domains?}

Energy metrics should also reflect the characteristics of emerging workloads. Energy per token may be insufficient for workflows involving retrieval, multiple reasoning steps, tool execution, or multimodal processing. This motivates a second RQ: \textit{Which energy metrics are most appropriate for evaluating end-to-end efficiency across emerging LLM workloads such as RAG, agentic AI, multimodal inference, and long-context processing?} Metrics such as energy per request, completed task, successful answer, or session may provide more meaningful measures in these settings.

Beyond benchmark design, consistent measurement procedures are needed to ensure that reported energy savings can be fairly compared across studies. Common experimental protocols should define factors such as warm-up procedures, idle-power treatment, measurement duration, repetition count, numerical precision, system utilization, and reporting requirements. This leads to a third RQ: \textit{How can standardized experimental protocols be designed to enable consistent and reproducible energy measurement across different LLM models, workloads, frameworks, and hardware platforms?} Evaluation should also consider latency, throughput, accuracy, memory consumption, carbon emissions, and implementation complexity alongside energy use to better capture practical deployment trade-offs.

%\subsubsection{Standardized Energy Evaluation}

%Standardized energy evaluation remains a key challenge for sustainable LLM inference. Energy savings observed in one environment may not generalize across different models, workloads, frameworks, hardware platforms, or deployment conditions, while variations in measurement methodology further complicate comparison. Future research should establish common experimental protocols covering factors such as warm-up procedures, batch size, context and output length, numerical precision, system utilization, idle-power treatment, measurement duration, and repetition count. This leads to an important RQ: \textit{How can standardized experimental protocols be designed to enable fair and reproducible energy evaluation across different LLM models, workloads, frameworks, and hardware platforms?} Standardized reporting would improve reproducibility and help assess whether optimization techniques generalize beyond specific settings. Evaluation should also consider latency, throughput, accuracy, memory consumption, carbon emissions, and implementation complexity alongside energy use. Although metrics such as the Energy-Delay Product combine energy and performance, broader evaluation methods are still needed to capture practical deployment trade-offs.

\subsubsection{Fine-Grained Energy Profiling}

Many studies report only aggregate energy consumption, providing limited insight into which inference components drive energy use. Future work should therefore explore fine-grained profiling across stages, modules, layers, and operations, including prefill, decoding, attention, feed-forward layers, KV-cache access, memory movement, sampling, and communication. This raises an important RQ: \textit{How can fine-grained energy profiling identify the contribution of individual inference components across different workloads and optimization settings?} Such profiling could also improve explainability by identifying which components are affected by an optimization and under what workload conditions it remains effective, supporting more targeted and generalizable energy-saving techniques.

\subsection{Emerging LLM Workloads}

\subsubsection{Energy Cost of Agentic AI}
The energy consumption of agentic AI remains largely unexplored. Unlike traditional LLM inference, agentic systems may involve planning, retrieval, reasoning, reflection, verification, tool execution, memory access, and inter-agent communication, causing a single request to trigger multiple LLM calls and external operations. Measuring only the main LLM call may therefore underestimate the true workflow energy cost. Future research should decompose energy consumption across individual agentic components and identify the dominant sources of energy use. This motivates an important RQ: \textit{How can the end-to-end energy consumption of agentic AI workflows be measured and attributed across planning, reasoning, retrieval, memory, tool execution, and inter-agent communication?} Another important direction is understanding whether additional computation actually provides sufficient benefit. This leads to a second RQ: \textit{How do reasoning depth, reflection, retrieval frequency, tool use, and agent interactions affect the trade-off between energy consumption and task performance?} Task-level metrics such as energy per successful task may therefore be more meaningful than energy per generated token. Future work should also investigate energy-aware agents that dynamically decide when additional computation is justified. This raises a third RQ: \textit{How can agentic systems dynamically select reasoning steps, tools, retrieval actions, and model sizes to minimize energy consumption while maintaining task quality?} For example, smaller models could handle routing, planning, or verification, while larger models are reserved for complex reasoning tasks.
%The energy consumption of agentic AI remains a largely unexplored research direction. Unlike traditional LLM inference, agentic systems may involve planning, retrieval, reasoning, reflection, verification, tool execution, memory access, and inter-agent communication. A single request can therefore trigger multiple LLM calls and external operations, with total energy depending on reasoning depth, context growth, retrieval frequency, model selection, tool use, and agent interactions. Measuring only the main LLM call may therefore underestimate the true energy cost of the workflow. Future research should decompose energy consumption across individual agentic components, such as planning, reasoning, retrieval, memory access, reflection, tool execution, and inter-agent communication. This would help identify the dominant sources of energy consumption and assess whether additional reasoning or reflection provides sufficient performance gains to justify its cost. Task-level metrics, such as energy per successful task, may therefore be more meaningful than energy per generated token. Energy-aware agents could also dynamically determine when additional computation is justified. For example, they could decide whether another reasoning step, retrieval, or tool call is likely to improve the result enough to justify its energy cost. Smaller models could also handle routing, planning, or verification, while larger models are reserved for complex reasoning tasks.

\subsubsection{Energy Implications of Multimodal LLMs}
Multimodal LLMs introduce additional sustainability challenges because they process text, images, audio, and video through modality-specific components, making their inference energy characteristics different from text-only models. Future research should examine how factors such as image resolution, visual-token count, video duration, frame rate, audio length, and fusion mechanisms influence energy and memory requirements. This leads to an important RQ: \textit{How do modality-specific input characteristics and fusion mechanisms affect the energy, memory, and performance trade-offs of multimodal inference?} Another promising direction is adaptive modality processing. Systems could reduce energy by selecting only relevant modalities, lowering image resolution, sampling fewer video frames, compressing visual tokens, or bypassing unnecessary encoders. This raises a third RQ: \textit{How can multimodal LLMs dynamically adapt modality selection and processing granularity to reduce energy consumption while preserving task performance?} Such evaluation should also consider continuous applications such as video understanding, voice assistants, embodied AI, robotics, and multimodal agents, where inference occurs repeatedly rather.

%Multimodal LLMs introduce additional sustainability challenges because they process text, images, audio, and video through modality-specific encoders, projection layers, and cross-modal attention. Their inference energy characteristics may therefore differ substantially from text-only models. Future research should characterize the energy contribution of individual multimodal components and examine how factors such as image resolution, visual-token count, video duration, frame rate, audio length, and fusion mechanisms affect computation and memory requirements. Adaptive modality processing is another promising direction. Systems could reduce energy by selecting only relevant modalities, lowering image resolution, sampling fewer video frames, compressing visual tokens, or bypassing unnecessary encoders. Energy evaluation should also consider continuous applications such as video understanding, voice assistants, embodied AI, robotics, and multimodal agents, where inference occurs repeatedly rather than as isolated requests.

\subsubsection{Long-Context and Retrieval-Augmented Inference}

Long-context and retrieval-augmented inference introduce important energy trade-offs. Longer contexts increase prefill computation, memory use, and KV-cache requirements, while RAG adds overhead from embedding generation, vector search, reranking, database access, and data transfer. This motivates an important RQ: \textit{Under what workload and system conditions is retrieval-augmented inference more energy efficient than directly processing longer contexts?} Future research should also explore context compression, selective retrieval, retrieval caching, and adaptive switching between retrieval-based and long-context processing. This leads to a second RQ: \textit{How can LLM systems dynamically choose between long-context processing, retrieval, and hybrid strategies to minimize energy consumption while maintaining task performance?}

\subsection{Hardware, Serving, and Distributed Inference}

\subsubsection{Transferability Across Hardware Platforms}
Many energy-efficient inference techniques are evaluated on a limited range of hardware, particularly NVIDIA GPUs, making it unclear whether reported savings generalize across CPUs, GPUs, NPUs, TPUs, edge accelerators, and emerging platforms. Differences in compute throughput, memory bandwidth, cache hierarchy, precision support, kernel implementations, and power management may significantly affect their effectiveness. This motivates an important RQ: \textit{To what extent do energy-efficient inference techniques generalize across heterogeneous hardware platforms, and which techniques require platform-specific adaptation?} Future research should also investigate hardware-aware inference systems that adapt models, numerical precision, and optimization strategies to accelerator characteristics. This leads to a second RQ: \textit{How can LLM inference systems dynamically select energy-efficient configurations based on the characteristics of the underlying hardware platform?}

%Many energy-efficient inference techniques are evaluated on a limited range of hardware, particularly NVIDIA GPUs, making it unclear whether reported savings generalize across CPUs, GPUs, NPUs, TPUs, edge accelerators, and emerging platforms. Their effectiveness may vary due to differences in compute throughput, memory bandwidth, cache hierarchy, numerical precision support, kernel implementations, and power-management mechanisms. Future research should systematically evaluate optimization techniques across heterogeneous hardware to identify which methods are broadly transferable and which require platform-specific adaptation. Such studies could also support hardware-aware inference systems that select models, precisions, and optimization strategies based on accelerator characteristics.

\subsubsection{Energy-Aware Open-Source Serving Frameworks}
Modern open-source serving frameworks such as vLLM, SGLang, TensorRT-LLM, and Ollama support mechanisms including continuous batching, KV-cache management, scheduling, prefix caching, parallelism, and quantization. While these frameworks primarily optimize throughput, latency, and memory utilization, many of these mechanisms could also improve energy efficiency. Future research should therefore treat energy as a first-class objective in LLM serving by using workload characteristics and real-time energy measurements to adapt batching, precision, model selection, decoding, caching, scheduling, and hardware allocation. This motivates an important RQ: \textit{How can LLM serving frameworks dynamically optimize runtime decisions using workload characteristics and energy telemetry to reduce energy consumption while maintaining performance and QoS requirements?}

%Modern open-source serving frameworks such as vLLM, SGLang, TensorRT-LLM, and Ollama support mechanisms including continuous batching, KV-cache management, scheduling, prefix caching, parallelism, and quantization. While these frameworks primarily optimize throughput, latency, and memory utilization, many of these mechanisms could also improve energy efficiency. Future research should therefore treat energy as a first-class objective in LLM serving. Frameworks could dynamically adjust batch size, precision, model selection, decoding, caching, and hardware allocation using workload characteristics and real-time energy measurements. Energy-aware schedulers could also consolidate requests during low-demand periods or route simpler requests to smaller models. Integrating energy telemetry and standardized energy-management APIs into serving frameworks could enable adaptive runtime optimization and improve reproducibility across studies.

\subsubsection{Sustainable Distributed and Disaggregated LLM Inference}
Distributed and disaggregated serving architectures are increasingly used to support large models and high request volumes. Techniques such as tensor, pipeline, and expert parallelism, remote KV-cache storage, and disaggregated prefill and decoding can improve scalability and hardware utilization, but they also introduce communication, synchronization, memory-transfer, and coordination overhead. These costs make it unclear whether distributed inference reduces total energy consumption or simply shifts energy use across system components. This motivates an important RQ: \textit{Under what workload and system conditions does distributed or disaggregated LLM inference reduce total system-level energy consumption compared with centralized inference?} Future research should also consider complete system-level energy costs, including accelerators, processors, memory, interconnects, and networks. Energy-aware systems could jointly optimize model placement, request routing, KV-cache placement, communication, accelerator utilization, and workload consolidation. This leads to a second RQ: \textit{How can distributed LLM serving systems jointly optimize resource placement, communication, and workload allocation to improve energy efficiency while maintaining scalability and quality-of-service requirements?}

\section{Limitations}

\subsection{Construct Validity}
Construct validity concerns whether concepts such as sustainable LLM inference and inference-time energy optimization were clearly defined and consistently applied. A key challenge is the lack of standardized terminology, as related work uses terms such as energy-efficient inference, green AI, carbon-aware inference, efficient serving, and resource-efficient deployment. Consequently, relevant studies may use terminology not captured by the search string. To mitigate this threat, we iteratively refined the search strategy using manually identified benchmark papers and additional keywords related to energy, deployment, measurement, and environmental impact. We also applied predefined inclusion and exclusion criteria and used structured data extraction forms to ensure consistent classification of optimization techniques, metrics, measurement approaches, and trade-offs.

\vspace{-0.3cm}
\subsection{Internal Validity}
Internal validity concerns whether the conclusions accurately reflect the evidence reported in the selected studies. The primary studies vary considerably in model families, datasets, hardware, inference frameworks, metrics, and quality of experimental reporting, limiting direct comparisons. To reduce this threat, we followed a predefined review protocol and consistently applied the selection criteria. Reviewer agreement during the pilot screening was assessed using Fleiss' Kappa, and disagreements were resolved before full screening. During data extraction, only explicitly reported information was recorded; missing details were marked as \textit{Not Specified} rather than inferred. Given the heterogeneous experimental settings, the review emphasizes common trends, challenges, and research directions rather than direct quantitative comparisons.

\vspace{-0.3cm}
\subsection{External Validity}
External validity concerns the extent to which the selected studies represent the broader literature on sustainable LLM inference. Since this is a rapidly evolving field, the review captures only studies available during the defined search period. Although four major digital libraries and relevant arXiv preprints were considered, non-English publications and grey literature such as technical reports, industrial white papers, blogs, and patents were excluded. Despite these limitations, the review follows a systematic and transparent methodology that can be extended as new studies, models, hardware platforms, and energy-evaluation methods emerge.

\section{Conclusion}

This paper presented the SLR of sustainable LLM inference, providing a comprehensive analysis of optimization techniques, their associated trade-offs, and current evaluation practices. We organized the existing literature according to four optimization levels, input, model, deployment, and hardware, and further synthesized studies that combine multiple optimization levels through cross-layer co-optimization. Based on this taxonomy, we analyzed energy-efficient optimization techniques, examined the trade-offs among energy consumption, latency, and accuracy, and reviewed the tools, datasets, and evaluation methodologies used to measure inference energy. The review shows that no single optimization technique is universally optimal; rather, its effectiveness depends on the optimization level, workload characteristics, deployment environment, and hardware platform. Cross-layer analysis further demonstrates that coordinating optimization across multiple levels offers the greatest potential for improving energy efficiency, latency, and overall system performance while preserving model quality. Finally, based on the identified gaps, we outline open research questions and future directions in areas such as standardized evaluation, emerging LLM workloads, adaptive optimization, and energy-aware serving. Overall, this review provides a structured foundation and practical insights for researchers and practitioners working toward more sustainable and energy-efficient LLM inference.

\bibliographystyle{ACM-Reference-Format}
\bibliography{references}

@inproceedings{niu2023psqs,
  author    = {Chenxu Niu and Wei Zhang and Suren Byna and Yong Chen},
  title     = {PSQS: Parallel Semantic Querying Service for Self-describing File Formats},
  booktitle = {2023 IEEE International Conference on Big Data (BigData)},
  pages     = {536--541},
  publisher = {IEEE},
  year      = {2023},
  doi       = {10.1109/BigData59044.2023.10386205}
}

@article{bai2401beyond,
  title={Beyond efficiency: A systematic survey of resource-efficient large language models. arXiv 2024},
  author={Bai, G and Chai, Z and Ling, C and Wang, S and Lu, J and Zhang, N and Shi, T and Yu, Z and Zhu, M and Zhang, Y and others},
  journal={arXiv preprint arXiv:2401.00625},
  year = {2024}
}

@article{zhou2024survey,
  title={A survey on efficient inference for large language models},
  author={Zhou, Zixuan and Ning, Xuefei and Hong, Ke and Fu, Tianyu and Xu, Jiaming and Li, Shiyao and Lou, Yuming and Wang, Luning and Yuan, Zhihang and Li, Xiuhong and others},
  journal={arXiv preprint arXiv:2404.14294},
  year={2024}
}

@article{liu2026efficient,
  title={Efficient inference for large reasoning models: A survey},
  author={Liu, Yue and Wu, Jiaying and He, Yufei and Gong, Ruihan and Xia, Jun and Li, Liang and Gao, Hongcheng and Chen, Hongyu and Bi, Baolong and Zhang, Jiaheng and others},
  journal={IEEE Transactions on Pattern Analysis and Machine Intelligence},
  year={2026},
  publisher={IEEE}
}

@article{cheng2025survey,
  title={Survey on Efficient Large Language Models: Principles, Algorithms, Applications, and Open Issues},
  author={Cheng, Jian and Kang, Haidong and Shao, Yuxin and Li, Nan and Chen, Pengjun and Wang, Rui and Long, Saiqin and Yang, Xiaochun and Ma, Lianbo},
  journal={IEEE Transactions on Neural Networks and Learning Systems},
  year={2025},
  publisher={IEEE}
}

@article{pan2025survey,
  title={A survey of llm inference systems},
  author={Pan, James and Li, Guoliang},
  journal={arXiv preprint arXiv:2506.21901},
  year={2025}
}

@article{park2025survey,
  title={A survey on inference engines for large language models: Perspectives on optimization and efficiency},
  author={Park, Sihyeong and Jeon, Sungryeol and Lee, Chaelyn and Jeon, Seokhun and Kim, Byung-Soo and Lee, Jemin},
  journal={ACM Transactions on Intelligent Systems and Technology},
  year={2025},
  publisher={ACM New York, NY}
}

@article{cai2026efficient,
  title={Efficient Inference for Edge Large Language Models: A Survey},
  author={Cai, Guanyu and Tian, Ruiming and Yang, Lang and Jia, Yunzhe and Li, Lingkun and Wang, Jiliang},
  journal={Tsinghua Science and Technology},
  volume={31},
  number={3},
  pages={1365--1380},
  year={2026},
  publisher={清华大学出版社}
}

@article{heisler2025llm,
  title={LLM Inference Scheduling: A Survey of Techniques, Frameworks, and Trade-offs},
  author={Heisler, Morgan and Yousefijamarani, Zahra and Wang, Xinglu and Wang, Qian and Shi, Ge and Sadri, Hanieh and Yu, Timothy and Li, Yifei and Li, Haley and Singh, Gursimran and others},
  journal={Authorea Preprints},
  year={2025},
  publisher={Authorea}
}

@article{pang2025optimizing,
  title={Optimizing LLM inference throughput via memory-aware and SLA-constrained dynamic batching},
  author={Pang, Bowen and Li, Kai and Wang, Feifan},
  journal={arXiv preprint arXiv:2503.05248},
  year={2025}
}

@article{hsu2024liger,
  title={Liger kernel: Efficient triton kernels for llm training},
  author={Hsu, Pin-Lun and Dai, Yun and Kothapalli, Vignesh and Song, Qingquan and Tang, Shao and Zhu, Siyu and Shimizu, Steven and Sahni, Shivam and Ning, Haowen and Chen, Yanning},
  journal={arXiv preprint arXiv:2410.10989},
  year={2024}
}

@article{li2024large,
  title={Large language model inference acceleration: A comprehensive hardware perspective},
  author={Li, Jinhao and Xu, Jiaming and Huang, Shan and Chen, Yonghua and Li, Wen and Liu, Jun and Lian, Yaoxiu and Pan, Jiayi and Ding, Li and Zhou, Hao and others},
  journal={arXiv preprint arXiv:2410.04466},
  year={2024}
}

@inproceedings{ma2025multi,
  title={Multi-tier multi-node scheduling of LLM for collaborative AI computing},
  author={Ma, Mulei and Gong, Chenyu and Zeng, Liekang and Yang, Yang},
  booktitle={IEEE INFOCOM 2025-IEEE Conference on Computer Communications},
  pages={1--10},
  year={2025},
  organization={IEEE}
}

@inproceedings{eliopoulos2025pruning,
  title={Pruning one more token is enough: Leveraging latency-workload non-linearities for vision transformers on the edge},
  author={Eliopoulos, Nicholas John and et al.},
  booktitle={Proceedings of the Winter Conference on Applications of Computer Vision},
  pages={7153--7162},
  year={2025}
}

@article{yuan2026understanding,
  title={Understanding and mitigating numerical sources of nondeterminism in llm inference},
  author={Yuan, Jiayi and Li, Hao and Ding, Xinheng and Xie, Wenya and Li, Yu-Jhe and Zhao, Wentian and Wan, Kun and Shi, Jing and Hu, Xia and Liu, Zirui},
  journal={Advances in Neural Information Processing Systems},
  volume={38},
  pages={169819--169851},
  year={2026}
}

@article{zhang2025attention,
  title={Attention Mechanism for LLM-based Agents Dynamic Diffusion under Information Asymmetry},
  author={Zhang, Yiwen and Wu, Yifu and Hua, Wenyue and Lu, Xiang and Hu, Xuming},
  journal={arXiv preprint arXiv:2502.13160},
  year={2025}
}

@article{lin2026qwen,
  title={Qwen-BIM: developing large language model for BIM-based design with domain-specific benchmark and dataset},
  author={Lin, Jia-Rui and Cai, Yun-Hong and Ni, Xiang-Rui and Zhou, Shaojie and Pan, Peng},
  journal={arXiv preprint arXiv:2602.20812},
  year={2026}
}

@inproceedings{DBLP:conf/dac/0009WWZLTY24,
  author       = {Xi Wang and
                  Gwok{-}Waa Wan and
                  Sam{-}Zaak Wong and
                  Layton Zhang and
                  Tianyang Liu and
                  Qi Tian and
                  Jianmin Ye},
  editor       = {Vivek De},
  title        = {ChatCPU: An Agile {CPU} Design and Verification Platform with {LLM}},
  booktitle    = {Proceedings of the 61st {ACM/IEEE} Design Automation Conference, {DAC}
                  2024, San Francisco, CA, USA, June 23-27, 2024},
  pages        = {212:1--212:6},
  publisher    = {{ACM}},
  year         = {2024},
  url          = {https://doi.org/10.1145/3649329.3658493},
  doi          = {10.1145/3649329.3658493}
}

@inproceedings{DBLP:conf/ssdbm/NiuZS025,
  author       = {Chenxu Niu and
                  Wei Zhang and
                  Mert Side and
                  Yong Chen},
  title        = {ICEAGE: Intelligent Contextual Exploration and Answer Generation Engine
                  for Scientific Data Discovery},
  booktitle    = {Proceedings of the 37th International Conference on Scalable Scientific
                  Data Management, {SSDBM} 2025, New York, NY, USA, July 9-11, 2025},
  pages        = {23:1--23:10},
  publisher    = {{ACM}},
  year         = {2025},
  url          = {https://doi.org/10.1145/3733723.3733731},
  doi          = {10.1145/3733723.3733731}
}

@inproceedings{DBLP:conf/hotcarbon/ChienLNRSW23,
  author       = {Andrew A. Chien et al.},
  title        = {Reducing the Carbon Impact of Generative {AI} Inference (today and
                  in 2035)},
  booktitle    = {Proceedings of the 2nd Workshop on Sustainable Computer Systems, HotCarbon
                  2023, Boston, MA, USA, 9 July 2023},
  pages        = {11:1--11:7},
  publisher    = {{ACM}},
  year         = {2023},
  url          = {https://doi.org/10.1145/3604930.3605705},
  doi          = {10.1145/3604930.3605705}
}

@article{jiang2024chatbots_lifecycle,
  author  = {Peng Jiang and Christian Sonne and Wangliang Li and Fengqi You and Siming You},
  title   = {Preventing the Immense Increase in the Life-Cycle Energy and Carbon Footprints of {LLM}-Powered Intelligent Chatbots},
  journal = {Engineering},
  volume  = {40},
  month   = sep,
  pages   = {202--210},
  year    = {2024},
  doi     = {10.1016/j.eng.2024.04.002}
}

@misc{hutt2019inferentia,
  author       = {Gadi Hutt and Vibhav Viswanathan and Adam Nadolski},
  title        = {Deliver High Performance {ML} Inference with {AWS} Inferentia},
  year         = {2019},
  howpublished = {AWS / talk or slide deck},
  note         = {Accessed online},
  url          = {https://speakerdeck.com/awsdevelopers/deliver-high-performance-ml-inference-with-aws-inferentia}
}

@inproceedings{vaswani2017attention,
  author    = {Ashish Vaswani and Noam Shazeer and Niki Parmar and Jakob Uszkoreit and Llion Jones and Aidan N. Gomez and {\L}ukasz Kaiser and Illia Polosukhin},
  title     = {Attention Is All You Need},
  booktitle = {Advances in Neural Information Processing Systems (NeurIPS)},
  year      = {2017}
}

@article{leviathan2022speculative,
  author        = {Yaniv Leviathan and Matan Kalman and Yossi Matias},
  title         = {Fast Inference from Transformers via Speculative Decoding},
  journal       = {arXiv preprint arXiv:2211.17192},
  year          = {2022},
  eprint        = {2211.17192},
  archivePrefix = {arXiv}
}

@inproceedings{samsi2023words_to_watts,
  author    = {Siddharth Samsi and Dan Zhao and Joseph McDonald and Baolin Li and Adam Michaleas and Michael Jones and William Bergeron and Jeremy Kepner and Devesh Tiwari and Vijay Gadepally},
  title     = {From Words to Watts: Benchmarking the Energy Costs of Large Language Model Inference},
  booktitle = {2023 IEEE High Performance Extreme Computing Conference (HPEC)},
  pages     = {1--9},
  year      = {2023},
  publisher = {IEEE},
  doi       = {10.1109/HPEC58863.2023.10363447}
}

@article{xu2024survey,
  title={A survey of resource-efficient llm and multimodal foundation models},
  author={Xu, Mengwei and Yin, Wangsong and Cai, Dongqi and Yi, Rongjie and Xu, Daliang and Wang, Qipeng and Wu, Bingyang and Zhao, Yihao and Yang, Chen and Wang, Shihe and others},
  journal={arXiv preprint arXiv:2401.08092},
  year={2024}
}

@article{fu2025moa,
  title        = {Mixture of Attention Spans: Optimizing {LLM} Inference Efficiency with Heterogeneous Sliding-Window Lengths},
  author       = {Fu, Tianyu and et al.},
  journal      = {arXiv preprint arXiv:2406.14909},
  year         = {2025},
  archivePrefix= {arXiv},
  eprint       = {2406.14909},
  primaryClass = {cs.LG},
  note         = {Published as a conference paper at COLM 2025}
}

@inproceedings{guo2026effectiveness,
  author    = {Guo, Zhinuan (Otto) and et al.},
  title     = {On the Effectiveness of Proposed Techniques to Reduce Energy Consumption in RAG Systems: A Controlled Experiment},
  booktitle = {Proceedings of the 2026 IEEE/ACM 48th International Conference on Software Engineering (ICSE-SEIS '26},
  year      = {2026},
  month     = apr,
  publisher = {ACM},
  numpages  = {12},
}

@inproceedings{potatmitis-etal-2025-cache-saver,
  title     = {Cache Saver: A Modular Framework for Efficient, Affordable, and Reproducible {LLM} Inference},
  author    = {Potatmitis, Nearchos and Klein, Lars and Mohammadi, Bardia and Xu, Chongyang and Mukherjee, Attreyee and Bindschaedler, Laurent and Tandon, Niket and Arora, Akhil},
  booktitle = {Findings of the Association for Computational Linguistics: EMNLP 2025},
  year      = {2025},
  month     = nov,
  pages     = {25703--25724},
  publisher = {Association for Computational Linguistics}
}

@misc{niu2025tokenpowerbench,
  title         = {TokenPowerBench: Benchmarking the Power Consumption of LLM Inference},
  author        = {Niu, Chenxu and Zhang, Wei and Li, Jie and Zhao, Yongjian and Wang, Tongyang and Wang, Xi and Chen, Yong},
  year          = {2025},
  eprint        = {2512.03024},
  archivePrefix = {arXiv},
  primaryClass  = {cs.LG},
  url           = {https://arxiv.org/abs/2512.03024},
  doi           = {10.48550/arXiv.2512.03024}
}

@inproceedings{xue2025hybridserve_icdcs,
  title     = {HYBRIDSERVE: Efficient Serving of Large AI Models with Confidence-Based Cascade Routing},
  author    = {Xue, Leyang and Fu, Yao and Mai, Luo and Marina, Mahesh K.},
  booktitle = {Proceedings of the 45th IEEE International Conference on Distributed Computing Systems (ICDCS)},
  year      = {2025},
  pages     = {1--6},
  publisher = {IEEE},
  note      = {Accepted/In press}
}

@misc{ziller2026greenserv,
  title         = {GreenServ: Energy-Efficient Context-Aware Dynamic Routing for Multi-Model LLM Inference},
  author        = {Ziller, Thomas and Ilager, Shashikant and Tundo, Alessandro and Bartocci, Ezio and Mariani, Leonardo and Brandic, Ivona},
  year          = {2026},
  eprint        = {2601.17551},
  archivePrefix = {arXiv},
  primaryClass  = {cs.PF},
  doi           = {10.48550/arXiv.2601.17551},
  url           = {https://arxiv.org/abs/2601.17551}
}

@misc{2025promptstopower,
  title         = {From Prompts to Power: Measuring the Energy Footprint of LLM Inference},
  author        = {Caravaca, Francisco and Cuevas, Angel and Cuevas, Ruben},
  year          = {2025},
  eprint        = {2511.05597},
  archivePrefix = {arXiv},
  primaryClass  = {cs.AI},
  doi           = {10.48550/arXiv.2511.05597},
  url           = {https://arxiv.org/abs/2511.05597}
}

@misc{delavande2026understandingefficiency,
  title         = {Understanding Efficiency: Quantization, Batching, and Serving Strategies in LLM Energy Use},
  author        = {Delavande, Julien and Pierrard, Regis and Luccioni, Sasha},
  year          = {2026},
  month         = jan,
  eprint        = {2601.22362},
  archivePrefix = {arXiv},
  primaryClass  = {cs.LG},
  note          = {arXiv preprint},
  url           = {https://arxiv.org/abs/2601.22362}
}

@misc{husom2025sustainable,
  title         = {Sustainable LLM Inference for Edge AI: Evaluating Quantized LLMs for Energy Efficiency, Output Accuracy, and Inference Latency},
  author        = {Husom, Erik Johannes and et al.},
  year          = {2025},
  month         = apr,
  eprint        = {2504.03360},
  archivePrefix = {arXiv},
  primaryClass  = {cs.CY},
  note          = {arXiv preprint},
  url           = {https://arxiv.org/abs/2504.03360}
}

@misc{shi2025systematic,
  title         = {Systematic Characterization of LLM Quantization: A Performance, Energy, and Quality Perspective},
  author        = {Shi, Yifan and Ding, Yufei},
  year          = {2025},
  eprint        = {2508.16712},
  archivePrefix = {arXiv},
  primaryClass  = {cs.LG},
  note          = {arXiv preprint},
  url           = {https://arxiv.org/abs/2508.16712}
}

@article{nik2026impact,
  title        = {Impact of decoding strategies on GPU energy usage in large language model text generation},
  author       = {Nik, Alireza and Riegler, Michael A. and Halvorsen, P{\aa}l},
  journal      = {Scientific Reports},
  volume       = {16},
  pages        = {2168},
  year         = {2026},
  doi          = {10.1038/s41598-025-31896-0},
  publisher    = {Springer Nature},
  url          = {https://doi.org/10.1038/s41598-025-31896-0}
}

@misc{qin2024mjsd,
  title         = {Multi-Token Joint Speculative Decoding for Accelerating Large Language Model Inference},
  author        = {Qin, Zongyue and Hu, Ziniu and He, Zifan and Prakriya, Neha and Cong, Jason and Sun, Yizhou},
  year          = {2024},
  month         = jul,
  eprint        = {2407.09722},
  archivePrefix = {arXiv},
  primaryClass  = {cs.CL},
  note          = {arXiv preprint},
  url           = {https://arxiv.org/abs/2407.09722}
}

@misc{solovyeva2026towards,
  title         = {Towards Green AI: Decoding the Energy of LLM Inference in Software Development},
  author        = {Solovyeva, Lola and Castor, Fernando},
  year          = {2026},
  howpublished  = {arXiv preprint},
  eprint        = {2602.05712},
  archivePrefix = {arXiv},
  primaryClass  = {cs.SE},
  doi           = {10.48550/arXiv.2602.05712},
  url           = {https://arxiv.org/abs/2602.05712}
}

@article{masciari2026toon,
  title         = {Are LLMs Ready for TOON? Benchmarking Structural Correctness--Sustainability Trade-offs in Novel Structured Output Formats},
  author        = {Masciari, Elio and et al.},
  journal       = {arXiv preprint arXiv:2601.12014},
  year          = {2026},
  eprint        = {2601.12014},
  archivePrefix = {arXiv},
  primaryClass  = {cs.AI}
}

@misc{stojkovic2024towards,
  title         = {Towards Greener {LLM}s: Bringing Energy-Efficiency to the Forefront of {LLM} Inference},
  author        = {Stojkovic, Jovan and Choukse, Esha and Zhang, Chaojie and Goiri, I{\~n}igo and Torrellas, Josep},
  year          = {2024},
  howpublished  = {arXiv preprint},
  eprint        = {2403.20306},
  archivePrefix = {arXiv},
  primaryClass  = {cs.AI},
  doi           = {10.48550/arXiv.2403.20306},
  url           = {https://arxiv.org/abs/2403.20306}
}

@misc{dao2022flashattention,
  title         = {FlashAttention: Fast and Memory-Efficient Exact Attention with IO-Awareness},
  author        = {Dao, Tri and Fu, Daniel Y. and Ermon, Stefano and Rudra, Atri and {R{\'e}}, Christopher},
  year          = {2022},
  howpublished  = {arXiv preprint},
  eprint        = {2205.14135},
  archivePrefix = {arXiv},
  primaryClass  = {cs.LG},
  doi           = {10.48550/arXiv.2205.14135},
  url           = {https://arxiv.org/abs/2205.14135}
}

@inproceedings{stojkovic2025dynamollm,
  title     = {DynamoLLM: Designing {LLM} Inference Clusters for Performance and Energy Efficiency},
  author    = {Stojkovic, Jovan and Zhang, Chaojie and Goiri, {\'I}{\~n}igo and Torrellas, Josep and Choukse, Esha},
  booktitle = {IEEE International Symposium on High Performance Computer Architecture (HPCA 2025)},
  year      = {2025},
  pages     = {1348--1362},
  publisher = {IEEE},
  doi       = {10.1109/HPCA61900.2025.00102},
  url       = {https://doi.org/10.1109/HPCA61900.2025.00102}
}

@inproceedings{kakolyris2025throttllem,
  title     = {throttLL'eM: Predictive {GPU} Throttling for Energy Efficient {LLM} Inference Serving},
  author    = {Kakolyris, Andreas Kosmas and Masouros, Dimosthenis and Vavaroutsos, Petros and Xydis, Sotirios and Soudris, Dimitrios},
  booktitle = {2025 IEEE International Symposium on High Performance Computer Architecture (HPCA)},
  year      = {2025},
  pages     = {1363--1378},
  publisher = {IEEE},
  doi       = {10.1109/HPCA61900.2025.00103},
  url       = {https://doi.org/10.1109/HPCA61900.2025.00103}
}

@misc{yuvarani2026switching,
  title         = {Sustainable LLM Inference using Context-Aware Model Switching},
  author        = {Yuvarani and Singh, Akashdeep and Fathanah, Zahra and Harlen, Salsabila and Zahari, Syeikha Syafura Al-Zahra binti and Subramaniam, Hema},
  year          = {2026},
  month         = feb,
  howpublished  = {arXiv preprint},
  eprint        = {2602.22261},
  archivePrefix = {arXiv},
  primaryClass  = {cs.LG},
  doi           = {10.48550/arXiv.2602.22261},
  url           = {https://arxiv.org/abs/2602.22261}
}

@inproceedings{wheeler-etal-2025-reducing,
  title     = {Reducing Environmental Costs whilst Maintaining Operational Effectiveness in Large Language Models through Chain Routing},
  author    = {Wheeler, Solomon and Torgbi, Melissa and Madabushi, Harish Tayyar},
  editor    = {Wartena, Christian and Heid, Ulrich},
  booktitle = {Proceedings of the 21st Conference on Natural Language Processing (KONVENS 2025): Workshops},
  month     = sep,
  year      = {2025},
  address   = {Hannover, Germany},
  publisher = {HsH Applied Academics},
  pages     = {250--273},
  url       = {https://aclanthology.org/2025.konvens-2.20/}
}

@misc{gar2026carbonaware,
  title        = {GAR: Carbon-Aware Routing for LLM Inference via Constrained Optimization},
  author       = {Anonymous authors},
  year         = {2026},
  note         = {Under review as a conference paper at ICLR 2026},
  howpublished = {Manuscript under review}
}

@inproceedings{fernandez2025energyconsiderations,
  title     = {Energy Considerations of Large Language Model Inference and Efficiency Optimizations},
  author    = {Fernandez, Jared and et al.},
  booktitle = {Proceedings of the 63rd Annual Meeting of the Association for Computational Linguistics (Volume 1: Long Papers)},
  pages     = {32556--32569},
  year      = {2025},
  publisher = {Association for Computational Linguistics}
}

@misc{liu2025greenllm,
  title         = {GreenLLM: SLO-Aware Dynamic Frequency Scaling for Energy-Efficient LLM Serving},
  author        = {Qunyou Liu and Darong Huang and Marina Zapater and David Atienza},
  year          = {2025},
  eprint        = {2508.16449},
  archivePrefix = {arXiv},
  primaryClass  = {cs.DC},
  url           = {https://arxiv.org/abs/2508.16449}
}

@inproceedings{yu2022orca,
  author    = {Gyeong-In Yu and
               Joo Seong Jeong and
               Geon-Woo Kim and
               Soojeong Kim and
               Byung-Gon Chun},
  title     = {{ORCA}: A Distributed Serving System for {Transformer-Based} Generative Models},
  booktitle = {16th USENIX Symposium on Operating Systems Design and Implementation (OSDI 22)},
  pages     = {521--538},
  year      = {2022},
  publisher = {USENIX Association}
}

@inproceedings{agrawal2024sarathi_serve,
  author    = {Amey Agrawal and Nitin Kedia and Ashish Panwar and Jayashree Mohan and Nipun Kwatra and Bhargav S. Gulavani and Alexey Tumanov and Ramachandran Ramjee},
  title     = {Taming Throughput-Latency Tradeoff in LLM Inference with Sarathi-Serve},
  booktitle = {18th USENIX Symposium on Operating Systems Design and Implementation (OSDI 24)},
  pages     = {117--134},
  year      = {2024},
  publisher = {USENIX Association},
  url       = {https://www.usenix.org/conference/osdi24/presentation/agrawal}
}

@article{shazeer2019fast,
  author  = {Noam Shazeer},
  title   = {Fast Transformer Decoding: One Write-Head is All You Need},
  journal = {CoRR},
  volume  = {abs/1911.02150},
  year    = {2019},
  url     = {https://arxiv.org/abs/1911.02150}
}

@article{zhu2023survey_compression_llms,
  author  = {X. Zhu and J. Li and Y. Liu and C. Ma and W. Wang},
  title   = {A Survey on Model Compression for Large Language Models},
  journal = {arXiv preprint arXiv:2308.07633},
  year    = {2023}
}

@article{park2024compression_language_models,
  author  = {S. Park and J. Choi and S. Lee and U. Kang},
  title   = {A Comprehensive Survey of Compression Algorithms for Language Models},
  journal = {arXiv preprint arXiv:2401.15347},
  year    = {2024}
}

@article{wang2024compression_efficient_inference_llms,
  author  = {W. Wang and W. Chen and Y. Luo and Y. Long and Z. Lin and L. Zhang and B. Lin and D. Cai and X. He},
  title   = {Model Compression and Efficient Inference for Large Language Models: A Survey},
  journal = {arXiv preprint arXiv:2402.09748},
  year    = {2024}
}

@article{tang2024transformer_compression_survey,
  author  = {Y. Tang and Y. Wang and J. Guo and Z. Tu and K. Han and H. Hu and D. Tao},
  title   = {A Survey on Transformer Compression},
  journal = {arXiv preprint arXiv:2402.05964},
  year    = {2024}
}

@article{ding2025efficiency,
  title={The efficiency spectrum of large language models: an algorithmic survey. arXiv; 2024},
  author={Ding, T and Chen, T and Zhu, H and Jiang, J and Zhong, Y and Zhou, J and Liang, L},
  journal={arXiv preprint arXiv:2312.00678},
  year={2025}
}

@article{miao2025towards,
  title={Towards efficient generative large language model serving: A survey from algorithms to systems},
  author={Miao, Xupeng and Oliaro, Gabriele and Zhang, Zhihao and Cheng, Xinhao and Jin, Hongyi and Chen, Tianqi and Jia, Zhihao},
  journal={ACM Computing Surveys},
  volume={58},
  number={1},
  pages={1--37},
  year={2025},
  publisher={ACM New York, NY}
}

@article{yuan2024llm_inference_unveiled,
  author  = {Z. Yuan and Y. Shang and Y. Zhou and Z. Dong and C. Xue and B. Wu and Z. Li and Q. Gu and Y. J. Lee and Y. Yan and others},
  title   = {LLM Inference Unveiled: Survey and Roofline Model Insights},
  journal = {arXiv preprint arXiv:2402.16363},
  year    = {2024}
}

@article{zhang2024speculative_execution_survey,
  author  = {C. Zhang and others},
  title   = {Beyond the Speculative Game: A Survey of Speculative Execution in Large Language Models},
  journal = {arXiv preprint arXiv:2404.14897},
  year    = {2024}
}

@article{xia2024speculative_decoding_survey,
  author  = {H. Xia and others},
  title   = {Unlocking Efficiency in Large Language Model Inference: A Comprehensive Survey of Speculative Decoding},
  journal = {arXiv preprint arXiv:2401.07851},
  year    = {2024}
}

@article{dong2022survey_in_context_learning,
  author        = {Q. Dong and et al.},
  title         = {A Survey for In-Context Learning},
  journal       = {arXiv preprint arXiv:2301.00234},
  year          = {2022}
}

@article{ning2023skeleton_of_thought,
  author        = {X. Ning and Z. Lin and Z. Zhou and H. Yang and Y. Wang},
  title         = {Skeleton-of-Thought: Large Language Models Can Do Parallel Decoding},
  journal       = {arXiv preprint arXiv:2307.15337},
  year          = {2023}
}

@article{jin2024adaptive_skeleton_graph_decoding,
  author        = {S. Jin and Y. Wu and H. Zheng and Q. Zhang and M. Lentz and Z. M. Mao and A. Prakash and F. Qian and D. Zhuo},
  title         = {Adaptive Skeleton Graph Decoding},
  journal       = {arXiv preprint arXiv:2402.12280},
  year          = {2024}
}

@article{liu2024apar,
  author        = {M. Liu and A. Zeng and B. Wang and P. Zhang and J. Tang and Y. Dong},
  title         = {APAR: LLMs Can Do Auto-Parallel Auto-Regressive Decoding},
  journal       = {arXiv preprint arXiv:2401.06761},
  year          = {2024}
}

@misc{cai2024medusa,
  author = {Tianle Cai and Yuhong Li and Zhengyang Geng and Hongwu Peng and James Demmel Lee and Tri Dao},
  title  = {Medusa: Simple LLM Inference Acceleration Framework with Multiple Decoding Heads},
  year   = {2024},
  eprint = {2401.10774},
  archivePrefix = {arXiv},
  primaryClass = {cs.LG}
}

@inproceedings{kwon2023pagedattention,
  author        = {W. Kwon and Z. Li and S. Zhuang and Y. Sheng and L. Zheng and C. H. Yu and J. Gonzalez and H. Zhang and I. Stoica},
  title         = {Efficient Memory Management for Large Language Model Serving with PagedAttention},
  booktitle     = {Proceedings of the 29th Symposium on Operating Systems Principles},
  pages         = {611--626},
  year          = {2023}
}

@article{zheng2023sglang,
  author        = {L. Zheng and L. Yin and Z. Xie and J. Huang and C. Sun and C. H. Yu and S. Cao and C. Kozyrakis and I. Stoica and J. E. Gonzalez and others},
  title         = {Efficiently Programming Large Language Models Using SGLang},
  journal       = {arXiv preprint arXiv:2312.07104},
  year          = {2023}
}

@inproceedings{chen2023frugalgpt,
  title={FrugalGPT: How to Use Large Language Models While Reducing Cost and Improving Performance},
  author={Chen, Lingjiao and Zaharia, Matei and Zou, James},
  booktitle={NeurIPS Workshop on Efficient Systems for Foundation Models},
  year={2023}
}

@article{kitchenham2002preliminary,
  title={Preliminary guidelines for empirical research in software engineering},
  author={Kitchenham, Barbara A and Pfleeger, Shari Lawrence and Pickard, Lesley M and Jones, Peter W and Hoaglin, David C. and El Emam, Khaled and Rosenberg, Jarrett},
  journal={IEEE Transactions on software engineering},
  volume={28},
  number={8},
  pages={721--734},
  year={2002},
  publisher={IEEE}
}

@book{fleiss1981measurement,
  title={Statistical Methods for Rates and Proportions},
  author={Fleiss, Joseph L. and Levin, Bruce and Paik, Myunghee Cho},
  edition={2},
  year={1981},
  publisher={John Wiley \& Sons},
  address={New York, USA},
  pages={212--236}
}

@article{verma2025energy,
  title={Energy-Aware Prompt Optimization for Large Language Models: Balancing Accuracy, Cost, and Sustainability},
  author={Verma, Ashish},
  journal={Authorea Preprints},
  year={2025},
  publisher={Authorea}
}

@article{tanaka2024adaptive,
  title={Adaptive neural token streaming: A novel approach for optimizing large language model efficiency and accuracy},
  author={Tanaka, Riwanami and Tsukada, Hiroto and Sakurai, Yusuke and Shigemura, Naoki and Hanabusa, Kenta and Takeyama, Ryota and Kamimura, Shohei},
  journal={Authorea Preprints},
  year={2024},
  publisher={Authorea}
}

@article{podder2026green,
  title={Green prompt engineering for sustainable generative AI},
  author={Podder, Sanjay and Date, Hema and Murthy, Shankar},
  journal={Environmental Science and Ecotechnology},
  pages={100684},
  year={2026},
  publisher={Elsevier}
}

@inproceedings{shriaarthy2025eco,
  title={Eco-Transformers: Carbon-Efficient and Accelerated Inference through Self-Pruning and Activation Freezing in Large Language Models},
  author={Shriaarthy, E and Sangeetha, S and Niranjana, RS and Rohini, R and Sivanesan, B and Vijayalakshmi, N},
  booktitle={2025 International Conference on Next Generation Computing Systems (ICNGCS)},
  pages={1--8},
  year={2025},
  organization={IEEE}
}

@article{wu2026crosskv,
  title={CrossKV: Accelerating Large Language Model Inference via Cross-Stage Dynamic Co-Optimization for KV Cache},
  author={Wu, Nan and Wang, Shenyu and Wang, Huizheng and Wang, Peng and Liu, Xiao and Wang, Zhihua and Hu, Yang and Jiang, Hanjun},
  journal={IEEE Transactions on Circuits and Systems I: Regular Papers},
  year={2026},
  publisher={IEEE}
}

@article{mandusia2025green,
  title={Green Transformers: Leveraging Neuromorphic Computing, Sparse Attention, and Adaptive Quantization for Energy-Efficient Inference},
  author={Mandusia, Ankit Kumar and Rana, Bhanu Pratap},
  journal={Authorea Preprints},
  year={2025},
  publisher={Authorea}
}

@inproceedings{rajput2024benchmarking,
  title={Benchmarking emerging deep learning quantization methods for energy efficiency},
  author={Rajput, Saurabhsingh and Sharma, Tushar},
  booktitle={2024 IEEE 21st International Conference on Software Architecture Companion (ICSA-C)},
  pages={238--242},
  year={2024},
  organization={IEEE}
}

@article{abstreiter2026sometimes,
  title={Sometimes Painful but Promising: Feasibility and Trade-Offs of On-Device Language Model Inference},
  author={Abstreiter, Maximilian and Tarkoma, Sasu and Morabito, Roberto},
  journal={ACM Transactions on Embedded Computing Systems},
  year={2026},
  publisher={ACM New York, NY}
}

@inproceedings{walkowiak2025energy,
  title={Energy Efficiency in Large Language Models: An Empirical Study},
  author={Walkowiak, Tomasz},
  booktitle={International Conference on Dependability and Complex Systems},
  pages={221--228},
  year={2025},
  organization={Springer}
}

@inproceedings{patel2024splitwise,
  title={Splitwise: Efficient generative llm inference using phase splitting},
  author={Patel, Pratyush and Choukse, Esha and Zhang, Chaojie and Shah, Aashaka and Goiri, {\'I}{\~n}igo and Maleki, Saeed and Bianchini, Ricardo},
  booktitle={2024 ACM/IEEE 51st Annual International Symposium on Computer Architecture (ISCA)},
  pages={118--132},
  year={2024},
  organization={IEEE}
}

@article{tian2026cache,
  title={Cache Your Prompt When It's Green—Carbon-Aware Caching for Large Language Model Serving},
  author={Tian, Yuyang and Sun, Desen and Ding, Yi and Liu, Sihang},
  journal={Proceedings of the ACM on Measurement and Analysis of Computing Systems},
  volume={10},
  number={1},
  pages={1--28},
  year={2026},
  publisher={ACM New York, NY, USA}
}

@inproceedings{zhang2024carbon,
  title={Carbon-aware workload shifting for mitigating environmental impact of generative AI models},
  author={Zhang, Eddie and Wu, Daniel and Boman, Jeffrey},
  booktitle={2024 IEEE International Conferences on Internet of Things (iThings) and IEEE Green Computing \& Communications (GreenCom) and IEEE Cyber, Physical \& Social Computing (CPSCom) and IEEE Smart Data (SmartData) and IEEE Congress on Cybermatics},
  pages={446--453},
  year={2024},
   organization={IEEE}
}

@inproceedings{mariaraj2025smart,
  title={Smart Meta-Controller: Energy Efficient Dynamic Routing for Large Language Model Inference Systems},
  author={Mariaraj, R Jeffrey and et al.},
  booktitle={2025 9th International Conference on Computational System and Information Technology for Sustainable Solutions (CSITSS)},
  pages={1--5},
  year={2025},
  organization={IEEE}
}

@article{shi2025disaggregated,
  title={Disaggregated Speculative Decoding for Carbon-Efficient LLM Serving},
  author={Shi, Tianyao and Wu, Yanran and Liu, Sihang and Ding, Yi},
  journal={IEEE Computer Architecture Letters},
  volume={24},
  number={2},
  pages={369--372},
  year={2025},
  publisher={IEEE}
}

@inproceedings{almeida2025idle,
  title={Idle Consumer GPUs as a Complement to Enterprise Hardware for LLM Inference: Performance, Cost and Carbon Analysis},
  author={Almeida, Aline Ribeiro de},
  booktitle={Proceedings of the 2025 6th International Artificial Intelligence and Blockchain Conference},
  pages={23--30},
  year={2025}
}

@article{lee2025tokens,
  title={From Tokens to Layers: Redefining Stall-Free Scheduling for LLM Serving with Layered Prefill},
  author={Lee, Gunjun and Kim, Jiwon and Park, Jaiyoung and Lee, Younjoo and Ahn, Jung Ho},
  journal={arXiv preprint arXiv:2510.08055},
  year={2025}
}

@inproceedings{qin2025dynamic,
  title={Dynamic-width speculative beam decoding for llm inference},
  author={Qin, Zongyue and He, Zifan and Prakriya, Neha and Cong, Jason and Sun, Yizhou},
  booktitle={Proceedings of the AAAI Conference on Artificial Intelligence},
  volume={39},
  number={23},
  pages={25056--25064},
  year={2025}
}

@inproceedings{wiesner2025carbon,
  title={Carbon-aware quality adaptation for energy-intensive services},
  author={Wiesner, Philipp and Grinwald, Dennis and Wei{\ss}, Philipp and Wilhelm, Patrick and Khalili, Ramin and Kao, Odej},
  booktitle={Proceedings of the 16th ACM International Conference on Future and Sustainable Energy Systems},
  pages={415--422},
  year={2025}
}

@inproceedings{wang2025wages,
  title={WAGES: Workload-Aware GPU Sharing System for Energy-Efficient Serverless LLM Serving},
  author={Wang, Tianyu and Rattihalli, Gourav and Dhakal, Aditya and Tang, Xulong and Milojicic, Dejan},
  booktitle={Proceedings of the SC'25 Workshops of the International Conference for High Performance Computing, Networking, Storage and Analysis},
  pages={508--515},
  year={2025}
}

@article{maliakel2025characterizing,
  title={Characterizing LLM Inference Energy-Performance Tradeoffs across Workloads and GPU Scaling},
  author={Maliakel, Paul Joe and Ilager, Shashikant and Brandic, Ivona},
  journal={arXiv preprint arXiv:2501.08219},
  year={2025}
}

@article{basit2026biscale,
  title={BiScale: Energy-efficient disaggregated LLM serving via phase-aware placement and DVFS},
  author={Basit, Omar and Liu, Yunzhao and Kong, Z Jonny and Hu, Y Charlie},
  journal={arXiv e-prints},
  pages={arXiv--2602},
  year={2026}
}

@article{he2025freesh,
  title={FREESH: Fair, Resource-and Energy-Efficient Scheduling for LLM Serving on Heterogeneous GPUs},
  author={He, Xuan and Fang, Zequan and Lian, Jinzhao and Tsang, Danny HK and Zhang, Baosen and Chen, Yize},
  journal={arXiv preprint arXiv:2511.00807},
  year={2025}
}

@article{hoxha2025deployment,
  title={A Deployment-Aware Framework for Carbon-and Water-Efficient LLM Serving},
  author={Hoxha, Julian and Thanasi-Bo{\c{c}}e, Marsela and Khalifa, Tarek},
  journal={Sustainability},
  volume={17},
  number={23},
  pages={10473},
  year={2025},
  publisher={MDPI}
}

@inproceedings{wang2025storellm,
  title={Storellm: Energy efficient large language model inference with permanently pre-stored attention matrices},
  author={Wang, Dan and Liu, Boan and Lu, Rui and Zhang, Zhaorui and Zhu, Shuntao},
  booktitle={Proceedings of the 16th ACM International Conference on Future and Sustainable Energy Systems},
  pages={398--406},
  year={2025}
}

@article{ps2024inference,
  title={Inference acceleration for large language models on CPUs},
  author={PS, Ditto and VG, Jithin and MS, Adarsh},
  journal={arXiv preprint arXiv:2406.07553},
  year={2024}
}

@article{kakolyris2024slo,
  title={Slo-aware gpu dvfs for energy-efficient llm inference serving},
  author={Kakolyris, Andreas Kosmas and Masouros, Dimosthenis and Xydis, Sotirios and Soudris, Dimitrios},
  journal={IEEE Computer Architecture Letters},
  volume={23},
  number={2},
  pages={150--153},
  year={2024},
  publisher={IEEE}
}

@inproceedings{li2025cim,
  title={CIM for Transformer Models: Enhancing Large Language Model Inference Efficiency},
  author={Li, Meng-Syuan and Ke, Jung-Fang and Huang, En-Ming and Liu, Zhi-Wei and Chen, Yu-Guang and Lee, Chun-Yi},
  booktitle={2025 IEEE Computer Society Annual Symposium on VLSI (ISVLSI)},
  volume={1},
  pages={1--6},
  year={2025},
  organization={IEEE}
}

@inproceedings{zhang2025energy,
  title={Energy Efficient GPU Frequency Scaling Policy for Inference Serving using Queue Model},
  author={Zhang, Yuhao and Zhang, Wenqian and Huang, Xiaowen and Zhang, Guanglin},
  booktitle={2025 International Conference on Sensor-Cloud and Edge Computing System (SCECS)},
  pages={57--62},
  year={2025},
  organization={IEEE}
}

@article{singh2025scalable,
  title={A Scalable and Energy-Efficient Processing-In-Memory Architecture for Gen-AI},
  author={Singh, Gian and Vrudhula, Sarma},
  journal={IEEE Journal on Emerging and Selected Topics in Circuits and Systems},
  year={2025},
  publisher={IEEE}
}

@article{wang2025low,
  title={Low-Latency PIM Accelerator for Edge LLM Inference},
  author={Wang, Xinyu and Sun, Xiaotian and Li, Wanqian and Min, Feng and Zhang, Xiaoyu and Zhang, Xinjiang and Han, Yinhe and Chen, Xiaoming},
  journal={IEEE Computer Architecture Letters},
  year={2025},
  publisher={IEEE}
}

@article{ortega2024pim,
  title={Pim-ai: A novel architecture for high-efficiency llm inference},
  author={Ortega, Cristobal and Falevoz, Yann and Ayrignac, Renaud},
  journal={arXiv preprint arXiv:2411.17309},
  year={2024}
}

@inproceedings{chen2025reis,
  title={REIS: A High-Performance and Energy-Efficient Retrieval System with In-Storage Processing},
  author={Chen, Kangqi and et al.},
  booktitle={Proceedings of the 52nd Annual International Symposium on Computer Architecture},
  pages={1171--1192},
  year={2025}
}

@article{li2025immutable,
  title={The Immutable Tensor Architecture: A Pure Dataflow Approach for Secure, Energy-Efficient AI Inference},
  author={Li, Fang},
  journal={arXiv preprint arXiv:2511.22889},
  year={2025}
}

@inproceedings{zhao2023sustainable,
  title={Sustainable supercomputing for AI: GPU power capping at HPC scale},
  author={Zhao, Dan and Samsi, Siddharth and McDonald, Joseph and Li, Baolin and Bestor, David and Jones, Michael and Tiwari, Devesh and Gadepally, Vijay},
  booktitle={Proceedings of the 2023 ACM symposium on cloud computing},
  pages={588--596},
  year={2023}
}

@inproceedings{zhu2025leveraging,
  title={Leveraging compute-in-memory for efficient generative model inference in tpus},
  author={Zhu, Zhantong and Li, Hongou and Ren, Wenjie and Wu, Meng and Ye, Le and Huang, Ru and Jia, Tianyu},
  booktitle={2025 Design, Automation \& Test in Europe Conference (DATE)},
  pages={1--7},
  year={2025},
  organization={IEEE}
}

@article{ji2026flash,
  title={E-Flash: Energy-Efficient LLM Mapping on NAND Flash-Based In-Storage Inference Computing},
  author={Ji, Gisan and Shin, Sanghun and Baik, Jangho and Shim, Wonbo and Ryu, Sungju},
  journal={IEEE Transactions on Very Large Scale Integration (VLSI) Systems},
  year={2026},
  publisher={IEEE}
}

@inproceedings{wang2025decoupled,
  title={Decoupled Analysis of DVFS Effects in Prefill and Decode Stages of Large Language Model Inference},
  author={Wang, Lei and Guo, Qinglai},
  booktitle={2025 IEEE 9th Conference on Energy Internet and Energy System Integration (EI2)},
  pages={2556--2561},
  year={2025},
  organization={IEEE}
}

@article{potkins2024improve,
  title={Improve the accuracy and efficiency of large language models via dynamic token compression and adaptive layer pruning},
  author={Potkins, Fabian and Langton, Octavian and Smirnov, Valentin and Gallacher, Cedric and Abernathy, Roderick},
  journal={Authorea Preprints},
  year={2024},
  publisher={Authorea}
}

@article{xu2026daypq,
  title={DayPQ: Dynamic Layerwise Pruning and Quantization for LLM Inference Acceleration},
  author={Xu, Yongchao and Li, Mingxuan and Zhuo, Ziying and Zhi, Qinzhe and Jia, Tianyu and Wang, Meiqi},
  journal={IEEE Transactions on Very Large Scale Integration (VLSI) Systems},
  year={2026},
  publisher={IEEE}
}

@article{yubeaton2025huff,
  title={Huff-llm: End-to-end lossless compression for efficient llm inference},
  author={Yubeaton, Patrick and Mahmoud, Tareq and Naga, Shehab and Taheri, Pooria and Xia, Tianhua and George, Arun and Khalil, Yasmein and Zhang, Sai Qian and Joshi, Siddharth and Hegde, Chinmay and others},
  journal={arXiv preprint arXiv:2502.00922},
  year={2025}
}

@article{zhang2024mixpe,
  title={Mixpe: Quantization and hardware co-design for efficient llm inference},
  author={Zhang, Yu and Wang, Mingzi and Zou, Lancheng and Liu, Wulong and Zhen, Hui-Ling and Yuan, Mingxuan and Yu, Bei},
  journal={arXiv preprint arXiv:2411.16158},
  year={2024}
}

@inproceedings{juneja2026halo,
  title={HALO: Hardware-aware quantization with low critical-path-delay weights for LLM acceleration},
  author={Juneja, Rohan and Aggarwal, Shivam and Huda, Safeen and Mitra, Tulika and Peh, Li-Shiuan},
  booktitle={Proceedings of the AAAI Conference on Artificial Intelligence},
  volume={40},
  number={27},
  pages={22472--22481},
  year={2026}
}

@article{moradifirouzabadi2025end,
  title={End-to-end acceleration of generative models with runtime regularized kv cache management},
  author={Moradifirouzabadi, Ashkan and Kang, Mingu},
  journal={IEEE Journal on Emerging and Selected Topics in Circuits and Systems},
  year={2025},
  publisher={IEEE}
}

@inproceedings{lee2025prime,
  title={PriME: PIM-Aware Efficient Compression for Memory-Bound Embedding Layers in sLLMs},
  author={Lee, Junghyeok and Jang, Jihoon and Kim, Hyun},
  booktitle={2025 IEEE 43rd International Conference on Computer Design (ICCD)},
  pages={221--228},
  year={2025},
  organization={IEEE}
}

@inproceedings{xu2025unicaim,
  title={UniCAIM: A Unified CAM/CIM Architecture with Static-Dynamic KV Cache Pruning for Efficient Long-Context LLM Inference},
  author={Xu, Weikai and Zeng, Wenxuan and Huang, Qianqian and Li, Meng and Huang, Ru},
  booktitle={2025 62nd ACM/IEEE Design Automation Conference (DAC)},
  pages={1--7},
  year={2025},
  organization={IEEE}
}

@inproceedings{qiao2026tellme,
  title={TeLLMe: An Efficient End-to-End Ternary LLM Prefill and Decode Accelerator with Table-Lookup Matmul on Edge FPGAs},
  author={Qiao, Ye and et al.},
  booktitle={Proceedings of the 2026 ACM/SIGDA International Symposium on Field Programmable Gate Arrays},
  pages={247--257},
  year={2026}
}

@article{hooper2025fgmp,
  title={FGMP: fine-grained mixed-precision weight and activation quantization for hardware-accelerated LLM inference},
  author={Hooper, Coleman and Sakr, Charbel and Keller, Ben and Venkatesan, Rangharajan and Keutzer, Kurt and Shao, Sophia and Khailany, Brucek},
  journal={arXiv preprint arXiv:2504.14152},
  year={2025}
}

@inproceedings{fu2025h,
  title={H 2 EAL: Hybrid-Bonding Architecture with Hybrid Sparse Attention for Efficient Long-Context LLM Inference},
  author={Fu, Zizhuo and et al.},
  booktitle={2025 IEEE/ACM International Conference On Computer Aided Design (ICCAD)},
  pages={1--9},
  year={2025},
  organization={IEEE}
}

@article{wu2025kllm,
  title={KLLM: Fast LLM Inference with K-Means Quantization},
  author={Wu, Xueying and Zhou, Baijun and Gao, Zhihui and Fu, Yuzhe and Zheng, Qilin and He, Yintao and Li, Hai},
  journal={arXiv preprint arXiv:2507.23035},
  year={2025}
}

@article{he2024hlstransform,
  title={HLSTransform: Energy-efficient LLaMA 2 inference on FPGAs via high level synthesis},
  author={He, Andy and Key, Darren and Bulling, Mason and Chang, Andrew and Shapiro, Skyler and Lee, Everett},
  journal={arXiv preprint arXiv:2405.00738},
  year={2024}
}

@article{jayanth2026fast,
  title={FAST-Prefill: FPGA Accelerated Sparse Attention for Long Context LLM Prefill},
  author={Jayanth, Rakshith and Prasanna, Viktor},
  journal={arXiv preprint arXiv:2602.20515},
  year={2026}
}

@article{yin2025tereffic,
  title={Tereffic: Highly efficient ternary llm inference on fpga},
  author={Yin, Chenyang and Bai, Zhenyu and Venkatram, Pranav and Aggarwal, Shivam and Li, Zhaoying and Mitra, Tulika},
  journal={arXiv preprint arXiv:2502.16473},
  year={2025}
}

@article{wang2026harmonia,
  title={Harmonia: Algorithm-Hardware Co-Design for Memory-and Compute-Efficient BFP-based LLM Inference},
  author={Wang, Xinyu and Li, Jieyu and Sun, Yanan and He, Weifeng},
  journal={arXiv preprint arXiv:2602.04595},
  year={2026}
}

@article{liang2026accllm,
  title={AccLLM: Accelerating long-context LLM inference via algorithm-hardware co-design},
  author={Liang, Yanbiao and Shi, Huihong and Shao, Haikuo and Wang, Zhongfeng},
  journal={IEEE Transactions on Very Large Scale Integration (VLSI) Systems},
  year={2026},
  publisher={IEEE}
}

@inproceedings{crafton2025finding,
  title={Finding the Pareto Frontier of Low-Precision Data Formats and MAC Architecture for LLM Inference},
  author={Crafton, Brian and Peng, Xiaochen and Sun, Xiaoyu and Lele, Ashwin and Zhang, Bo and Khwa, Win-San and Akarvardar, Kerem},
  booktitle={2025 62nd ACM/IEEE Design Automation Conference (DAC)},
  pages={1--7},
  year={2025},
  organization={IEEE}
}

@inproceedings{kim20254,
  title={A 4.21 TFLOPS/W Memory-Efficient LLM Inference Accelerator with Bit-Layered Non-Uniform Quantization},
  author={Kim, Byeongcheol and Kim, Sangjin and Ha, Sangwoo and Um, Soyeon and Sohn, Kyomin and Yoo, Hoi-Jun},
  booktitle={2025 IEEE International Symposium on Circuits and Systems (ISCAS)},
  year={2025},
  organization={IEEE}
}

@article{monteiro2026reducing,
  title={Reducing Energy Footprint of LLM Inference Through FPGA-Based Heterogeneous Computing Platforms},
  author={Monteiro, Thiago Cormie and Guerrieri, Andrea},
  journal={Electronics},
  volume={15},
  number={5},
  pages={1052},
  year={2026},
  publisher={MDPI},
  doi={10.3390/electronics15051052},
  url={https://doi.org/10.3390/electronics15051052}
}

@inproceedings{wang2025mcbp,
  title={MCBP: A Memory-Compute Efficient LLM Inference Accelerator Leveraging Bit-Slice-enabled Sparsity and Repetitiveness},
  author={Wang, Huizheng and Wang, Zichuan and Yue, Zhiheng and Long, Yousheng and Wei, Taiquan and Yang, Jianxun and Wang, Yang and Li, Chao and Wei, Shaojun and Hu, Yang and Yin, Shouyi},
  booktitle={Proceedings of the 58th IEEE/ACM International Symposium on Microarchitecture (MICRO)},
  pages={1592--1608},
  year={2025},
  publisher={ACM},
  doi={10.1145/3725843.3756037},
  url={https://doi.org/10.1145/3725843.3756037}
}

@inproceedings{wang2025lad,
  title={LAD: Efficient Accelerator for Generative Inference of LLM with Locality Aware Decoding},
  author={Wang, H. and Li, Y. and Xu, H. and Wang, Y. and Liu, L. and Yang, J. and Han, Y.},
  booktitle={2025 IEEE International Symposium on High Performance Computer Architecture (HPCA)},
  year={2025},
  publisher={IEEE},
  url={https://ieeexplore.ieee.org}
}

@incollection{du2024energy,
  title={Energy Efficient Spike Transformer Accelerator at the Edge},
  author={Du, C. and Wen, Q. and Wei, Z. and Zhang, H.},
  booktitle={Intelligent Marine Technology and Systems},
  year={2024},
  publisher={Springer}
}

@inproceedings{xu2024llamaf,
  title={Llamaf: An Efficient Llama2 Architecture Accelerator on Embedded FPGAs},
  author={Xu, H. and Li, Y. and Ji, S.},
  booktitle={2024 IEEE 10th World Forum on Internet of Things (WF-IoT)},
  year={2024},
  publisher={IEEE},
  url={https://ieeexplore.ieee.org}
}

@inproceedings{fang2025anda,
  title={Anda: Unlocking Efficient LLM Inference with a Variable-Length Grouped Activation Data Format},
  author={Fang, C. and Shi, M. and Geens, R. and Symons, A. and Wang, Z. and Verhelst, M.},
  booktitle={2025 IEEE International Symposium on High Performance Computer Architecture (HPCA)},
  year={2025},
  publisher={IEEE},
  url={https://ieeexplore.ieee.org}
}

@inproceedings{ardakani2025llmpi,
  title={LLMPi: Optimizing LLMs for High-Throughput on Raspberry Pi},
  author={Ardakani, M. and Malekar, J. and Zand, R.},
  booktitle={Proceedings of the IEEE/CVF Conference on Computer Vision and Pattern Recognition Workshops (CVPRW)},
  year={2025},
  publisher={IEEE/CVF},
  url={https://openaccess.thecvf.com}
}

@inproceedings{kang2025dynamic,
  title={Dynamic Offloading Optimization for Multi-PIM LLM Inference},
  author={Kang, J. Y. and Ko, J. H. and Hwang, T. and Choi, K. H.},
  booktitle={2025 IEEE/IEIE International Conference on Consumer Electronics (ICCE)},
  year={2025},
  publisher={IEEE},
  url={https://ieeexplore.ieee.org}
}

@article{jiang2026rapid,
  title={Power Aware Dynamic Reallocation For Inference},
  author={Jiang, Y. and Chowdhary, S. and Morris, N. and Jain, R. and Manne, S. and Bayliss, S.},
  journal={arXiv preprint arXiv:2601.12241},
  year={2026},
  url={https://arxiv.org/abs/2601.12241}
}

@article{scherer2024deeploy,
  title={Deeploy: Enabling Energy-Efficient Deployment of Small Language Models on Heterogeneous Microcontrollers},
  author={Scherer, M. and Macan, L. and Jung, V. J. B. and Wiese, P. and Bompani, L. and Burrello, A. and Conti, F. and Benini, L.},
  journal={IEEE Transactions on Computer-Aided Design of Integrated Circuits and Systems},
  year={2024},
  publisher={IEEE},
  url={https://ieeexplore.ieee.org}
}

@inproceedings{fan2025ellie,
  title={ELLIE: Energy-Efficient LLM Inference at the Edge Via Prefill-Decode Splitting},
  author={Fan, H. and Lin, Y. C. and Prasanna, V.},
  booktitle={2025 IEEE 36th International Conference on Application-specific Systems, Architectures and Processors (ASAP)},
  year={2025},
  publisher={IEEE},
  url={https://ieeexplore.ieee.org}
}

@article{feng2026ellm,
  title={E$^2$LLM: Structure-Guided Efficient Inference for LLMs in Distributed Edge-IoT Environments},
  author={Feng, X. and Yang, H. and Chen, Z. and Luo, C. and Wang, Y. and Zhou, Z. and Xu, W. and Leung, V. C. M.},
  journal={IEEE Transactions on Mobile Computing},
  year={2026},
  publisher={IEEE},
  url={https://ieeexplore.ieee.org}
}

@inproceedings{xu2026spactndp,
  title={SpAct-NDP: Efficient LLM Inference via Sparse Activation on NDP-GPU Heterogeneous Architecture},
  author={Xu, J. and Xie, T. and Zhou, Y. and Li, J. and Lian, Y. and Zhu, Z. and Wang, Y. and Dai, G.},
  booktitle={2026 31st Asia and South Pacific Design Automation Conference (ASP-DAC)},
  year={2026},
  publisher={IEEE},
  url={https://ieeexplore.ieee.org}
}

@article{xu2025camel,
  title={Camel: Energy-aware LLM Inference on Resource-Constrained Devices},
  author={Xu, H. and Peng, L. and Song, S. and Liu, X. and Jun, M. and Li, S. and Yu, J. and Mao, X.},
  journal={arXiv preprint arXiv:2508.09173},
  year={2025},
  url={https://arxiv.org/abs/2508.09173}
}

@inproceedings{appuswamy2024northpole,
  title={Breakthrough Low-Latency, High-Energy-Efficiency LLM Inference Performance Using NorthPole},
  author={Appuswamy, R. and others},
  booktitle={2024 IEEE High Performance Extreme Computing Conference (HPEC)},
  year={2024},
  publisher={IEEE},
  url={https://ieeexplore.ieee.org}
}

@inproceedings{yang2025hades,
  title={Hades: Hardware accelerated decoding for efficient speculation in large language models},
  author={Yang, Ze and Jin, Yihong and Xu, Xinhe},
  booktitle={2025 6th International Conference on Computer Engineering and Application (ICCEA)},
  pages={01--05},
  year={2025},
  organization={IEEE}
}

@article{jareno2026energyefficient,
  title={Energy-efficient Large Language Models},
  author={Jare{\~n}o, J. and Arag{\'o}n-Jurado, J. M. and De La Torre, J. C. and Ruiz, P. and Dorronsoro, B.},
  journal={Future Generation Computer Systems},
  year={2026},
  publisher={Elsevier},
  url={https://www.sciencedirect.com/journal/future-generation-computer-systems}
}

@inproceedings{wang2025leap,
  title={LEAP: LLM Inference on Scalable PIM-NoC Architecture with Balanced Dataflow and Fine-Grained Parallelism},
  author={Wang, Y. and Chong, Y. J. and Fong, X.},
  booktitle={2025 IEEE/ACM International Conference on Computer-Aided Design (ICCAD)},
  year={2025},
  publisher={IEEE/ACM},
  url={https://ieeexplore.ieee.org}
}

@inproceedings{moon2025hybe,
  title={Hybe: GPU-NPU Hybrid System for Efficient LLM Inference with Million-Token Context Window},
  author={Moon, S. and Cha, J. and Park, H. and Kim, J. Y.},
  booktitle={Proceedings of the 52nd Annual International Symposium on Computer Architecture (ISCA)},
  year={2025},
  publisher={ACM},
  url={https://dl.acm.org}
}

@inproceedings{everman2023carbon,
  title={Evaluating the Carbon Impact of Large Language Models at the Inference Stage},
  author={Everman, B. and Villwock, T. and Chen, D. and Soto, N. and Zhang, O. and Zong, Z.},
  booktitle={2023 IEEE International Performance, Computing, and Communications Conference (IPCCC)},
  year={2023},
  publisher={IEEE},
  url={https://ieeexplore.ieee.org}
}

@article{fan2025sparse,
  title={Sparse Attention Remapping with Clustering for Efficient LLM Decoding on PIM},
  author={Fan, Zehao and Gagnon, Garrett and Liu, Zhenyu and Liu, Liu},
  journal={arXiv preprint arXiv:2505.05772},
  year={2025}
}

@inproceedings{wang2025asymserve,
  title={AsymServe: Demystifying and Optimizing LLM Serving Efficiency on CPU Acceleration Units},
  author={Wang, Xinkai and Zhuansun, Yiming and Li, Chao and Wang, Jing and Hou, Xiaofeng and Sun, Lingyu and Wang, Luping and Guo, Minyi},
  booktitle={Advanced Parallel Processing Technologies},
  series={Lecture Notes in Computer Science},
  volume={16062},
  pages={231--245},
  year={2025},
  publisher={Springer},
  doi={10.1007/978-981-96-7300-6_16},
  url={https://link.springer.com/chapter/10.1007/978-981-96-7300-6_16}
}

@inproceedings{moore2025slit,
  title={Sustainable Carbon-Aware and Water-Efficient LLM Scheduling in Geo-Distributed Cloud Datacenters},
  author={Moore, H. and Qi, S. and Hogade, N. and Milojicic, D. and Bash, C. and Pasricha, S.},
  booktitle={Proceedings of the Great Lakes Symposium on VLSI 2025},
  year={2025},
  publisher={ACM},
  url={https://dl.acm.org}
}

@inproceedings{xu2025llmnpu,
  title={Fast On-Device LLM Inference with NPUs},
  author={Xu, D. and Zhang, H. and Yang, L. and Liu, R. and Huang, G. and Xu, M. and Liu, X.},
  booktitle={Proceedings of the 30th ACM International Conference on Architectural Support for Programming Languages and Operating Systems (ASPLOS)},
  year={2025},
  publisher={ACM},
  url={https://dl.acm.org}
}

@inproceedings{xia2025kelle,
  title={Kelle: Co-design KV Caching and eDRAM for Efficient LLM Serving in Edge Computing},
  author={Xia, T. and Zhang, S. Q.},
  booktitle={Proceedings of the 58th IEEE/ACM International Symposium on Microarchitecture (MICRO)},
  year={2025},
  publisher={ACM/IEEE},
  url={https://dl.acm.org}
}

@article{yu2025voltanallm,
  title={VoltanaLLM: Feedback-Driven Frequency Control and State-Space Routing for Energy-Efficient LLM Serving},
  author={Yu, J. and Taneja, A. and Lin, J. and Zhang, M.},
  journal={arXiv preprint arXiv:2509.04827},
  year={2025},
  url={https://arxiv.org/abs/2509.04827}
}

@article{li2025ecoserve,
  title={EcoServe: Designing Carbon-Aware AI Inference Systems},
  author={Li, Y. and Hu, Z. and Choukse, E. and Fonseca, R. and Suh, G. E. and Gupta, U.},
  journal={arXiv preprint arXiv:2502.05043},
  year={2025},
  url={https://arxiv.org/abs/2502.05043}
}

@article{lu2025thermalaware,
  title={A Thermal-aware Workload Scheduler for High-performance LLM Inference in Cooling-regulated Datacenters},
  author={Lu, R. and Wang, D.},
  journal={ACM SIGENERGY Energy Informatics Review},
  year={2025},
  publisher={ACM},
  url={https://dl.acm.org}
}

@article{gosmar2026prompt,
  title={Prompt Injection Mitigation with Agentic AI, Nested Learning, and AI Sustainability via Semantic Caching},
  author={Gosmar, D. and Dahl, D. A.},
  journal={arXiv preprint arXiv:2601.13186},
  year={2026},
  url={https://arxiv.org/abs/2601.13186}
}

@article{li2024sprout,
  title={Toward Sustainable GenAI Using Generation Directives for Carbon-Friendly Large Language Model Inference},
  author={Li, Baolin and Jiang, Yankai and Gadepally, Vijay and Tiwari, Devesh},
  journal={arXiv preprint arXiv:2403.12900},
  year={2024},
  doi={10.48550/arXiv.2403.12900},
  url={https://doi.org/10.48550/arXiv.2403.12900}
}

@inproceedings{kalzhan2025energyefficient,
  title={An Energy-Efficient Vision Language Model Inference with Importance-aware Token Pruning},
  author={Kalzhan, Z. and Hong, S. and Yoo, H. J.},
  booktitle={2025 IEEE Asia Pacific Conference on Circuits and Systems (APCCAS)},
  year={2025},
  publisher={IEEE},
  url={https://ieeexplore.ieee.org}
}

@inproceedings{khan2025optimizing,
  title={Optimizing Large Language Models: Metrics, Energy Efficiency, and Case Study Insights},
  author={Khan, T. and Motie, S. and Kocak, S. A. and Raza, S.},
  booktitle={2025 IEEE Conference on Artificial Intelligence (CAI)},
  year={2025},
  publisher={IEEE},
  url={https://ieeexplore.ieee.org}
}

@article{argerich2024measuring,
  title={Measuring and Improving the Energy Efficiency of Large Language Models Inference},
  author={Argerich, M. F. and Pati{\~n}o-Mart{\'\i}nez, M.},
  journal={IEEE Access},
  volume={12},
  pages={},
  year={2024},
  publisher={IEEE},
  url={https://ieeexplore.ieee.org}
}

@article{chadli2025pearl,
  title={PEARL: Performance and Energy Aware Routing for LLMs},
  author={Chadli, K. and Botterweck, G. and Saber, T.},
  journal={Future Generation Computer Systems},
  year={2025},
  publisher={Elsevier},
  url={https://www.sciencedirect.com/journal/future-generation-computer-systems}
}

@article{peng2024m2cache,
  title={Harnessing Your DRAM and SSD for Sustainable and Accessible LLM Inference with Mixed-Precision and Multi-Level Caching},
  author={Peng, J. and Cao, Z. and Qu, H. and Zhang, Z. and Guo, C. and Zhang, Y. and Cao, Z. and Chen, T.},
  journal={arXiv preprint arXiv:2410.14740},
  year={2024},
  url={https://arxiv.org/abs/2410.14740}
}

@inproceedings{raha2025llmnpu,
  title={LLM-NPU: Towards Efficient Foundation Model Inference on Low-Power Neural Processing Units},
  author={Raha, A. and et al.},
  booktitle={2025 IEEE International Conference on Omni-layer Intelligent Systems (COINS)},
  year={2025},
}

@article{tanaka2026aellm,
  title={AE-LLM: Adaptive Efficiency Optimization for Large Language Models},
  author={Tanaka, K. and Ito, M. and Nishimura, Y. and Matsuda, K. and Nakayama, A.},
  journal={arXiv preprint arXiv:2603.20492},
  year={2026},
  url={https://arxiv.org/abs/2603.20492}
}

@techreport{belkahla2025measurements,
  title={Measurements and Optimizations of the Energy Consumption of AI Systems},
  author={Belkahla, M. and Hains, G.},
  institution={HAL Open Science},
  year={2025},
}

@article{qiao2025tellme,
  title={TeLLMe: An Energy-Efficient Ternary LLM Accelerator for Prefilling and Decoding on Edge FPGAs},
  author={Qiao, Y. and Chen, Z. and Zhang, Y. and Wang, Y. and Huang, S.},
  journal={arXiv preprint arXiv:2504.16266},
  year={2025},
  url={https://arxiv.org/abs/2504.16266}
}

@article{xie2025energyaware,
  title={An Energy-Aware Generative AI Edge Inference Framework for Low-Power IoT Devices},
  author={Xie, Y. and Fang, Q.},
  journal={Electronics},
  volume={14},
  number={},
  year={2025},
  publisher={MDPI},
  url={https://www.mdpi.com/journal/electronics}
}

@article{hisaharo2024optimizing,
  title={Optimizing LLM Inference Clusters for Enhanced Performance and Energy Efficiency},
  author={Hisaharo, S. and Nishimura, Y. and Takahashi, A.},
  journal={Authorea Preprints},
  year={2024},
  url={https://www.techrxiv.org}
}

@article{chen2026onewlaw,
  title={The 1/W Law: An Analytical Study of Context-Length Routing Topology and GPU Generation Gains for LLM Inference Energy Efficiency},
  author={Chen, H. and Liu, X. and Liu, Y. and Jiang, J. and He, B. and Liu, X.},
  journal={arXiv preprint arXiv:2603.17280},
  year={2026},
  url={https://arxiv.org/abs/2603.17280}
}

@inproceedings{li2026ecothink,
  title={EcoThink: A Green Adaptive Inference Framework for Sustainable and Accessible Agents},
  author={Li, L. and Lu, Z.},
  booktitle={Proceedings of the ACM Web Conference 2026},
  year={2026},
  publisher={ACM},
  url={https://dl.acm.org}
}

@article{nguyen2025towards,
  title={Towards Sustainable Large Language Model Serving},
  author={Nguyen, Sophia and Zhou, Beihao and Ding, Yi and Liu, Sihang},
  journal={ACM SIGENERGY Energy Informatics Review},
  volume={4},
  number={5},
  pages={134--140},
  year={2025},
  publisher={ACM},
  doi={10.1145/3727200.3727220},
  url={https://doi.org/10.1145/3727200.3727220}
}

@manual{nvidia_nvml,
  author       = {{NVIDIA Corporation}},
  title        = {{NVIDIA Management Library (NVML) API Reference Guide}},
  organization = {NVIDIA Corporation},
  note         = {Accessed: 2026-07-21}
}

@manual{nvidia_smi,
  author       = {{NVIDIA Corporation}},
  title        = {{NVIDIA System Management Interface}},
  organization = {NVIDIA Corporation},
  year = {[n.\,d.]},
  note         = {Accessed: 2026-07-21}
}

@inproceedings{david2010rapl,
  author    = {David, Howard and Gorbatov, Eugene and Hanebutte,
               Ulf R. and Khanna, Rahul and Le, Christian},
  title     = {{RAPL: Memory Power Estimation and Capping}},
  booktitle = {Proceedings of the International Symposium on
               Low Power Electronics and Design},
  pages     = {189--194},
  year      = {2010}
}

@manual{nvidia_dcgm,
  author       = {{NVIDIA Corporation}},
  title        = {{NVIDIA Data Center GPU Manager Documentation}},
  organization = {NVIDIA Corporation},
  year = {[n.\,d.]},
  note         = {Accessed: 2026-07-21}
}

@manual{linux_perf,
  author = {{Linux Kernel Community}},
  title  = {{perf: Performance Analysis Tools for Linux}},
  note   = {Linux perf manual, accessed: 2026-07-21}
}

@manual{apple_powermetrics,
  author       = {{Apple Inc.}},
  title        = {{powermetrics Manual Page}},
  organization = {Apple Inc.},
  note         = {macOS system power and hardware activity
                  monitoring utility}
}

@misc{hwinfo,
  author = {{REALiX}},
  title  = {{HWiNFO: Hardware Analysis, Monitoring and Reporting}},
  note   = {Software documentation, accessed: 2026-07-21}
}

@article{courty2024codecarbon,
  author  = {Benoit Courty and Victor Schmidt and Sasha Luccioni and
             Goyal-Kamal and Marion Coutarel and Boris Feld and
             Jérémy Lecourt and Liam Connell and others},
  title   = {{CodeCarbon}: Estimate and Track Carbon Emissions from
             Machine Learning Computing},
  journal = {Journal of Open Source Software},
  volume  = {9},
  number  = {96},
  pages   = {6311},
  year    = {2024},
  doi     = {10.21105/joss.06311}
}

@article{anthony2020carbontracker,
  author  = {Anthony, Lasse F. Wolff and Kanding, Benjamin
             and Selvan, Raghavendra},
  title   = {{Carbontracker: Tracking and Predicting the Carbon
              Footprint of Training Deep Learning Models}},
  journal = {arXiv preprint arXiv:2007.03051},
  year    = {2020}
}

@inproceedings{you2023zeus,
  author    = {You, Jie and Chung, Jae-Won and Chowdhury,
               Mosharaf},
  title     = {{Zeus: Understanding and Optimizing GPU Energy
                Consumption of DNN Training}},
  booktitle = {Proceedings of the 20th USENIX Symposium on
               Networked Systems Design and Implementation},
  year      = {2023}
}

@manual{monsoon_power_monitor,
  author       = {{Monsoon Solutions Inc.}},
  title        = {{Monsoon High Voltage Power Monitor User Guide}},
  organization = {Monsoon Solutions Inc.},
  year         = {2023},
  note         = {Available: https://www.msoon.com}
}

@manual{nvidia_nsight,
  author       = {{NVIDIA Corporation}},
  title        = {NVIDIA Nsight Systems User Guide},
  organization = {NVIDIA Corporation},
  year         = {2026},
  note         = {Accessed: July 21, 2026},
  url          = {https://docs.nvidia.com/nsight-systems/}
}

@manual{nvidia_tegrastats,
  author       = {{NVIDIA Corporation}},
  title        = {Jetson Linux Developer Guide: Tegrastats Utility},
  organization = {NVIDIA Corporation},
  year         = {2026},
  note         = {Accessed: July 21, 2026},
  url          = {https://docs.nvidia.com/jetson/archives/r36.2/DeveloperGuide/AT/JetsonLinuxDevelopmentTools/TegrastatsUtility.html}
}

@manual{monsoon2023,
  author       = {{Monsoon Solutions Inc.}},
  title        = {High Voltage Power Monitor User Guide},
  organization = {Monsoon Solutions Inc.},
  year         = {2023},
  note         = {Accessed: July 21, 2026},
  url          = {https://www.msoon.com/hvpm-product-documentation}
}

@manual{makerhawk_um34c,
  author       = {{MakerHawk}},
  title        = {UM34C USB Power Meter User Manual},
  organization = {MakerHawk},
  year         = {2023},
  note         = {USB voltage/current measurement instrument}
}

@manual{keysight_n6715c,
  author       = {{Keysight Technologies}},
  title        = {N6715C DC Power Analyzer User Guide},
  organization = {Keysight Technologies},
  year         = {2024},
  note         = {Accessed: July 21, 2026},
  url          = {https://www.keysight.com/gb/en/product/N6715C/dc-power-analyzer-custom-configured-modular-600-w-4-slots.html}
}

@manual{joulescope_js110,
  author       = {{Jetperch LLC}},
  title        = {Joulescope JS110 User Guide},
  organization = {Jetperch LLC},
  year         = {2023},
  note         = {Accessed: July 21, 2026},
  url          = {https://www.joulescope.com}
}

@standard{iso30134pue2026,
  author       = {{International Organization for Standardization}},
  title        = {{ISO/IEC 30134-2:2026}: Information Technology---Data
                  Centres Key Performance Indicators---Part 2:
                  Power Usage Effectiveness ({PUE})},
  organization = {ISO},
  year         = {2026}
}

@standard{iso30134wue2018,
  author       = {{International Organization for Standardization}},
  title        = {{ISO/IEC 30134-9:2018}: Information Technology---Data
                  Centres---Key Performance Indicators---Part 9:
                  Water Usage Effectiveness ({WUE})},
  organization = {ISO},
  year         = {2018}
}

@article{gonzalez1996energy,
  author  = {Ricardo Gonzalez and Mark Horowitz},
  title   = {Energy Dissipation in General Purpose Microprocessors},
  journal = {IEEE Journal of Solid-State Circuits},
  volume  = {31},
  number  = {9},
  pages   = {1277--1284},
  year    = {1996},
  doi     = {10.1109/4.535411}
}

@article{banbury2021mlperftiny,
  author  = {Colby Banbury and Vijay Janapa Reddi and Peter Torelli
             and Jeremy Holleman and Nat Jeffries and Csaba Kiraly
             and others},
  title   = {{MLPerf Tiny} Benchmark},
  journal = {Proceedings of the Neural Information Processing Systems
             Track on Datasets and Benchmarks},
  volume  = {1},
  year    = {2021}
}

@article{wan2024efficient,
  author  = {Zhongwei Wan and Xin Wang and Che Liu and Samiul Alam
             and Yu Zheng and Jiachen Liu and Zhongnan Qu and
             Shen Yan and Yi Zhu and Quanlu Zhang and
             Mosharaf Chowdhury and Mi Zhang},
  title   = {Efficient Large Language Models: A Survey},
  journal = {Transactions on Machine Learning Research},
  year    = {2024}
}

@inproceedings{xia2024unlocking,
  author    = {Heming Xia and
               Zhe Yang and
               Qingxiu Dong and
               Peiyi Wang and
               Yongqi Li and
               Tao Ge and
               Tianyu Liu and
               Wenjie Li and
               Zhifang Sui},
  title     = {Unlocking Efficiency in Large Language Model Inference:
               A Comprehensive Survey of Speculative Decoding},
  booktitle = {Findings of the Association for Computational Linguistics:
               ACL 2024},
  pages     = {7655--7671},
  year      = {2024},
  address   = {Bangkok, Thailand},
  publisher = {Association for Computational Linguistics},
  doi       = {10.18653/v1/2024.findings-acl.456},
  url       = {https://aclanthology.org/2024.findings-acl.456/}
}

@article{koilia2024hardware,
  author  = {Nikoletta Koilia and Christoforos Kachris},
  title   = {Hardware Acceleration of {LLM}s: A Comprehensive Survey and Comparison},
  journal = {arXiv preprint arXiv:2409.03384},
  year    = {2024},
  doi     = {10.48550/arXiv.2409.03384},
  url     = {https://arxiv.org/abs/2409.03384}
}

@article{li2024hardware,
  author  = {Jinhao Li and et al.},
  title   = {Large Language Model Inference Acceleration: A Comprehensive Hardware Perspective},
  journal = {arXiv preprint arXiv:2410.04466},
  year    = {2024},
  doi     = {10.48550/arXiv.2410.04466},
  url     = {https://arxiv.org/abs/2410.04466}
}

@techreport{kitchenham2007guidelines,
  author      = {Barbara Kitchenham and Stuart Charters},
  title       = {Guidelines for Performing Systematic Literature Reviews in Software Engineering},
  institution = {Keele University and Durham University},
  type        = {EBSE Technical Report},
  number      = {EBSE-2007-01},
  year        = {2007}
}

@article{page2021prisma,
  author  = {Matthew J. Page and Joanne E. McKenzie and Patrick M. Bossuyt and Isabelle Boutron and Tammy C. Hoffmann and Cynthia D. Mulrow and Larissa Shamseer and Jennifer M. Tetzlaff and Elie A. Akl and Sue E. Brennan and others},
  title   = {The {PRISMA} 2020 Statement: An Updated Guideline for Reporting Systematic Reviews},
  journal = {BMJ},
  volume  = {372},
  pages   = {n71},
  year    = {2021},
  doi     = {10.1136/bmj.n71}
}

@inproceedings{strubell2019energy,
  author    = {Emma Strubell and Ananya Ganesh and Andrew McCallum},
  title     = {Energy and Policy Considerations for Deep Learning in NLP},
  booktitle = {Proceedings of the 57th Annual Meeting of the Association for Computational Linguistics (ACL)},
  pages     = {3645--3650},
  year      = {2019},
  publisher = {Association for Computational Linguistics},
  doi        = {10.18653/v1/P19-1355},
  url        = {https://doi.org/10.18653/v1/P19-1355}
}

\newpage
\appendix

\section{Data Extraction}
We present the analysis of data from the final set of 123
primary studies. This section outlines the extracted data, providing a structured basis for analysis and enabling comparisons across all studies. 

\vspace{-0.4cm}
\subsection{Optimization Strategy}
\subsubsection{Inference Impact}

We categorize the reviewed studies based on the inference stage affected by the reported optimization strategy. Table~\ref{tab:data_stage_impact} summarizes the stage impact categories considered across the selected studies. It can be observed that the majority of studies focus on optimizations affecting both the prefill and decode stages, reflecting the growing emphasis on end-to-end inference efficiency in LLM systems. Several studies specifically target decode-stage optimization because of its significant contribution to inference latency and energy consumption during token generation. In contrast, only a limited number of studies focus exclusively on prefill-stage optimization, which is primarily associated with prompt processing and context encoding overheads. Additionally, some of the studies that are categorized as “Not Specified” do not explicitly identify whether the proposed optimization affects the prefill stage, decode stage, or both.

\begin{table*}[h!]
\centering
\caption{Stage impact categories in the reviewed studies}
\label{tab:data_stage_impact}
\small

\renewcommand{\arraystretch}{1.01}

\begin{tabular}{p{1.8cm}p{11.5cm}p{0.5cm}}
\hline

\textbf{Stage} &
\textbf{Primary Studies (PS)} &
\textbf{Total} \\
\hline

Prefill Only &
\cite{solovyeva2026towards}, \cite{mandusia2025green},
\cite{shi2025systematic},
\cite{walkowiak2025energy},  \cite{wang2025decoupled}, \cite{jayanth2026fast}, \cite{feng2026ellm},
\cite{hoxha2025deployment}
& 8 \\
\hline

Decode Only &
\cite{masciari2026toon}, \cite{guo2026effectiveness},  \cite{fu2025moa}, \cite{nik2026impact}, 
\cite{qin2024mjsd},
\cite{delavande2026understandingefficiency},
\cite{tian2026cache}, \cite{mariaraj2025smart}, \cite{ziller2026greenserv},
\cite{chen2026onewlaw},
\cite{xue2025hybridserve_icdcs}, \cite{gar2026carbonaware}, \cite{zhang2025energy}, \cite{ji2026flash},
\cite{liu2025greenllm}, 
\cite{wu2026crosskv},
\cite{wang2025wages},  \cite{zhang2024mixpe}, \cite{xu2025unicaim}, \cite{wu2025kllm}, \cite{kim20254}, \cite{xu2024llamaf}, \cite{scherer2024deeploy}, \cite{appuswamy2024northpole}, \cite{fan2025sparse}, \cite{yu2025voltanallm}, \cite{khan2025optimizing}  & 27 \\
\hline

Both &
\cite{
samsi2023words_to_watts,
niu2025tokenpowerbench,
stojkovic2024towards,
verma2025energy,
tanaka2024adaptive,
podder2026green,
shriaarthy2025eco,
rajput2024benchmarking,
abstreiter2026sometimes,
patel2024splitwise,
zhang2024carbon,
stojkovic2025dynamollm,
shi2025disaggregated,
fernandez2025energyconsiderations,
yuvarani2026switching,
almeida2025idle,
lee2025tokens,
potatmitis-etal-2025-cache-saver,
qin2025dynamic,
wiesner2025carbon,
maliakel2025characterizing,
basit2026biscale,
he2025freesh,
kakolyris2025throttllem,
wheeler-etal-2025-reducing,
wang2025storellm,
ps2024inference,
kakolyris2024slo,
li2025cim,
singh2025scalable,
wang2025low,
ortega2024pim,
li2025immutable,
zhao2023sustainable,
zhu2025leveraging,
potkins2024improve,
xu2026daypq,
yubeaton2025huff,
juneja2026halo,
moradifirouzabadi2025end,
lee2025prime,
qiao2026tellme,
hooper2025fgmp,
fu2025h,
he2024hlstransform,
yin2025tereffic,
wang2026harmonia,
liang2026accllm,
crafton2025finding,
monteiro2026reducing,
wang2025mcbp,
wang2025lad,
du2024energy,
fang2025anda,
kang2025dynamic,
jiang2026rapid,
fan2025ellie,
xu2026spactndp,
xu2025camel,
yang2025hades,
jareno2026energyefficient,
wang2025leap,
moon2025hybe,
everman2023carbon,
wang2025asymserve,
moore2025slit,
xu2025llmnpu,
xia2025kelle,
li2025ecoserve,
lu2025thermalaware,
li2024sprout,
kalzhan2025energyefficient,
husom2025sustainable,
argerich2024measuring,
chadli2025pearl,
peng2024m2cache,
raha2025llmnpu,
tanaka2026aellm,
belkahla2025measurements,
qiao2025tellme,
xie2025energyaware,
li2026ecothink,
nguyen2025towards,
2025promptstopower}
& 84 \\
\hline

Not Specified &
\cite{ardakani2025llmpi},\cite{chen2025reis},  \cite{gosmar2026prompt}, \cite{hisaharo2024optimizing}
& 4 \\
\hline

\end{tabular}
\end{table*}

\subsection{Model}

\subsubsection{Model Category}

Table~\ref{tab:data_model_categories} presents the model categories considered across the reviewed literature. To enable systematic analysis, the models are categorized into small ($\leq$7B), medium (8--30B), and large ($>$30B) parameter-scale groups. Studies that evaluate models across multiple parameter-scale categories are classified as mixed-scale studies to accurately reflect the scope of their evaluation. Studies that do not explicitly report model sizes and those based on unspecified architectures are categorized as unspecified. 

\begin{table*}[h!]
\centering
\caption{Model categories evaluated in the reviewed studies}
\label{tab:data_model_categories}
\small
\renewcommand{\arraystretch}{1.02}

\begin{tabular}{p{2.3cm}p{11cm}p{0.5cm}}
\hline

\textbf{Model Category} &
\textbf{Primary Studies (PS)} &
\textbf{Total} \\
\hline

Small ($\leq$7B) &
\cite{solovyeva2026towards,
shriaarthy2025eco,
mandusia2025green,
rajput2024benchmarking, 
zhang2024carbon, 
yuvarani2026switching, 
wang2025storellm, 
kakolyris2024slo, 
chen2025reis,
potkins2024improve, 
juneja2026halo,
lee2025prime,
xu2025unicaim,
wu2025kllm,
he2024hlstransform,
jayanth2026fast,
wang2026harmonia,
liang2026accllm,
crafton2025finding,
kim20254,
du2024energy,
fang2025anda, 
jiang2026rapid, 
xu2026spactndp, 
xu2025camel,
yang2025hades,moon2025hybe, everman2023carbon,xu2025llmnpu, kalzhan2025energyefficient, 
husom2025sustainable, 
chadli2025pearl, 
qiao2025tellme,
xie2025energyaware, 
nguyen2025towards}
& 35 \\
\hline

Medium (8--30B) &
\cite{guo2026effectiveness}, 
\cite{qin2025dynamic}, \cite{li2025cim}, \cite{ji2026flash}, 
\cite{liu2025greenllm}, \cite{kang2025dynamic}, \cite{lu2025thermalaware}
& 7 \\
\hline

Large ($>$30B) &
\cite{patel2024splitwise}, \cite{potatmitis-etal-2025-cache-saver},
\cite{stojkovic2024towards},
\cite{li2025immutable}, \cite{zhao2023sustainable},
\cite{he2025freesh},
 \cite{basit2026biscale},  \cite{hoxha2025deployment} 
& 8 \\
\hline

Mixed-scale Models &

\cite{
masciari2026toon,
podder2026green,
wu2026crosskv,
fu2025moa,
nik2026impact,
qin2024mjsd,
abstreiter2026sometimes,
walkowiak2025energy,
tian2026cache,
stojkovic2025dynamollm,
mariaraj2025smart,
shi2025disaggregated,
fernandez2025energyconsiderations,
ziller2026greenserv,
almeida2025idle,
lee2025tokens,
wiesner2025carbon,
wang2025wages,
maliakel2025characterizing,
xue2025hybridserve_icdcs,
kakolyris2025throttllem,
wheeler-etal-2025-reducing,
gar2026carbonaware,
samsi2023words_to_watts,
ps2024inference,
zhang2025energy,
singh2025scalable,
wang2025low,
ortega2024pim,
zhu2025leveraging,
xu2026daypq,
yubeaton2025huff,
zhang2024mixpe,
moradifirouzabadi2025end,
qiao2026tellme,
hooper2025fgmp,
fu2025h,
yin2025tereffic,
monteiro2026reducing,
wang2025mcbp,
xu2024llamaf,
scherer2024deeploy,
fan2025ellie,
feng2026ellm,
appuswamy2024northpole,
jareno2026energyefficient,
wang2025leap,
fan2025sparse,
wang2025asymserve,
moore2025slit,
xia2025kelle,
niu2025tokenpowerbench,
yu2025voltanallm,
li2025ecoserve,
gosmar2026prompt,
li2024sprout,
khan2025optimizing,
shi2025systematic,
argerich2024measuring,
delavande2026understandingefficiency,
peng2024m2cache,
raha2025llmnpu,
tanaka2026aellm,
belkahla2025measurements,
chen2026onewlaw,
li2026ecothink,
2025promptstopower}
& 67 \\
\hline

Unspecified &
\cite{verma2025energy}, \cite{tanaka2024adaptive}, 
\cite{ardakani2025llmpi},
\cite{wang2025decoupled}, \cite{wang2025lad},  \cite{hisaharo2024optimizing}
& 6 \\
\hline

\end{tabular}
\end{table*}

\subsubsection{Model Family}
To understand the diversity of models evaluated in the selected studies, 
Table~\ref{tab:data_model_groups} lists the model groups considered across the reviewed literature. The models are categorized according to their architectural family or primary design characteristics, as reported by the original authors, to accurately reflect the experimental settings used in each work. Studies that do not explicitly report the evaluated models or rely on unspecified architectures are categorized as unspecified. 

\begin{table*}[h!]
\centering
\caption{Model groups evaluated in the reviewed studies}
\label{tab:data_model_groups}
\small
\renewcommand{\arraystretch}{1.02}

\begin{tabular}{p{3.8cm}p{9.6cm}p{0.4cm}}
\hline

\textbf{Model Group} &
\textbf{Primary Studies} &
\textbf{Total} \\
\hline

LLaMA-family &
\cite{
verma2025energy,
solovyeva2026towards,
wu2026crosskv,
nik2026impact,
qin2024mjsd,
abstreiter2026sometimes,
patel2024splitwise,
ps2024inference,
mariaraj2025smart,
fernandez2025energyconsiderations,
liu2025greenllm,
wiesner2025carbon,
basit2026biscale,
hoxha2025deployment,
gar2026carbonaware,
li2025cim,
ortega2024pim,
zhao2023sustainable,
potkins2024improve,
zhang2024mixpe,
xu2025unicaim,
wu2025kllm,
wang2026harmonia,
wang2025mcbp,
fang2025anda,
scherer2024deeploy,
xu2025camel,
moon2025hybe,
stojkovic2024towards,
yu2025voltanallm,
li2024sprout,
argerich2024measuring,
raha2025llmnpu,
chen2026onewlaw,
nguyen2025towards}
& 35 \\
\hline

Qwen-family &
 
\cite{podder2026green, 
nik2026impact,
zhang2024carbon,
wang2025wages, 
samsi2023words_to_watts,
jayanth2026fast, 
fan2025sparse} 
& 7 \\
\hline

Mistral / Mixtral &
 \cite{shi2025disaggregated, 
 wheeler-etal-2025-reducing, juneja2026halo,
 shi2025systematic}
& 4 \\
\hline

Code-specialized LLMs &
\cite{mandusia2025green, yuvarani2026switching, 
kakolyris2024slo, 
hooper2025fgmp, 
wang2025leap}
& 5 \\
\hline

GPT / OPT / BLOOM / Falcon &
\cite{
masciari2026toon,
guo2026effectiveness,
fu2025moa,
tian2026cache,
ziller2026greenserv,
qin2025dynamic,
xue2025hybridserve_icdcs,
zhang2025energy,
ji2026flash,
yin2025tereffic,
xu2024llamaf,
appuswamy2024northpole,
lu2025thermalaware,
qiao2025tellme}
& 14 \\
\hline

BERT / T5 / Encoder-style &
\cite{
tanaka2024adaptive,
shriaarthy2025eco,
walkowiak2025energy,
lee2025tokens,
kakolyris2025throttllem,
li2025immutable,
lee2025prime,
kim20254,
feng2026ellm,
xu2025llmnpu,
delavande2026understandingefficiency}
& 11 \\
\hline

Gemma-family &
\cite{
maliakel2025characterizing,
du2024energy,
xia2025kelle,
tanaka2026aellm}
& 4 \\
\hline

Phi-family &
\cite{
he2025freesh,
yubeaton2025huff,
kang2025dynamic,
li2025ecoserve}
& 4 \\
\hline

Multimodal / Vision-Language &
\cite{
rajput2024benchmarking,
potatmitis-etal-2025-cache-saver,
wang2025decoupled,
wang2025lad,
husom2025sustainable}
& 5 \\
\hline

BitNet / Binary-style &
\cite{
wang2025storellm,
he2024hlstransform,
jareno2026energyefficient,
xie2025energyaware}
& 4 \\
\hline

Hardware-specific Transformers &
\cite{
stojkovic2025dynamollm,
singh2025scalable,
crafton2025finding}
& 3 \\
\hline

Generic Transformer &
\cite{
almeida2025idle,
wang2025low,
xu2026daypq,
qiao2026tellme,
monteiro2026reducing,
jiang2026rapid,
yang2025hades,
moore2025slit,
kalzhan2025energyefficient,
belkahla2025measurements,
nguyen2025towards,
2025promptstopower}
& 12 \\
\hline

Unspecified &
\cite{
niu2025tokenpowerbench,
chen2025reis,
zhu2025leveraging,
moradifirouzabadi2025end,
fu2025h,
liang2026accllm,
ardakani2025llmpi,
fan2025ellie,
xu2026spactndp,
everman2023carbon,
wang2025asymserve,
gosmar2026prompt,
khan2025optimizing,
chadli2025pearl,
peng2024m2cache,
hisaharo2024optimizing,
li2026ecothink}
& 17 \\
\hline

\end{tabular}
\end{table*}

\vspace{-0.4cm}
\subsection{Evaluation}

\subsubsection{Energy Metrics}
Table~\ref{tab:data_energy_metrics}  summarizes the energy metrics reported across the selected studies. Power consumption metrics capture the instantaneous or average power draw, typically reported in watts, GPU power, or power draw. Energy consumption metrics quantify the cumulative energy usage using units such as joules, watt-hours, kilowatt-hours, or milliwatt-hours. Energy efficiency metrics measure how effectively computational work is performed per unit of energy or power, including tokens/J, performance/W, and throughput/W. Token-level metrics capture energy consumed per generated or processed token, while query-level metrics report energy per request, prompt, response, or inference run. 

Carbon-related metrics estimate the environmental impact using CO$_2$ emissions or carbon footprint. A smaller number of studies consider broader environmental-support metrics such as water usage, Power Usage Effectiveness (PUE)\cite{iso30134pue2026}, and Water Usage Effectiveness (WUE)\cite{iso30134wue2018}. Energy-delay product metrics \cite{gonzalez1996energy} combine energy and latency into a single trade-off measure, while battery-level energy is mainly used in device-level evaluation\cite{banbury2021mlperftiny}. Simulation-based energy estimation is used in studies that rely on analytical models, architectural simulators, or hardware estimation tools rather than direct runtime measurement. Overall, power consumption and energy consumption are the most frequently reported metrics. In contrast, only one paper \cite{fang2025anda} reported battery-level energy, while fine-grained metrics such as energy per query and energy-delay product were reported less frequently.

 \begin{table*}[!t]
\centering
\caption{Categorization of energy metrics reported in the selected studies}
\label{tab:data_energy_metrics}
\small

\renewcommand{\arraystretch}{1.03}

\begin{tabular}{p{3cm}p{10.5cm}p{0.4cm}}
\hline

\textbf{Energy Metric} &
\textbf{Primary Studies} &
\textbf{Total} \\
\hline

Power Consumption &
\cite{
wu2026crosskv,
fu2025moa,
nik2026impact,
abstreiter2026sometimes,
walkowiak2025energy,
patel2024splitwise,
zhang2024carbon,
stojkovic2025dynamollm,
fernandez2025energyconsiderations,
basit2026biscale,
he2025freesh,
kakolyris2025throttllem,
samsi2023words_to_watts,
ps2024inference,
kakolyris2024slo,
wang2025low,
ortega2024pim,
li2025immutable,
ji2026flash,
wang2025decoupled,
potkins2024improve,
yubeaton2025huff,
zhang2024mixpe,
xu2025unicaim,
wu2025kllm,
jayanth2026fast,
yin2025tereffic,
liang2026accllm,
crafton2025finding,
kim20254,
wang2025mcbp,
wang2025lad,
du2024energy,
fang2025anda,
kang2025dynamic,
jiang2026rapid,
fan2025ellie,
feng2026ellm,
xu2025camel,
appuswamy2024northpole,
yang2025hades,
jareno2026energyefficient,
wang2025leap,
moon2025hybe,
fan2025sparse,
wang2025asymserve,
stojkovic2024towards,
niu2025tokenpowerbench,
li2025ecoserve,
lu2025thermalaware,
husom2025sustainable,
shi2025systematic,
belkahla2025measurements,
qiao2025tellme,
hisaharo2024optimizing,
chen2026onewlaw}
& 57 \\
\hline

Energy Consumption &
\cite{
verma2025energy,
tanaka2024adaptive,
solovyeva2026towards,
podder2026green,
guo2026effectiveness,
fu2025moa,
rajput2024benchmarking,
nik2026impact,
qin2024mjsd,
abstreiter2026sometimes,
walkowiak2025energy,
patel2024splitwise,
stojkovic2025dynamollm,
mariaraj2025smart,
fernandez2025energyconsiderations,
ziller2026greenserv,
almeida2025idle,
liu2025greenllm,
lee2025tokens,
potatmitis-etal-2025-cache-saver,
qin2025dynamic,
wiesner2025carbon,
wang2025wages,
maliakel2025characterizing,
basit2026biscale,
he2025freesh,
xue2025hybridserve_icdcs,
kakolyris2025throttllem,
kakolyris2024slo,
li2025cim,
zhang2025energy,
singh2025scalable,
ortega2024pim,
li2025immutable,
zhu2025leveraging,
wang2025decoupled,
juneja2026halo,
moradifirouzabadi2025end,
lee2025prime,
qiao2026tellme,
wu2025kllm,
kim20254,
monteiro2026reducing,
wang2025mcbp,
xu2024llamaf,
ardakani2025llmpi,
jiang2026rapid,
scherer2024deeploy,
fan2025ellie,
xu2026spactndp,
moon2025hybe,
xu2025llmnpu,
yu2025voltanallm,
li2025ecoserve,
lu2025thermalaware,
li2024sprout,
kalzhan2025energyefficient,
khan2025optimizing,
belkahla2025measurements,
2025promptstopower}
& 61 \\
\hline

Energy Efficiency &
\cite{
masciari2026toon,
wu2026crosskv,
nik2026impact,
kakolyris2025throttllem,
zhang2025energy,
singh2025scalable,
wang2025low,
chen2025reis,
zhao2023sustainable,
zhu2025leveraging,
potkins2024improve,
xu2025unicaim,
hooper2025fgmp,
fu2025h,
he2024hlstransform,
jayanth2026fast,
yin2025tereffic,
wang2026harmonia,
liang2026accllm,
crafton2025finding,
monteiro2026reducing,
wang2025mcbp,
wang2025lad,
du2024energy,
xu2024llamaf,
fang2025anda,
kang2025dynamic,
jiang2026rapid,
scherer2024deeploy,
feng2026ellm,
xu2025camel,
appuswamy2024northpole,
jareno2026energyefficient,
wang2025leap,
fan2025sparse,
xia2025kelle,
raha2025llmnpu,
tanaka2026aellm,
qiao2025tellme,
chen2026onewlaw}
& 40 \\
\hline

Energy per token &
\cite{
niu2025tokenpowerbench,
solovyeva2026towards,
mandusia2025green,
fu2025moa,
abstreiter2026sometimes,
walkowiak2025energy,
ziller2026greenserv,
almeida2025idle,
lee2025tokens,
qin2025dynamic,
maliakel2025characterizing,
basit2026biscale,
wang2025storellm,
ortega2024pim,
wu2025kllm,
jiang2026rapid,
everman2023carbon,
wang2025asymserve,
husom2025sustainable,
shi2025systematic,
argerich2024measuring,
delavande2026understandingefficiency,
nguyen2025towards,
2025promptstopower}
& 24 \\
\hline

Energy per request &
\cite{
niu2025tokenpowerbench,
mariaraj2025smart,
yuvarani2026switching,
xue2025hybridserve_icdcs,
hoxha2025deployment,
ortega2024pim,
ji2026flash,
xu2026spactndp,
wang2025asymserve,
kalzhan2025energyefficient,
husom2025sustainable,
chadli2025pearl,
delavande2026understandingefficiency,
xie2025energyaware,
hisaharo2024optimizing,
li2026ecothink,
nguyen2025towards,
2025promptstopower} & 18 \\
\hline

Carbon / CO$_2$ Metrics &
\cite{
masciari2026toon,
podder2026green,
shriaarthy2025eco,
mandusia2025green,
tian2026cache,
zhang2024carbon,
stojkovic2025dynamollm,
mariaraj2025smart,
shi2025disaggregated,
yuvarani2026switching,
almeida2025idle,
potatmitis-etal-2025-cache-saver,
wiesner2025carbon,
he2025freesh,
hoxha2025deployment,
wheeler-etal-2025-reducing,
gar2026carbonaware,
moon2025hybe,
moore2025slit,
li2024sprout,
li2026ecothink}
& 21 \\
\hline

Water / PUE / WUE &
\cite{
mandusia2025green,
hoxha2025deployment,
moore2025slit,
gosmar2026prompt,
hisaharo2024optimizing}
& 5 \\
\hline

Energy-Delay Product &
\cite{
maliakel2025characterizing,
zhang2025energy,
singh2025scalable,
scherer2024deeploy,
xu2026spactndp,
belkahla2025measurements}
& 6 \\
\hline

Battery / Device Energy &
\cite{fang2025anda}
& 1 \\
\hline

Simulation-based Estimation &
\cite{
samsi2023words_to_watts,
singh2025scalable,
wang2025low,
ortega2024pim,
chen2025reis,
zhao2023sustainable,
xu2026daypq,
qiao2026tellme}
& 8 \\
\hline

\end{tabular}
\end{table*}

\subsubsection{Tools/libraries for Measurement}

To analyze how energy consumption is measured across the selected studies, Table~\ref{tab:data_measurement_tools} summarizes the software libraries, runtime utilities, and hardware-based monitoring tools.
Each tool is listed exactly as reported in the original studies and mapped to the corresponding primary studies.
It can be observed that the majority of studies rely on software-based measurement frameworks, particularly GPU monitoring interfaces such as NVIDIA Management Library (NVML/PyNVML) and command-line utilities such as NVIDIA-SMI. Many studies additionally employ CPU-level power interfaces, including Intel RAPL and Linux perf, while carbon-estimation frameworks such as CodeCarbon and CarbonTracker are also frequently adopted.

\begin{table}[t!]
\centering
\caption{Tools and libraries used for energy measurement}
\label{tab:data_measurement_tools}
\small
\renewcommand{\arraystretch}{1.05}

\begin{tabular}{p{3.5cm}p{10cm}p{0.3cm}}
\hline
\textbf{Tool / Library} & \textbf{Primary Studies} & \textbf{Total} \\
\hline
PyNVML \cite{nvidia_nvml} & 
\cite{
niu2025tokenpowerbench,
verma2025energy,
solovyeva2026towards,
guo2026effectiveness,
fu2025moa,
tian2026cache,
shi2025disaggregated,
fernandez2025energyconsiderations,
yuvarani2026switching,
liu2025greenllm,
lee2025tokens,
maliakel2025characterizing,
basit2026biscale,
xue2025hybridserve_icdcs,
zhang2025energy,
wu2025kllm,
moon2025hybe,
xia2025kelle,
yu2025voltanallm,
li2025ecoserve,
lu2025thermalaware,
shi2025systematic,
argerich2024measuring,
delavande2026understandingefficiency,
tanaka2026aellm,
belkahla2025measurements,
nguyen2025towards,
2025promptstopower} & 28 \\
\hline
NVIDIA-SMI \cite{nvidia_smi} & 
\cite{
wu2026crosskv,
nik2026impact,
qin2024mjsd,
walkowiak2025energy,
stojkovic2025dynamollm,
maliakel2025characterizing,
he2025freesh,
kakolyris2024slo,
li2025cim,
li2025immutable,
he2024hlstransform,
wang2025leap,
wang2025asymserve,
li2024sprout,
argerich2024measuring}
& 15 \\
\hline
CodeCarbon \cite{courty2024codecarbon} &
\cite{
masciari2026toon,
shriaarthy2025eco,
mandusia2025green,
rajput2024benchmarking,
fernandez2025energyconsiderations,
wheeler-etal-2025-reducing,
moradifirouzabadi2025end,
wu2025kllm,
delavande2026understandingefficiency,
li2026ecothink,
2025promptstopower}
& 11 \\
\hline
Intel RAPL \cite{david2010rapl} & 
\cite{
verma2025energy,
guo2026effectiveness,
tian2026cache,
niu2025tokenpowerbench,
li2025ecoserve,
argerich2024measuring,
delavande2026understandingefficiency}
& 7 \\
\hline
CarbonTracker \cite{anthony2020carbontracker} & 
\cite{potatmitis-etal-2025-cache-saver, li2024sprout} & 2 \\
\hline
ZeusMonitor \cite{you2023zeus} & \cite{ziller2026greenserv}  & 1 \\
\hline
DGCM \cite{nvidia_dcgm} & 
\cite{nik2026impact, wang2025asymserve, niu2025tokenpowerbench, lu2025thermalaware} & 4 \\
\hline
Linux perf \cite{linux_perf} & \cite{abstreiter2026sometimes,fan2025sparse} & 2 \\
\hline
Monsoon power monitor\cite{monsoon_power_monitor} & \cite{abstreiter2026sometimes} & 1 \\
\hline
macOS Powermetrics \cite{apple_powermetrics} &  \cite{yang2025hades} & 1 \\
\hline
HWiNFO /HWINFO \cite{hwinfo} & \cite{kim20254, raha2025llmnpu} & 2\\
\hline
EcoLogits  & \cite{podder2026green} & 1 \\
\hline
AMD-SMI  & \cite{kang2025dynamic} & 1 \\
\hline
qMeter & \cite{shi2025systematic} & 1 \\
\hline
EnergyMeter & \cite{argerich2024measuring} & 1 \\
\hline
NVIDIA Nsight \cite{nvidia_nsight} & \cite{feng2026ellm} & 1 \\
\hline
Jetson tegrastats \cite{nvidia_tegrastats} & \cite{xie2025energyaware} & 1 \\
\hline
I2C monitoring & \cite{xu2026spactndp} & 1 \\
\hline

Not Reported & Remaining papers & 63 \\
\hline
\end{tabular}
\end{table}

\subsubsection{External Hardware for Energy Monitoring}

Table~\ref{tab:data_external_hardware}  summarizes the external hardware devices used for energy measurement across the selected studies. Only a small number of studies employ dedicated physical power monitoring devices, such as Monsoon High Voltage Power Monitor, Joulescope JS110 Hardware Power Meter, and Keysight DC Power Analyzer, to obtain direct measurements of system-level energy consumption. 
\begin{table}[h!]
\centering
\small
\caption{External hardware devices used for energy measurement}
\label{tab:data_external_hardware}

\renewcommand{\arraystretch}{1.05}

\begin{tabular}{p{7cm}p{2.3cm}p{0.3cm}}
\hline

\textbf{External Hardware} &
\textbf{Primary Studies} &
\textbf{Total} \\
\hline

Monsoon High Voltage Power Monitor\cite{monsoon2023} &
\cite{abstreiter2026sometimes} &
1 \\
\hline

MakerHawk UM34C USB Multimeter\cite{makerhawk_um34c} &
\cite{fang2025anda} &
1 \\
\hline

Keysight N6715C DC Power Analyzer/Supply\cite{keysight_n6715c} &
\cite{jiang2026rapid} &
1 \\
\hline

External Hardware Power Meter &
\cite{scherer2024deeploy} &
1 \\
\hline

Q-Meter &
\cite{shi2025systematic} &
1 \\
\hline

Joulescope JS110 Hardware Power Meter\cite{joulescope_js110} &
\cite{husom2025sustainable} &
1 \\
\hline 

Not Reported &
Remaining papers &
117 \\
\hline

\end{tabular}
\end{table}

\subsubsection{Benchmark Analysis}

We categorize the evaluation data based on the type of benchmark or dataset used in the selected studies. Table~\ref{tab:data_dataset_groups} summarizes the dataset groups considered across the reviewed studies. Studies categorized as “Unspecified” do not clearly report the dataset, benchmark, or evaluation workload used.

\begin{table*}[!t]
\centering
\caption{Dataset groups used in the reviewed studies}
\label{tab:data_dataset_groups}
\small

\renewcommand{\arraystretch}{1.05}

\begin{tabular}{p{5cm}p{8.5cm}p{0.4cm}}
\hline

\textbf{Dataset Group} &
\textbf{Primary Studies} &
\textbf{Total} \\
\hline

QA / Reasoning Benchmarks &
\cite{
verma2025energy,
walkowiak2025energy,
tian2026cache,
ziller2026greenserv,
qin2025dynamic,
maliakel2025characterizing,
wheeler-etal-2025-reducing,
gar2026carbonaware,
chen2025reis,
zhu2025leveraging,
xu2026daypq,
moradifirouzabadi2025end,
lee2025prime,
qiao2026tellme,
fu2025h,
liang2026accllm,
monteiro2026reducing,
wang2025mcbp,
du2024energy,
fang2025anda,
fan2025ellie,
everman2023carbon,
wang2025asymserve,
xia2025kelle,
li2024sprout,
chadli2025pearl,
raha2025llmnpu,
belkahla2025measurements,
li2026ecothink}
& 29 \\
\hline

General NLP Benchmarks &
\cite{
tanaka2024adaptive,
mandusia2025green,
rajput2024benchmarking,
nik2026impact,
xue2025hybridserve_icdcs,
wang2025storellm,
li2025cim,
zhang2025energy,
wang2025low,
ortega2024pim,
wang2025decoupled,
potkins2024improve,
zhang2024mixpe,
juneja2026halo,
xu2025unicaim,
jayanth2026fast,
wang2026harmonia,
xu2024llamaf,
appuswamy2024northpole,
yang2025hades,
hisaharo2024optimizing}
& 21 \\
\hline

Production Traces / Real-world Logs &
\cite{
patel2024splitwise,
zhang2024carbon,
stojkovic2025dynamollm,
almeida2025idle,
liu2025greenllm,
lee2025tokens,
wiesner2025carbon,
wang2025wages,
basit2026biscale,
he2025freesh,
hoxha2025deployment,
kakolyris2025throttllem,
kakolyris2024slo,
scherer2024deeploy,
feng2026ellm,
moore2025slit,
yu2025voltanallm,
chen2026onewlaw}
& 18 \\
\hline

Code Generation / Programming Datasets &
\cite{
solovyeva2026towards,
wu2026crosskv,
qin2024mjsd,
shi2025disaggregated,
fernandez2025energyconsiderations,
potatmitis-etal-2025-cache-saver,
xu2025camel,
husom2025sustainable,
shi2025systematic,
peng2024m2cache,
tanaka2026aellm}
& 11 \\
\hline

Custom / Synthetic / Prompt-based Dataset &
\cite{
podder2026green,
shriaarthy2025eco,
yuvarani2026switching,
moon2025hybe,
lu2025thermalaware,
gosmar2026prompt,
argerich2024measuring,
delavande2026understandingefficiency,
qiao2025tellme,
2025promptstopower}
& 10 \\
\hline

Long-context Evaluation Datasets &
\cite{
niu2025tokenpowerbench,
hooper2025fgmp,
yin2025tereffic,
kang2025dynamic,
xu2025llmnpu,
li2025ecoserve}
& 6 \\
\hline

Vision / Multimodal Datasets &
 \cite{zhao2023sustainable,xie2025energyaware}
& 2 \\
\hline

Summarization / Text Generation Datasets &
\cite{fu2025moa}
& 1 \\
\hline

Unspecified &
\cite{
samsi2023words_to_watts,
stojkovic2024towards,
masciari2026toon,
guo2026effectiveness,
mariaraj2025smart,
ps2024inference,
singh2025scalable,
li2025immutable,
ji2026flash,
yubeaton2025huff,
wu2025kllm,
he2024hlstransform,
crafton2025finding,
kim20254,
wang2025lad,
ardakani2025llmpi,
jiang2026rapid,
xu2026spactndp,
jareno2026energyefficient,
wang2025leap,
fan2025sparse,
kalzhan2025energyefficient,
khan2025optimizing,
nguyen2025towards,
abstreiter2026sometimes}
& 25 \\
\hline

\end{tabular}
\end{table*}

\subsection{Trade-off Analysis}

We categorize the studies that primarily investigate the relationship between energy consumption and other performance, system, and sustainability objectives. Table \ref{tab:data_tradeoff_categories} lists the trade-off categories identified in the literature. The reviewed studies commonly evaluate trade-offs involving accuracy, latency, throughput, hardware complexity, memory overhead, and environmental impact. In particular, studies focusing on deployment- and hardware-level optimization frequently evaluate the balance between energy efficiency and resource utilization metrics. Notably, the majority of studies focus on balancing energy efficiency with inference latency and model accuracy, reflecting the rising focus on achieving efficient LLM inference without significantly degrading performance. Among these categories, energy–latency and energy–accuracy trade-offs are the most frequently evaluated, whereas comparatively fewer studies investigate trade-offs involving throughput, memory and hardware resources, or environmental impact.

\begin{table*}[h!]
\centering
\small
\caption{Trade-off categories in the reviewed studies}
\label{tab:data_tradeoff_categories}

\renewcommand{\arraystretch}{1.0}

\begin{tabular}{p{3.8cm}p{9.7cm}p{0.4cm}}
\hline

\textbf{Trade-off} &
\textbf{Primary Studies (PS)} &
\textbf{Total} \\
\hline

Energy vs Accuracy &
\cite{
masciari2026toon,
verma2025energy,
tanaka2024adaptive,
solovyeva2026towards,
podder2026green,
guo2026effectiveness,
shriaarthy2025eco,
wu2026crosskv,
mandusia2025green,
fu2025moa,
rajput2024benchmarking,
nik2026impact,
abstreiter2026sometimes,
mariaraj2025smart,
ziller2026greenserv,
yuvarani2026switching,
potatmitis-etal-2025-cache-saver,
qin2025dynamic,
wiesner2025carbon,
maliakel2025characterizing,
xue2025hybridserve_icdcs,
wheeler-etal-2025-reducing,
gar2026carbonaware,
wang2025storellm,
chen2025reis,
wang2025decoupled,
potkins2024improve,
yubeaton2025huff,
zhang2024mixpe,
juneja2026halo,
moradifirouzabadi2025end,
lee2025prime,
xu2025unicaim,
qiao2026tellme,
hooper2025fgmp,
fu2025h,
he2024hlstransform,
yin2025tereffic,
wang2026harmonia,
liang2026accllm,
crafton2025finding,
monteiro2026reducing,
wang2025mcbp,
wang2025lad,
du2024energy,
xu2024llamaf,
fang2025anda,
fan2025ellie,
moon2025hybe,
everman2023carbon,
xu2025llmnpu,
xia2025kelle,
li2024sprout,
kalzhan2025energyefficient,
khan2025optimizing,
husom2025sustainable,
shi2025systematic,
argerich2024measuring,
chadli2025pearl,
raha2025llmnpu,
tanaka2026aellm,
belkahla2025measurements,
xie2025energyaware,
hisaharo2024optimizing,
li2026ecothink}
& 65 \\
\hline

Energy vs Latency &
\cite{
niu2025tokenpowerbench,
stojkovic2024towards,
verma2025energy,
tanaka2024adaptive,
guo2026effectiveness,
shriaarthy2025eco,
wu2026crosskv,
mandusia2025green,
nik2026impact,
qin2024mjsd,
abstreiter2026sometimes,
walkowiak2025energy,
patel2024splitwise,
tian2026cache,
zhang2024carbon,
stojkovic2025dynamollm,
mariaraj2025smart,
shi2025disaggregated,
fernandez2025energyconsiderations,
yuvarani2026switching,
almeida2025idle,
liu2025greenllm,
lee2025tokens,
potatmitis-etal-2025-cache-saver,
wang2025storellm,
kakolyris2024slo,
li2025cim,
zhang2025energy,
wang2025low,
ortega2024pim,
chen2025reis,
li2025immutable,
zhao2023sustainable,
ji2026flash,
wang2025decoupled,
xu2026daypq,
yubeaton2025huff,
moradifirouzabadi2025end,
lee2025prime,
hooper2025fgmp,
fu2025h,
wu2025kllm,
he2024hlstransform,
jayanth2026fast,
kim20254,
monteiro2026reducing,
wang2025mcbp,
du2024energy,
xu2024llamaf,
fang2025anda,
ardakani2025llmpi,
kang2025dynamic,
jiang2026rapid,
scherer2024deeploy,
fan2025ellie,
feng2026ellm,
xu2026spactndp,
xu2025camel,
appuswamy2024northpole,
yang2025hades,
wang2025leap,
everman2023carbon,
fan2025sparse,
wang2025asymserve,
moore2025slit,
xu2025llmnpu,
xia2025kelle,
yu2025voltanallm,
li2025ecoserve,
gosmar2026prompt,
husom2025sustainable,
shi2025systematic,
argerich2024measuring,
chadli2025pearl,
delavande2026understandingefficiency,
peng2024m2cache,
raha2025llmnpu,
tanaka2026aellm,
belkahla2025measurements,
qiao2025tellme,
xie2025energyaware,
hisaharo2024optimizing,
nguyen2025towards,
2025promptstopower}
& 85
\\
\hline

Energy vs Throughput / Perf. &
\cite{
niu2025tokenpowerbench,
stojkovic2024towards,
masciari2026toon,
fu2025moa,
rajput2024benchmarking,
abstreiter2026sometimes,
patel2024splitwise,
fernandez2025energyconsiderations,
lee2025tokens,
ps2024inference,
singh2025scalable,
li2025immutable,
zhao2023sustainable,
ji2026flash,
yubeaton2025huff,
zhang2024mixpe,
lee2025prime,
xu2025unicaim,
hooper2025fgmp,
fu2025h,
wu2025kllm,
he2024hlstransform,
wang2026harmonia,
monteiro2026reducing,
wang2025lad,
jiang2026rapid,
jareno2026energyefficient,
fan2025sparse,
li2025ecoserve,
lu2025thermalaware,
delavande2026understandingefficiency,
belkahla2025measurements,
qiao2025tellme,
hisaharo2024optimizing,
chen2026onewlaw,
nguyen2025towards,
2025promptstopower}
& 37 \\
\hline

Energy vs Environmental Impact&
\cite{
niu2025tokenpowerbench,
masciari2026toon,
shriaarthy2025eco,
mandusia2025green,
tian2026cache,
zhang2024carbon,
stojkovic2025dynamollm,
shi2025disaggregated,
almeida2025idle,
potatmitis-etal-2025-cache-saver,
wiesner2025carbon,
he2025freesh,
hoxha2025deployment,
wheeler-etal-2025-reducing,
gar2026carbonaware,
wang2025lad,
moon2025hybe,
moore2025slit,
li2025ecoserve,
gosmar2026prompt,
li2024sprout,
khan2025optimizing,
peng2024m2cache,
nguyen2025towards}
& 24 \\
\hline

Energy vs Memory / Hardware &
\cite{
tanaka2024adaptive,
fu2025moa,
shi2025disaggregated,
fernandez2025energyconsiderations,
qin2025dynamic,
ps2024inference,
li2025cim,
singh2025scalable,
wang2025low,
zhao2023sustainable,
zhu2025leveraging,
wang2025decoupled,
potkins2024improve,
xu2026daypq,
yubeaton2025huff,
juneja2026halo,
lee2025prime,
qiao2026tellme,
jayanth2026fast,
wang2026harmonia,
liang2026accllm,
crafton2025finding,
kim20254,
monteiro2026reducing,
ardakani2025llmpi,
jiang2026rapid,
fan2025ellie,
feng2026ellm,
appuswamy2024northpole,
jareno2026energyefficient,
wang2025leap,
lu2025thermalaware,
kalzhan2025energyefficient,
argerich2024measuring,
peng2024m2cache,
tanaka2026aellm,
chen2026onewlaw,
2025promptstopower}
& 38 \\
\hline

\end{tabular}
\end{table*}

%To enable consistent comparison, the evaluated models are grouped into categories such as LLaMA-family, Qwen-family, Mistral/Mixtral-family, code-specialized LLMs, encoder-style transformers, multimodal models, and hardware-efficient transformer variants. Some studies evaluate multiple model groups and are therefore associated with more than one category to accurately reflect their evaluation scope. 
%This classification provides a clearer view of the model trends explored in the literature and highlights the wide range of architectures investigated for improving energy-efficient LLM inference.

\end{document}